\documentclass[12pt]{JHEP3}
\usepackage{mathrsfs}
\usepackage{amsmath,amssymb}
\usepackage{epsfig}
\usepackage{relsize}
\usepackage{graphicx}
\input epsf

\def\x'{\mathaccent 19 x}
\def\y'{\mathaccent 19 y}
\def\n'{\mathaccent 19 n}
\def\u'{\mathaccent 19 u}

\def\et'{\mathaccent 19 \eta}
\def\th'{\mathaccent 19 \theta}
\def\lam'{\mathaccent 19 \lambda}
\def\varet'{\mathaccent 19 \vartheta}
\def\rh'{\mathaccent 19 \rho}
\def\ph'{\mathaccent 19 \phi}
\def\xb'{\mathaccent 19 {\bar{x}}}

\def\tH{\widetilde H}
\def\bQ{\overline{Q}}
\def\bC{C^\dagger}

\def\hP{\hat{P}}

\def\cA{{\cal A}}

\def\l{{\lambda}}

\def\sl(2){\alg{sl}(2)}

\def\be{\begin{equation}}
\def\ee{\end{equation}}

\newcommand{\bea}{\begin{eqnarray}}
\newcommand{\eea}{\end{eqnarray}}

\def\a {\alpha}
\def\b {\beta}
\def\s {\sigma}
\def\pa {\partial}
\def\P {\mathscr{P}}
\def\Q {\mathscr{Q}}
\def\g {\gamma}
\def\om {\omega}

\def\la{\label}
\def\e{\epsilon}
\def\ov{\over}

\def\Tr{\text{Tr}}

\def\S{\Sigma}
\def\H{{\cal H}}

\def\dx{\dot{x}}

\def\hP{{\bf P}}

\def\bJ{{\bf J}}
\def\bJ{{\bf J}}

\def\bR{{\bf R}}
\def\bL{{\bf L}}
\def\bQ{{\bf Q}}
\def\tbQ{{\widetilde{ \bf Q}}}
\def\bC{{\bf C}}

\def\bH{{\bf H}}
\def\tbH{{\widetilde {\bf H}}}
\def\tp{{\widetilde p}}

\def\vp{\varphi}

\newcommand{\alg}[1]{\mathfrak{#1}}
\newcommand{\su}{\alg{su}}

\newcommand{\psu}{\alg{psu}}

\newcommand{\AdS}{{\rm  AdS}_5\times {\rm S}^5}
\newcommand{\ads}{{\rm  AdS}_5\times {\rm S}^5}

\def\L{\mathscr L}

\newcommand{\atopfrac}[2]{\genfrac{}{}{0pt}{}{#1}{#2}}
\newcommand{\sfrac}[2]{{\textstyle\frac{#1}{#2}}}

\newcommand{\bem}{\left (\begin{matrix}}
\newcommand{\eem}{\end{matrix} \right )}
\newcommand{\Str}{\mbox{Str}}

\def\S{{\cal S}}
\def\P{{\cal P}}
\def\Q{{\cal Q}}
\def\I{{\cal I}}

\newcommand{\sn}{\mathop{\mathrm{sn}}\nolimits}
\newcommand{\cn}{\mathop{\mathrm{cn}}\nolimits}
\newcommand{\dn}{\mathop{\mathrm{dn}}\nolimits}
\def\am{\mathrm{am}}

\author{Gleb Arutyunov$^a$\footnote{Email: G.Arutyunov@phys.uu.nl, frolovs@maths.tcd.ie} {}\footnote{Correspondent fellow at Steklov
Mathematical Institute, Moscow.} \  and Sergey Frolov$^{b\, \dagger}$
 \\ $^{a}$ {\it Institute for Theoretical
Physics and Spinoza Institute,\\ ~~Utrecht University, 3508 TD
Utrecht, The Netherlands} \\ $^b$ {\it School of Mathematics,
Trinity College, Dublin 2, Ireland}}

\abstract{The study of finite $J$ effects for the light-cone
$\AdS$ superstring by means of the Thermodynamic Bethe Ansatz
requires an understanding of a companion 2d theory which we call
the mirror model. It is obtained from the original string model by
the double Wick rotation. The S-matrices describing the scattering
of physical excitations in the string and mirror models are
related to each other by an analytic continuation. We show that
the unitarity requirement for the mirror S-matrix fixes the
S-matrices of both theories essentially uniquely. The resulting
string S-matrix $S(z_1,z_2)$ satisfies the generalized unitarity
condition and, up to a scalar factor, is a meromorphic function on
the elliptic curve  associated to each variable $z$. The double
Wick rotation is then accomplished by shifting the variables $z$
by quarter of the imaginary period of the torus. We discuss the
apparent bound states of the string and mirror models, and show
that depending on a choice of the physical region there are one,
two or $2^{M-1}$ solutions of the $M$-particle bound state
equations sharing the same conserved charges. For very large but
finite values of $J$,  most of these solutions, however, exhibit
various signs of pathological behavior. In particular, they might
receive a finite $J$ correction to their energy which is complex,
or the energy correction might exceed corrections arising due to
finite $J$ modifications of the Bethe equations thus making the
asymptotic Bethe ansatz inapplicable. }

\title{
On String S-matrix, Bound States and TBA}

\preprint{
          \smaller{\smaller{\smaller{ITP-UU-07-50}}}\\[-.5ex]
          \smaller{\smaller{\smaller{SPIN-07-37}}}\\[-.5ex]
          \smaller{\smaller{\smaller{TCDMATH 07-15}}}}

\begin{document}

\renewcommand{\thefootnote}{\arabic{footnote}}
\setcounter{footnote}{0}

\section{Introduction and summary}

The conjectured duality between the maximally supersymmetric
Yang-Mills theory in four dimensions and type IIB superstring  in
the $\AdS$ background \cite{M} is the subject of active research.
Integrability emerging on both sides of the gauge/string
correspondence \cite{MZ,BPR} proved to be an indispensable tool in
matching the spectra of gauge and string theories. Namely, it was
shown that the problem of determining the spectra in the large
volume (charge) limit, can be reduced to the problem of solving a
set of algebraic (Bethe) equations. The corresponding Bethe
equations are based on the knowledge of the S-matrix which
describes the scattering of world-sheet excitations of the
gauge-fixed string sigma-model, or alternatively, the excitations
of a certain spin chain in the dual gauge theory
\cite{Kazakov:2004qf}-\cite{B}.

\smallskip

Remarkably, the S-matrix is severely restricted by the requirement
of invariance under the global symmetry of the model, the
centrally extended $\psu(2|2)\oplus \psu(2|2)$ superalgebra
\cite{B,AFPZ}.  The invariance condition  fixes its matrix form
almost uniquely up to an overall phase \cite{B,Bn}. The
constraints on the overall phase were derived in \cite{Janik} by
demanding the S-matrix to satisfy crossing symmetry. Recently, a
physically relevant solution to the crossing relation, which
interpolates between the weak (gauge) and strong (string) coupling
regimes   was conjectured \cite{BHL,BES}, building on the previous
work \cite{Beisert:2005cw}-\cite{AF06}. This solution successfully
passed
 a number of non-trivial tests
\cite{Bern}-\cite{Korch}.

\medskip

So far the main focus of research was on determining the spectrum
of string theory in the limiting case in which at least one of the
global charges carried by a string state (and by the corresponding
gauge theory operator) is large. Our ultimate goal, however, is to
understand how the energies of string states (the conformal
dimensions of  dual gauge theory operators) depend on the coupling
constant for finite values of all the other global symmetry
charges. Although the conjectured S-matrix \cite{BHL,BES} provides
an important starting point in addressing this issue, by now there
is firm evidence that the corresponding Bethe equations \cite{BS}
fail to correctly reproduce the finite-size effects, neither in
string \cite{SZZ} nor in gauge theory \cite{Kotikov:2007cy}.
Indeed, already in the semi-classical string theory deviations
from the controllable exact spectrum arise which are exponentially
small in the effective string length playing the role of a large
symmetry charge. One of the reasons behind this is that the
interactions on the world-sheet are not ultra-local, typically the
scattered states are solitons of finite size \cite{HM,magnon}.
Also, as is common to many field-theoretic models, vacuum
polarization effects smear bare point-like interactions and lead
to exponential corrections to the energy levels in the
large-volume limit \cite{AJK}. Complementary, in the spin chain
description of the dual gauge theory, the obstruction to the
validity of the Bethe ansatz comes from the wrapping effects.
Hence, the Bethe ansatz for the planar AdS/CFT system in its
present form  is merely of  asymptotic type. Obviously, solving a
quantum sigma-model in finite volume is much harder. As an
illustrative example, we mention the Sinh-Gordon model for which
the equations describing the finite-size spectrum have been
recently obtained in \cite{Teschner:2007ng}.

\smallskip

Determination of finite-size effects in the context of integrable
models is a wide area of research.  Basically, there are  three
different but related ways in which this problem could be
addressed: the thermodynamic Bethe ansatz approach (TBA)
\cite{za}, nonlinear integral equations (NLIE)
\cite{Destri:1992qk} and functional relations for commuting
transfer-matrices \cite{Bazhanov:1996aq}.

\smallskip

The aim of the present paper is to investigate the structure of
the string S-matrix which is the necessary step for constructing
the TBA equations for the quantum string sigma-model on $\AdS$
background. This would eventually allow one to describe the
finite-size spectrum of the corresponding model.

\smallskip

The TBA approach was initially developed for studying
thermodynamic properties of non-relativistic quantum mechanics in
one dimension  \cite{Yang:1968rm} and further applied to the
computation of the ground state energy in integrable relativistic
field theories in finite volume \cite{za}. The method also was
later extended to account for energies of excited states
\cite{Dorey:1996re} (see also \cite{Martins:1991hw}).

\smallskip

Implementation of the TBA approach consists of several steps. The
primary goal is to obtain an expression for the ground state
energy of a Lorentzian theory compactified on a circle of
circumference $L$ and at zero temperature. The starting point is
the \emph{Euclidean extension} of the original theory, put on a
torus generated by two orthogonal circles of circumferences $L$
and $R$. The partition function of this theory can be viewed as
originating from two different Lorentzian theories: the
\emph{original} one, which lives on a circle of length $L$ at
temperature $T$ and has the Hamiltonian $H$, or the \emph{mirror}
theory which is defined on a circle of length $R=1/T$ at
temperature $\tilde{T}=1/L$ and has the Hamiltonian $\tilde{H}$.
For Lorentz-invariant theories, the original and the mirror
Hamiltonians are the same. However, in general and in particular
for the case of interest here, the two theories need not be the
same. Taking the thermodynamic limit $R \rightarrow \infty$  one
ends up with the mirror theory on a line and at finite
temperature, for which the exact (mirror) Bethe equations can be
written. Thus, computation of finite-size effects in the original
theory translates into the problem of solving the infinite volume
mirror theory at finite temperature.

\smallskip

Although taking the thermodynamic limit simplifies the system, a
serious complication arises due to the fact that the mirror theory
could have bound states which manifest themselves as poles of the
two-particle mirror S-matrix. Thus, the complete spectrum would
consist of particles and their bound states; the latter should be
thought of as new asymptotic particles. Having identified the
spectrum,  one has to determine the S-matrix which scatters {\it
all} asymptotic particles. It is this S-matrix which should be
used to formulate the system of TBA equations.

\smallskip As is clear from the discussion above, the mirror
theory plays a crucial role in the TBA approach. In this paper we
will analyze the mirror theory in some detail. First, we will
explain its relation to the original theory. Indeed, given that
the model in question is not Lorentz invariant, the mirror and the
original Hamiltonians are not the same. However, since the
dispersion relation and the S-matrix  can be deduced from the
2-point and 4-point correlation functions on the world-sheet, and
since the correlation functions in the mirror theory are inherited
from the original model by performing a double Wick rotation, it
follows that the mirror dispersion relation and the mirror
S-matrix can be obtained from the original ones by the double Wick
rotation. Here we will meet an important subtlety. To explain it,
we first have to recall the basic properties of the string
S-matrix.

\smallskip

As was shown in \cite{AFZ}, the $\psu(2|2)\oplus
\psu(2|2)$-invariant S-matrix ${\cal S}(p_1,p_2)$, which depends
on real momenta $p_1$ and $p_2$ of scattering particles, obeys
\begin{itemize}
\item the Yang-Baxter equation
\begin{equation}
\nonumber
 {\cal S}_{23}{\cal S}_{13}{\cal S}_{12}={\cal S}_{12}{\cal S}_{13}{\cal S}_{23}\,
\end{equation}

\item the unitarity condition
\begin{equation}
\nonumber {\cal S}_{12}(p_1,p_2){\cal S}_{21}(p_2,p_1)={\mathbb
I}\,
\end{equation}

\item the physical unitarity condition
$$
{\cal S}_{12}(p_1,p_2){\cal S}_{12}^{\dagger}(p_1,p_2) ={\mathbb
I}\,
$$

\item the requirement of crossing symmetry
\bea \nonumber \mathscr{C}_1^{-1}{\cal S}_{12}^{t_1}(p_1,p_2)
\mathscr{C}_1
 {\cal S}_{12}(-p_1,p_2)={\mathbb I}\, , \eea
where $\mathscr{C}$ is the charge conjugation matrix.

\end{itemize}
The first three properties naturally follow from the consistency
conditions of the associated Zamolodchikov-Faddeev (ZF) algebra
\cite{Zam,Fad}, while the last one reflects the fact that the
particle-to-anti-particle transformation is an automorphism of the
ZF algebra \cite{AFZ}. The unitarity and physical unitarity
conditions imply the following property
\begin{equation}
\nonumber {\cal S}_{21}(p_2,p_1)= {\cal
S}_{12}^{\dagger}(p_1,p_2)\, .
\end{equation}

\smallskip

One should bear in mind that the S-matrix is defined up to
unitary equivalence only: unitary transformations (depending on
the particle momentum) of a basis of one-particle states
correspond to unitary transformations of the scattering matrix
without spoiling any of the properties listed above.

\smallskip

The mirror S-matrix $\tilde{\cal S}(\tilde{p}_1,\tilde{p_2})$ is
obtained from ${\cal S}(p_1,p_2)$ by the double Wick rotation. The
above-mentioned subtlety lies in the fact that only for {\it a
very special choice} of the one-particle basis the corresponding
mirror S-matrix remains unitary. As we will show, this problem can
be naturally attributed to the properties of the double Wick
rotation for fermionic variables. Upon the basis is properly
chosen to guarantee unitarity of the mirror S-matrix, the only
freedom in the {\it matrix structure} of ${\cal S}(p_1,p_2)$
reduces to constant, i.e. momentum-independent, unitary
transformations\footnote{Of course, there is always a freedom of
multiplying the S-matrix by an overall (momentum-dependent)
phase.}.

\smallskip

There is another interesting explanation of the interrelation
between the original theory and its mirror. As was shown in
\cite{Janik}, the dispersion relation between the energy and
momentum of a single particle can be naturally uniformized in
terms of a complex variable $z$ living on a torus with real and
imaginary periods equal to $2\om_1$ and $2\om_2$, respectively.
Since $z$ plays the role of the generalized rapidity variable, it
is quite natural to think about the S-matrix as the function
${\cal S}(z_1,z_2)$, which for real values of the generalized
rapidity variables coincides with ${\cal S}(p_1,p_2)$. In other
words, the S-matrix admits an analytic continuation to the complex
values of momenta. It appears that the unitary momentum-dependent
freedom in the matrix structure of the S-matrix gets  fixed if we
require the analytic continuation to be compatible with the
requirement of

\begin{itemize}
\item generalized unitarity
$${\cal S}_{12}(z_1^*,z_2^*)\big[{\cal
S}_{12}(z_1,z_2)\big]^{\dagger}={\mathbb I}\, ,$$
\end{itemize}
which can be thought of as the physical unitarity condition
extended to the generalized rapidity torus. The unitarity and the
generalized unitarity further imply
$$
{\cal S}_{21}(z_2^*,z_1^*)= \big[{\cal
S}_{12}(z_1,z_2)\big]^{\dagger}\,
$$
 In fact, the last
equation is equivalent to the standard requirement of hermitian
analyticity for an S-matrix in two-dimensional relativistic
quantum field theories.

\smallskip

Thus, the S-matrix  which admits the analytic continuation to the
generalized rapidity torus compatible with the requirement of
hermitian analyticity is essentially unique. Of course, it
satisfies all the other properties listed above, including
crossing symmetry. As we will show, the mirror S-matrix is
obtained from ${\cal S}(z_1,z_2)$ considered for real values of
$z_1,z_2$ by shifting $z_1,z_2$ by quarter of the imaginary period
$$
\tilde{{\cal S}}(z_1,z_2)={\cal
S}\big(z_1+\sfrac{\om_2}{2},z_2+\sfrac{\om_2}{2}\big)\, .
$$
There is a close analogy with what happens in relativistic models.
In the latter case the physical region is defined as the strip
$0\leq{\rm Im}\,\theta\leq \pi$, where $\theta=\theta_2-\theta_1$
is the rapidity variable. A passage to the mirror theory
corresponds to the shift $\theta_k\to \theta_k+i\frac{\pi}{2}$,
i.e. to the shift by the quarter of imaginary period\footnote{The
shift of $\theta$ by the half-period corresponds to the crossing
transformation.}. Of course, for relativistic models, due to
Lorentz invariance, the S-matrix depends on the difference of
rapidities and, therefore, it remains unchanged under the double
Wick rotation transformation. Also, in our present case the notion
of the physical region is not obvious and its identification
requires further analysis of the analytic properties of the string
S-matrix.

\smallskip

Having identified the mirror S-matrix, we can investigate the
question about the bound states. We first discuss the Bethe equations for
the gauge-fixed string theory where the existence of the BPS bound states
is known \cite{D}. No non-BPS bound states exist, according to \cite{D}-\cite{DHM}.
We find out, however, that  the number of solutions of the BPS bound state equations
depends on the choice of the physical region of the model, and for a given value of the bound state momentum
there could be 1, 2 or $2^{M-1}$ $M$-particle bound states sharing
the same set of global conserved charges.
 It is unclear to us whether this indicates that the
actual physical region is the one that contains only a single
$M$-particle bound state or it hints on a hidden symmetry of the
model responsible for the degeneracy of the spectrum. These
solutions behave, however, differently for very large but finite
values of $L$; most of them exhibit various signs of pathological
behavior. In particular, they might have complex finite $L$
correction to the energy, or the correction would exceed the
correction due to finite $L$ modifications of the Bethe equations
thus making the asymptotic Bethe ansatz inapplicable. In the weak
coupling limit, i.e. in perturbative gauge theory, and for small
enough values of the bound state momentum only one solution
reduces to the well-known Bethe string solution of the Heisenberg
spin chain. It is also the only solution that behaves reasonably
well for finite values of $L$. Therefore, it is tempting to
identify the physical region of the string model as the one that
contains this solution only.

\smallskip

By analyzing the Bethe equations for the mirror theory, we show
that bound states exist and that they can be regarded as ``mirror
reflections'' of the BPS bound states in the original theory. No
other bound states exist, in agreement with the results by
\cite{DHM}. Given the knowledge of bound states, the next step
would be to construct the S-matrix which describes scattering of
all asymptotic particles including the ones which correspond to
bound states. In principle, such an S-matrix can be obtained by
the fusion procedure \cite{Dorey2,Roiban:2006gs} applied to the
``fundamental" S-matrix we advocate here. This is the bootstrap
program whose discussion we will postpone for the future.

\smallskip

The paper is organized as follows. The next section contains the
discussion of the double Wick rotation, the mirror dispersion
relation and the mirror magnon. In section 3 we discuss the
supersymmetry algebra and the construction of the mirror S-matrix.
In section 4 we analyze the double Wick rotation on the
generalized rapidity torus as well as various possible definitions
for the physical region. In section 5 the properties of the string
S-matrix defined on the generalized rapidity torus are discussed.
We also prove here the unitarity of the scalar factor in the
mirror theory. In section 6 we present various versions of the
Bethe equations in the original and  mirror theory pointing out
that the Bethe equations based on the
$\su(2|2)\oplus\su(2|2)$-invariant string S-matrix should be
modified in the odd winding number sector since for this case the
fermions of the gauge-fixed string sigma model are anti-periodic.
Sections 7 and 8 contain an analysis of the bound states of the
$\AdS$ gauge-fixed model and its mirror theory. Section 9 consists
of several appendices.

\section{Generalities}
In this section we discuss how the vacuum energy of a
two-dimensional field theory on a circle can be found by
considering the Thermodynamic Bethe Ansatz for a mirror model
obtained from the field theory by a double Wick rotation. We
follow the approach developed in \cite{za}.
\subsection{Double Wick rotation and mirror Hamiltonian}
Consider any two-dimensional field-theoretic model defined on a
circle of circumference  $L$. Let \bea\la{H} H = \int_0^L {\rm
d}\s \, \H(p,x, x') \eea be the Hamiltonian of the model,  where
$p$ and $x$ are canonical momenta and coordinates. They may also
include fermions but in this section we confine ourselves to
bosonic fields only. We will refer to the action corresponding to
the Hamiltonian $H$ as to the Minkowski action, however  it does
not have to be relativistic invariant.

\smallskip

We want to compute the partition function of the model defined as
follows \bea\la{Z0} Z(R,L) \equiv \sum_n \langle \psi_n| e^{- H
R}|\psi_n\rangle =\sum_n  e^{- E_n R} \,, \eea where
$|\psi_n\rangle$ is the complete set of eigenstates of $H$. By
using the standard path integral representation \cite{SF}, we get
\bea\la{Z1} Z(R,L) = \int {\cal D}p\, {\cal D} x\, e^{\int_0^R
d\tau \int_0^L\, d\s ( i p \dx - \H)}\,, \eea where the
integration is taken over $x$ and $p$ periodic in both $\tau$ and
$\s$. Formula (\ref{Z1}) shows that $- \int_0^R d\tau \int_0^L\,
d\s ( i p \dx - \H)$ can be understood as the Euclidean action
written in the first-order formalism. Indeed, integrating over $p$
in the usual first-order action $ \int_0^R d\tau \int_0^L\, d\s (
p \dx - \H)$,   we get the Minkowski-type action, and the
Euclidean action is obtained from it by replacing  $\dx \to i\dx$
which is equivalent to the Wick rotation $\tau\to -i\tau$.

\smallskip

Let us now take the Euclidean action, and replace $x'\to -ix'$ or,
equivalently, do the Wick rotation of the $\s$-coordinate $\s\to
i\s$. As a result we get the action where $\s$ can be considered
as the new time coordinate. Let $\widetilde H$ be the Hamiltonian
with respect to $\s$ \bea\la{Ht} \widetilde H = \int_0^R {\rm
d}\tau \, \widetilde{\H}({\widetilde p},x,\dx)\,, \eea where
${\widetilde p}$ are canonical momenta of the coordinates $x$ with
respect to $\s$.

\smallskip
We will refer to the model with the Hamiltonian $\tH$ as to the mirror theory.
If the original model is not Lorentz-invariant then the mirror Hamiltonian is not
equal to the original one, and the Hamiltonians $H$ and $\tH$  describe {\it different} Minkowski theories.

\smallskip
The partition function of the mirror model is given by
\bea\la{Z0d} \widetilde Z(R,L) \equiv \sum_n \langle\widetilde
\psi_n| e^{- \widetilde H L}|\widetilde \psi_n\rangle =\sum_n e^{-
\widetilde E_n L} \,, \eea where $|\widetilde \psi_n\rangle$ is
the complete set of eigenstates of $\widetilde H$. Again, by using
the path integral representation, we obtain \bea\la{Z1a}
\widetilde Z(R,L) = \int {\cal D}\widetilde p\, {\cal D} x\,
e^{\int_0^R d\tau \int_0^L\, d\s ( i \widetilde p x' - \widetilde
\H)}\,. \eea Finally, integrating over $\widetilde p$, we get the
same Euclidean action and, therefore, we conclude that \bea
\widetilde Z(R,L) =Z(R,L)\,. \eea Now, if we take the limit
$R\to\infty$, then $\log Z(R,L) \sim - R E(L)$, where $E(L)$ is
the ground state energy. On the other hand, $\log  \widetilde
Z(R,L)\sim - RL f(L)$, where $f(L)$ is the bulk free energy of the
system at  temperature $T=1/L$ with $\s$ considered as the time
variable. This leads to the relation \bea E(L) = L f(L)\,. \eea To
find the free energy we can use the thermodynamic Bethe ansatz
because $R>>1$. This requires, however, the knowledge of the
S-matrix and the asymptotic Bethe equations for the mirror system
with the Hamiltonian $\widetilde H$. Although the light-cone
gauge-fixed string theory on $\AdS$ is not Lorentz invariant,
$\widetilde H\neq H$, it is still natural to expect that there is
a close relation between the two systems because their Euclidean
versions coincide.

\smallskip
A potential problem with the proof that $ \widetilde Z(R,L)
=Z(R,L)$ is that the integration over $p$ and $\widetilde p$
produces additional  measure factors which may be nontrivial. The
contribution of such a factor is however local, and one usually
does not have to take it into account. We will assume throughout
the paper that this would not cause any problem.


\subsection{Mirror dispersion relation}

The dispersion relation in any quantum field theory can be found
by analyzing the pole structure of the corresponding two-point
correlation function. Since the correlation function can be
computed in Euclidean space, both dispersion relations in the
original theory with $H$ and in the mirror one with $ \widetilde
H$ are obtained from the following expression \bea\la{disrele}
H_{{\rm E}}^2 + 4g^2\sin^2{p_{{\rm E}}\ov 2} +1\,, \eea which
appears in the pole of the 2-point correlation function. Here and
in what follows we consider the light-cone gauge-fixed string
theory on $\AdS$ which has the Euclidean dispersion relation
(\ref{disrele}) in the decompactification limit $L\equiv
P_+\to\infty$ \cite{BDS,AFS,B,AFPZ}. The parameter $g$ is  the
string tension, and is related to the 't Hooft coupling $\lambda$
of the dual gauge theory as $g={\sqrt\l\ov 2\pi}$.

\smallskip

Then the dispersion relation in the original theory follows from
the analytic continuation (see also \cite{AJK})
 \bea H_{{\rm E}}\to -i H\,,~~\quad
p_{{\rm E}}\to p\ ~~\Rightarrow~~\
H^2 = 1+4g^2\sin^2{p\ov 2} \,, \label{ordis}
\eea
and the mirror one from
\bea \label{mirror}~~~~~~~~H_{{\rm E}}\to \widetilde p\,,~~~\quad
p_{{\rm E}}\to i\widetilde H\ ~~\Rightarrow\ ~~\widetilde H = 2\,
{\rm arcsinh}\Big( {1\ov 2g}\sqrt{1+\widetilde p^2}\Big)\,.
\eea Comparing these formulae, we see that $p$ and $\widetilde p$
are related by the following analytic continuation
\bea\la{ancon}
p \to  2i\, {\rm arcsinh}\Big( {1\ov 2g}\sqrt{1+\widetilde
p^2}\Big)\,,~~\quad~~ H=\sqrt{1+4g^2\sin^2{p\ov 2}}\to
i\widetilde p\,. \eea
We note that the plane-wave type limit
corresponds to taking $g\to\infty$ with $\tp\,$ fixed, in which
case we get the standard relativistic dispersion relation \bea
\widetilde H_{{\rm pw}} = {1\ov g}\sqrt{1+\widetilde
p^2}\, . \eea
The expression above suggests that in this limit it
is natural to rescale $\widetilde{H}$ by $1/g$ or,
equivalently, to rescale $\widetilde\tau=i\s$ by $g$.
This also indicates that the semi-classical limit in the mirror
theory should correspond to $g\to\infty$ with $\tp/g$
fixed, so that the dispersion relation acquires the form \bea
\label{dmd} \widetilde H_{\rm sc} = 2\, {\rm arcsinh}\left(
{|\tp|\ov 2g}\right)\,. \eea We will show in the next
subsection that the mirror theory admits a one-soliton solution
whose energy exactly reproduces eq.(\ref{dmd}).

\smallskip

In what follows we need to know how the parameters $x^\pm$
introduced in \cite{BDS} are expressed through $\tp$. By using
formulae (\ref{ancon}), we find \bea\la{xpmtp} x^\pm(p)\to {1\ov
2g}\left(\sqrt{1 +{4g^2\ov 1+\widetilde p^2}}\ \mp\ 1\right)\left(
\widetilde p -i \right)\, \eea and, as a consequence,
$$
ix^- -ix^+\to {1\ov g} \left(1+i\widetilde p\right)\,.
$$
Note that these relations are well-defined for real $p$, but one
should use them with caution for complex values of $p$. In section
\ref{torus} we introduce a more convenient parametrization of the
physical quantities in terms of a complex rapidity variable $z$
living on a torus \cite{Janik}. In  this parametrization the
analytic continuation would simply correspond to the shift of $z$
by the quarter of the imaginary period of the torus.


\subsection{Mirror magnon}
In this section we will derive the dispersion relation for the
``giant magnon" in the semi-classical mirror theory. This will
provide further evidence for the validity of the proposed
dispersion relation (\ref{dmd}).

\smallskip

Consider the classical string sigma-model on $\AdS$ and fix the
generalized uniform light-cone gauge as in \cite{We,FPZ}. The
gauge choice depends continuously on a parameter $a$ with the
range $0\leq a\leq 1$. The gauge-fixed Lagrangian in the
generalized $a$-gauge can be obtained either from the
corresponding Hamiltonian \cite{Arutyunov:2004yx,FPZ} by using the
canonical formalism or by T-dualizing the action in the direction
canonically conjugate to the light-cone momentum $P_+$
\cite{KMRZ}. Its explicit form in terms of the world-sheet fields
is given in appendix 9.1. To keep the discussion simple, in what
follows we will restrict our analysis to the $a=1$
gauge\footnote{Recall that unlike to the case of finite $P_+$ the
dispersion relation of the giant magnon in the infinite volume
limit $P_+=\infty$ was shown to be gauge independent
\cite{magnon}.}.

\smallskip

We are interested in finding a soliton solution in the mirror
theory, which is obtained from the original theory via the double
Wick rotation with further exchange of the time and spacial
directions \bea\tilde{\sigma}= - i \tau\, , ~~~~~~~~~\tilde{\tau}
= i \sigma\, , \label{dwr}\eea where $\sigma,\tau$ are the
variables parametrizing the world-sheet  of the original theory.

\smallskip

Recall that the giant magnon  can be thought of as a solution of
the light-cone gauge-fixed string sigma-model described by a
solitonic profile $y\equiv y(\sigma-v\tau)$, where $y$ is one of
the fields parametrizing the five-sphere and $v$ is the velocity
of the soliton\footnote{See \cite{magnon} and appendix 9.1 for
more details.}. In the infinite $P_+$ limit this soliton exhibits
the dispersion relation (\ref{ordis}), where $p$ coincides, in
fact, with the total world-sheet momentum $p_{\rm ws}$ carried by
the soliton. Owing to the same form of the dispersion relation in
the dual gauge theory, this gives a reason to call this soliton a
``giant magnon" \cite{HM}. For our further discussion it is
important to realize that if, instead of taking the field $y$ from
the five-sphere, we would make a solitonic ansatz $z\equiv
z(\sigma-v\tau)$, where $z$ is one of the fields parametrizing
${\rm AdS}_5$, we would find no solutions exhibiting the
dispersion (\ref{ordis}). As we will now show, in the mirror
theory the situation is reversed: this time the giant magnon
propagates in the AdS part, while there is no soliton solution
associated to the five-sphere.

\smallskip

Take the string Lagrangian (\ref{Lgf}) in the gauge $a=1$ and put
all the fields to zero except a single excitation $z$ from ${\rm
AdS}_5$. Upon making the double Wick rotation (\ref{dwr}), the
corresponding mirror action can be written as follows
\begin{equation}
\label{SAdS} S =g \int_{- r}^{ r}\, {\rm
d}\tilde{\s}{\rm d}\tilde{\tau}\, \left( - 1+ \frac{\sqrt{1+z^2 -
z'^2 + (1+ z^2) \dot{z}^2}}{1+z^2} \right)\equiv \int_{- r}^{ r}\, {\rm
d}\tilde{\s}{\rm d}\tilde{\tau}\, \L \, .
\end{equation}
Here $r$ is an integration bound for $\tilde{\sigma}$ and $\dot{z}
\equiv \partial_{\tilde{\tau}} z\, ,z'
 \equiv\partial_{\tilde{\sigma}} z $.
Although our goal is to identify the mirror magnon configuration
in the decompactification limit, i.e. when $r\to \infty$, for the
moment we prefer to keep $r$ finite.

\smallskip

To construct a one-soliton solution of the equations of motions
corresponding to the action (\ref{SAdS}), we make the following
ansatz
\begin{equation}\nonumber
z = z(\tilde{\sigma} - v \tilde{\tau}) \, .
\end{equation}
Our further discussion follows closely \cite{magnon}. Plugging the
ansatz into (\ref{SAdS}), we obtain the reduced Lagrangian, $L_{red}=L_{red}(z,z')$, which
describes a one-particle mechanical system with $\tilde{\sigma}$ treated as a
time variable. Introducing the canonical momentum $\pi$ conjugate
to $z$, we construct the corresponding reduced Hamiltonian
\begin{equation}\nonumber
H_{\rm red} = \pi z' -  L_{red} \, ,
\end{equation}
which is a conserved quantity with respect to time
$\tilde{\sigma}$. Fixing $H_{\rm red}=1-\omega$, where $\omega$ is
a constant, we get the following equation to determine the
solitonic profile \bea \label{sp} (z')^2 = \frac{1+ z^2
-\frac{1}{\omega^2} }{ 1- v^2 - v^2
  z^2} \, .
\eea The minimal value of $z$ corresponds to the point where the
derivative of $z$ vanishes, while the maximum value is achieved at
the point where the derivative diverges
\begin{equation}\nonumber
z_{\rm min} = \sqrt{\frac{1}{\omega^2} - 1}\, ,  \quad \quad
z_{\rm max} = \sqrt{\frac{1}{v^2} -1} \, ,  \quad v < \omega  < 1
\, .
\end{equation}
The range of $\tilde{\sigma}$ is determined  from the equation
\begin{eqnarray}
r &=& \int_0^r {\rm d}\tilde{\sigma} = \int_{z_{\rm min}}^{z_{\rm
max}} \frac{{\rm d} z}{|z'|} = \sqrt{1 - v^2}\, ~{\rm E}\left(
{\rm arcsin} (z \sqrt{ \omega^2/(1
  -\omega^2)}),\eta\right)\bigg|_{z_{\rm min}}^{z_{\rm max}}\, , \nonumber
\end{eqnarray}
where we have introduced $\eta = \frac{v^2}{\omega^2}
\frac{1-\omega^2}{1 - v^2}$. Here E stands for the elliptic
integral of the second kind. We see that the range  of $\sigma$
tends to infinity when $\omega \rightarrow 1$. Thus,  $\omega\to
1$ corresponds to taking the decompactification limit.

\smallskip

The density of the world-sheet Hamiltonian is given by \bea
\widetilde{\cal{H}}= p_z\dot{z} - \L\, , \nonumber\eea where
$p_z=\frac{\partial \L}{\partial \dot{z}}$ is the momentum
conjugate to $z$ with respect to time $\tilde{\tau}$. For our
solution in the limiting case $\om=1$ we find \bea \nonumber
p_{z}&=&- \frac{v |z|}{\sqrt{1 - v^2(1 + z^2)}}\, . \eea The
energy of the soliton is then \bea \widetilde{H}=g
\int_{-\infty}^{\infty}{\rm d }\sigma\, \widetilde{\cal{H}}=
2g\int_{z_{\rm min}}^{z_{\rm max}} \frac{{\rm d}z}{|z'|}\,
\widetilde{\cal{H}} = 2g\,  {\rm arcsinh}
\frac{\sqrt{1-v^2}}{|v|}\, .\nonumber \eea
\smallskip

To find the dispersion relation, we also need to compute the
world-sheet momentum $p_{\rm ws}$, the latter coincides with the
momentum $\tilde{p}$ of the mirror magnon considered as a point
particle. It is given by \bea \tilde{p}=p_{\rm
ws}=-\int_{-\infty}^{\infty}{\rm d }\sigma\, p_zz'= 2 \int_{z_{\rm
min}=0}^{z_{\rm max}} {\rm d}z |p_z| = 2 \frac{\sqrt{1 -
     v^2}}{|v|}\, . \nonumber\eea
Finally, eliminating $v$ from the expressions for $\widetilde{H}$
and $\tilde{p}$ we find the following dispersion relation \bea
\widetilde{H}=2g\,  {\rm arcsinh}\, \frac{|\tilde{p}|}{2}\,
.\nonumber \eea To consider the semi-classical limit $g\to\infty$,
one has to rescale the time as $\tilde{\tau}\to \tilde{\tau}/g $
so that the energy $\widetilde{H}\to g\widetilde{H}$ will be
naturally measured in units of $1/g$. Under this rescaling the
momentum $\tilde{p}$ scales as well, so that the dispersion
relation takes the form \bea \widetilde{H}=2\, {\rm arcsinh}\,
\frac{|\tilde{p}|}{2g}\, ,\eea which is precisely the previously
announced expression (\ref{dmd}) for the energy of the mirror
magnon.


\section{Mirror S-matrix and supersymmetry algebra}\la{algebra}

The S-matrix in field theory can be obtained from four-point
correlation functions  by using the LSZ reduction formula. Since
the correlation functions can be computed by means of the Wick
rotation, it is natural to expect that the mirror S-matrix is
related to the original one by the same analytic continuation
\bea\la{tSS} \widetilde  S (\widetilde p_1,\widetilde  p_2) =
S(p_1,p_2)\,, \eea where we replace $p_i$ in the original S-matrix
by $\widetilde p_i$ by using formulas (\ref{ancon}). Just as the
original S-matrix, the resulting mirror S-matrix should satisfy
the Yang-Baxter equation, unitarity, physical unitarity, and
crossing relations for real $\widetilde p_k$.

\smallskip

On the other hand, the original S-matrix is
$\su(2|2)\oplus\su(2|2)$ invariant and the states of the
light-cone gauge-fixed $\AdS$ string theory carry  unitary
representations of the symmetry algebra $\su(2|2)\oplus\su(2|2)$.
Therefore, if the relation (\ref{tSS}) is correct then the mirror
S-matrix should possess the same symmetry, and the states of the
mirror theory also should carry unitary representations of
$\su(2|2)\oplus\su(2|2)$. This indicates that there should exist a
way to implement the double Wick rotation on the symmetry algebra
level, and that is what we discuss in this section.


\subsection{Double Wick rotation for fermions}

It is obvious that the double Wick rotation preserves the bosonic
symmetry ${\rm SU}(2)^4$. To understand what happens with the
supersymmetry generators it is instructive to apply the double
Wick rotation to fermions. We consider the quadratic part of the
light-cone gauge-fixed Green-Schwarz action depending on the
fermions $\eta$ in the form given in \cite{FPZ} \bea\la{Lfm}
{\mathscr L} = i\eta_a^\dagger \dot\eta_a -{1\ov 2}
\left(\eta_a\eta'_{5-a} - \eta_a^\dagger
\eta'{}_{5-a}^\dagger\right) - \eta_a^\dagger\eta_a =
i\eta_a^\dagger \dot\eta_a -\H\,. \eea Here, $a=1,2,3,4$, and we
set $\kappa=1$ and rescale $\s$ in the action from \cite{FPZ} so
that $\widetilde\lambda$ disappears.

\smallskip

Computing again the partition function of the model and using the
path integral representation, we get \bea\la{Zf1} Z(R,L) = \int
{\cal D}\eta^\dagger {\cal D}\eta e^{\int_0^R d\tau \int_0^L\, d\s
(-\eta_a^\dagger \dot\eta_a - \H)}\,. \eea We note that fermionic
variables here are anti-periodic in the time direction:
$$\mbox{$\eta(\tau+R)=-\eta(\tau)$}.$$ Would fermions be periodic
in the time direction, the corresponding path integral would
coincide with Witten's index ${\rm Tr}(-1)^Fe^{-HR}$, where $F$ is
the fermion number \cite{Witten}. Since in the mirror model
$\tau$ plays the role of the spatial direction, the mirror
fermions  are always anti-periodic in the spacial direction of the
mirror model. On the other hand, the periodicity condition in the
time direction of the mirror model coincides with a fermion
periodicity condition in the spacial direction of the original
model. In particular, if the fermions of the original model are
periodic then the partition function of the original model is
equal to the Witten's index of the mirror model.

\smallskip

After the first Wick rotation the Lagrangian takes the form
\bea\la{Lf} {\mathscr L} = -\eta_a^\dagger \dot\eta_a -{1\ov 2}
\left(\eta_a\eta'_{5-a} - \eta_a^\dagger
\eta'{}_{5-a}^\dagger\right) - \eta_a^\dagger\eta_a\,. \eea Note
that the fermions in this Euclidean action are not anymore
hermitian conjugate to each other.

\smallskip

Let us now perform the following change of the fermionic variables
\bea\la{cf} \eta_a = {i\ov \sqrt{2}}\left( \psi_{5-a}^\dagger -
\psi_a \right)\,,\quad \eta_a^\dagger = {i\ov \sqrt{2}}\left(
\psi_a^\dagger + \psi_{5-a} \right)\,. \eea Computing (\ref{Lf}),
we get \bea\la{Lf2} {\mathscr L} = -\psi_a^\dagger \psi'_a -{1\ov
2} \left(\psi_a\dot\psi_{5-a} - \psi_a^\dagger
\dot\psi{}_{5-a}^\dagger\right) - \psi_a^\dagger\psi_a\,. \eea It
is the same action as (\ref{Lf}) after the interchange
$\tau\leftrightarrow\sigma$ and $\psi\to\eta$, and this shows that
the double Wick rotation should be accompanied by the change of
variables (\ref{cf}). Note, that in terms of $\psi$'s the
supersymmetry algebra has the standard form with the usual
unitarity condition. Thus, we expect that the supersymmetry
generators will be linear combinations of the original ones. One
may assume that in the interacting theory (beyond the quadratic
level) one would take the same linear combinations.

\smallskip

To summarize, the consideration above seems to indicate that the
symmetry algebra of the mirror theory should correspond to a
different real slice of the complexified  $\su(2|2)\oplus\su(2|2)$
algebra. Moreover, one might expect that the unitary
representation of the $\AdS$ string model could be chosen in such
a way that its analytic continuation by means of formulae
(\ref{ancon})  would produce a unitary representation of the
mirror model.


\subsection{Changing the basis of supersymmetry generators}

Let us recall that the centrally extended $\su(2|2)$ algebra
consists of the bosonic rotation generators
$\bL_a{}^b\,,\ \bR_\a{}^\b$,
the supersymmetry generators $\bQ_\a{}^a,\,\
\bQ_a^{\dagger}{}^\a$, and three central elements $\bH$, $\bC$ and
$\bC^\dagger$. The algebra relations are \bea \label{su22} &&
\left[\bL_a{}^b,\bJ_c\right]=\delta_c^b \bJ_a - {1\ov 2}\delta_a^b
\bJ_c\,,\qquad~~~~ \left[\bR_\a{}^\b,\bJ_\g\right]=\delta^\b_\g
\bJ_\a - {1\ov 2}\delta^\b_\a \bJ_\g\,,
 \nonumber \\
&& \left[\bL_a{}^b,\bJ^c\right]=-\delta_a^c \bJ^b + {1\ov 2}\delta_a^b \bJ^c\,,\qquad
~~\left[\bR_\a{}^\b,\bJ^\g\right]=-\delta_\a^\g \bJ^\b + {1\ov 2}\delta_\a^\b \bJ^\g\,, \nonumber \\
&& \{ \bQ_\a{}^a, \bQ_b^\dagger{}^\b\} = \delta_b^a \bR_\a{}^\b + \delta_\a^\b \bL_b{}^a +{1\ov 2}\delta_b^a\delta^\b_\a  \bH\,, \nonumber \\
&& \{ \bQ_\a{}^a, \bQ_\b{}^b\} = \epsilon_{\a\b}\epsilon^{ab}~\bC\,
, ~~~~~~~~ \{ \bQ_a^\dagger{}^\a, \bQ_b^\dagger{}^\b\} =
\epsilon_{ab}\epsilon^{\a\b}~\bC^\dagger \,. \eea Here in the
first two lines we indicate how the indices $c$ and $\gamma$ of
any Lie algebra generator transform under the action of
$\bL_a{}^b$ and $\bR_\a{}^\b$. For the $\AdS$ string model the
supersymmetry generators $\bQ_\a{}^a$ and $\bQ_a^{\dagger}{}^\a$,
and the central elements $\bC$ and $\bC^\dagger$ are hermitian
conjugate to each other: $\left(
\bQ_\a{}^a\right)^\dagger=\bQ_a^{\dagger}{}^\a$. The central
element $\bH$ is hermitian and is identified with the world-sheet
light-cone Hamiltonian. It was shown in \cite{AFPZ} that the
central elements $\bC$ and $\bC^\dagger$ are expressed through the
world-sheet momentum $\hP$ as follows
\bea \label{Cc} \bC={i\ov
2}g\,(e^{i\hP}-1)e^{2i\xi}\,,\quad \bC^\dagger=-{i\ov
2}g\,(e^{-i\hP}-1)e^{-2i\xi}\,,\quad  g ={\sqrt\l\ov 2 \pi}\, .
\eea
The phase $\xi$ is an arbitrary function of the central
elements, and reflects the obvious ${\rm U(1)}$ automorphism of
the algebra (\ref{su22}): $\bQ\to e^{i\xi}\bQ\,,\ \bC\to
e^{2i\xi}\bC$. In our previous paper \cite{AFZ} we fixed the phase
$\xi$ to be zero to match the gauge theory spin chain convention
\cite{B} and to simplify the comparison with the explicit string
theory computation of the S-matrix performed in \cite{KMRZ}. As we
will see in a moment, if we want to implement the double Wick
rotation under which $\hP \to  i\tbH\,,\quad \bH\to  i\widetilde
\hP$ on the algebra level then we should choose $\xi=-\hP/4$. This
choice makes the central elements $\bC$ and $\bC^\dagger$ to be
hermitian and equal to each other\footnote{A possibility of this
choice was noticed in \cite{AFPZ}.}
\bea\la{Ccs} \bC = \bC^\dagger
=-g\sin{\hP\ov 2}\,. \eea

\smallskip

As we discussed above, the symmetry algebra of the mirror theory
should correspond to a different real slice of the complexified
$\su(2|2)\oplus\su(2|2)$ algebra. This means that we should give
up the hermiticity condition  for the algebra generators and
consider a linear transformation of the generators which is an
automorphism of the complexified  $\su(2|2)\oplus\su(2|2)$
algebra. The transformation (\ref{cf}) suggests to consider the
following change of the supersymmetry generators which manifestly
preserves the bosonic ${\rm SU}(2)^4$ symmetry
\bea\la{nq}
\tbQ_\a{}^a = {1\ov \sqrt 2}\left( \bQ_\a{}^a -
i\,\epsilon^{ac}\,\bQ_c^\dagger{}^\g\,
\epsilon_{\g\a}\right)\,,\quad \tbQ^\dagger_a{}^\a = {1\ov \sqrt
2}\left( \bQ^\dagger_a{}^\a - i\,\epsilon^{\a\b}\,\bQ_\b{}^b\,
\epsilon_{ba}\right)\,. \eea
Then, by using the commutation
relations (\ref{su22}),  we find
\bea
\label{su22n}
\hspace{-1.2cm} \{ \tbQ_\a{}^a, \tbQ_b^\dagger{}^\b\} &=&
\delta_b^a \bR_\a{}^\b + \delta_\a^\b \bL_b{}^a +{i\ov
2}\delta_b^a\delta^\b_\a\, (\bC +\bC^\dagger)\,,\\ \nonumber
\hspace{-1.2cm}\{ \tbQ_\a{}^a, \tbQ_\b{}^b\} &=&
\epsilon_{\a\b}\epsilon^{ab}~{1\ov 2}(\bC -\bC^\dagger +i\bH)\, ,
~~ \{ \tbQ_a^\dagger{}^\a, \tbQ_b^\dagger{}^\b\} =
\epsilon_{ab}\epsilon^{\a\b}~{1\ov 2}(\bC^\dagger -\bC+i\bH) \,.
\eea
Now we see that if we want to interpret the change of the
supersymmetry generators as a result of the double Wick rotation
then we should choose the central elements $\bC\, , \bC^\dagger$
to be of the form (\ref{Ccs}) because with this choice the algebra
relations (\ref{su22n}) take the form
\bea
\label{su22n2}\begin{aligned} \hspace{-1.2cm} \{ \tbQ_\a{}^a,
\tbQ_b^\dagger{}^\b\} &=\delta_b^a \bR_\a{}^\b + \delta_\a^\b
\bL_b{}^a
-{1\ov 2}\delta_b^a\delta^\b_\a\, 2ig\sin{\hP\ov 2}\,, \\
\hspace{-1.2cm}\{ \tbQ_\a{}^a, \tbQ_\b{}^b\} &=
\epsilon_{\a\b}\epsilon^{ab}~{i\ov 2}\bH\, , ~~ \{
\tbQ_a^\dagger{}^\a, \tbQ_b^\dagger{}^\b\} =
\epsilon_{ab}\epsilon^{\a\b}~{i\ov 2}\bH \,, \end{aligned}
\eea
and
performing the analytic continuation
\bea\nonumber \hP \to
i\tbH\,,\quad \bH\to i\widetilde \hP\,, \eea
we obtain the mirror
algebra
\bea \label{su22n3}\begin{aligned} \hspace{-1.2cm} \{
\tbQ_\a{}^a, \tbQ_b^\dagger{}^\b\} &= \delta_b^a \bR_\a{}^\b +
\delta_\a^\b \bL_b{}^a
+g\,\delta_b^a\delta^\b_\a\, \sinh{\tbH\ov 2}\,,  \\
\hspace{-1.2cm}\{ \tbQ_\a{}^a, \tbQ_\b{}^b\} &=-
\epsilon_{\a\b}\epsilon^{ab}~{\widetilde \hP\ov 2}\, , ~~ \{
\tbQ_a^\dagger{}^\a, \tbQ_b^\dagger{}^\b\} =-
\epsilon_{ab}\epsilon^{\a\b}~{\widetilde \hP\ov 2} \,.
\end{aligned}\eea
 Note that after the analytic continuation has been
done we can impose on the new supersymmetry generators $\tbQ$ and
new central elements $\tbH\,, \widetilde \hP$ the same hermiticity
condition as was assumed for the original generators. It is also
clear that the algebra (\ref{su22n3}) implies the mirror
dispersion relation (\ref{mirror}).

\subsection{Mirror S-matrix}

The symmetric choice of the central charges (\ref{Ccs}) differs
from the one we made in \cite{AFZ}. The S-matrix corresponding to
the symmetric choice (\ref{Ccs}) coincides, however, with the
string S-matrix in \cite{AFZ}. Indeed, this choice simply
corresponds to multiplication of $\bQ$ and $\bQ^\dagger$  by
$e^{-i\hP/4}$ and $e^{i\hP/4}$, respectively,  which apparently
does not change the invariance condition for the S-matrix. On the
other hand, the string S-matrix also depends on the parameters
$\eta$'s which reflect the freedom in the choice of a basis of
two-particle states. This freedom was partially fixed in
\cite{AFZ} by requiring the string S-matrix to satisfy the
standard Yang-Baxter equation. This still allows one to change the
basis of one-particle states, or, in other words to change the
basis of the fundamental representation of $\su(2|2)$. We will see
that the requirement that the representation remains unitary after
the analytic continuation fixes the parameters $\eta$'s basically
uniquely.

\smallskip

To this end, we compute the action of the generators $\tbQ$ and
$\tbQ^\dagger$ on the fundamental representation of $\su(2|2)$,
see \cite{B,Bn,AFZ} for details. Starting with \bea
\begin{aligned} \bQ_\a{}^a |e_b\rangle &=& a\,
\delta_b^a|e_\a\rangle \,,\qquad \bQ_\a{}^a
|e_\b\rangle = b\,\e_{\a\b}\e^{ab}|e_b\rangle \,,\\
\bQ_a^\dagger{}^\a  |e_\b\rangle &=&
d\,\delta_{\b}^{\a}|e_a\rangle \,,\qquad \bQ_a^\dagger{}^\a
|e_b\rangle =c\, \e_{ab}\e^{\a\b}|e_\b\rangle
 \, \end{aligned} \la{repg}
\eea we get
 \bea
 \begin{aligned} \tbQ_\a{}^a |e_b\rangle &=&
\widetilde a\, \delta_b^a|e_\a\rangle \,,\qquad \tbQ_\a{}^a
|e_\b\rangle =
\widetilde b\,\e_{\a\b}\e^{ab}|e_b\rangle \,,\\
\tbQ_a^\dagger{}^\a  |e_\b\rangle &=& \widetilde
d\,\delta_{\b}^{\a}|e_a\rangle \,,\qquad \tbQ_a^\dagger{}^\a
|e_b\rangle =\widetilde c\, \e_{ab}\e^{\a\b}|e_\b\rangle
 \, , \end{aligned} \la{repgn}
\eea
where
\bea \label{rot1} \widetilde  a= {1\ov\sqrt 2}(a +i
c)\,,\quad \widetilde  b= {1\ov\sqrt 2}(b +i d)\,,\quad \widetilde
c= {1\ov\sqrt 2}(c +i a)\,,\quad \widetilde  d= {1\ov\sqrt 2}(d +i
b)\,.~~~~~ \eea
and $\tbQ_\a{}^a$, $\tbQ_a^\dagger{}^\a $ are
defined by eqs.(\ref{nq}). The unitarity of the representation
after the analytic continuation requires
\bea\label{ucn} (c+i a) =
b^* -i d^*\,. \eea The parameters of the original unitary
representation before the analytic continuation are given by \bea
\label{ad}\begin{aligned} &a =\sqrt{ig x^- - ig x^+\ov 2} e^{i(
\xi+ \vp )} \,,\quad ~~~~b=-{1\ov
x^-}\sqrt{ig x^- - igx^+\ov 2} e^{i(\xi-\vp )}\,,\\
 &d = \sqrt{ig x^- - ig x^+\ov 2} e^{-i( \xi+ \vp )} \,,\quad
~~c=-{1\ov x^+}\sqrt{ig x^- - igx^+\ov 2}e^{-i(\xi-\vp )}\,,
\end{aligned}\eea where $\xi\sim p$ and $\vp\sim p$ are real, and
the parameters $x^\pm$ satisfy the following complex conjugation
rule
\begin{equation}
\label{con1}
(x^+)^* = x^- \, .
\end{equation}
 After the analytic continuation, $\xi\,, \vp$ and $p$ become imaginary (so that $\tp$ is real) and
\begin{equation}
\label{con2} (x^+)^* = \frac{1}{x^-}\, .
\end{equation}
Taking this into account and computing (\ref{ucn}), we find that
the  analytically  continued representation is unitary for any
choice of $\xi$ if \bea e^{2i\vp} = \sqrt{{x^+\ov
x^-}}=e^{\frac{i}{2}p}\,. \eea This means that the S-matrix which
is unitary for real $p$ and real $\widetilde p$ is obtained from
the string S-matrix, see eq. (8.7) in \cite{AFZ},  by choosing
\bea &&\eta_1=\eta(p_1)e^{\frac{i}{2}p_2}\, , ~~~
\eta_2=\eta(p_2)\, , ~~ \widetilde{\eta}_1=\eta(p_1)\, ,
~~\widetilde{\eta}_2=\eta(p_2)e^{\frac{i}{2}p_1}\, ,
~~~~~~~~~\label{etaun}\eea where we have introduced \bea
\eta(p)=e^{\frac{i}{4}p}\sqrt{ix^-(p)-ix^+(p)}. \la{phaseeta}\eea
Up to a scalar factor the S-matrix reads as  \cite{AFZ}

{\footnotesize
 \bea
 \nonumber
 S(p_1,p_2)&=&\frac{x^-_2-x^+_1}{x^+_2-x^-_1}\frac{\eta_1\eta_2}{\tilde{\eta}_1\tilde{\eta}_2}
\Big(E_{1}^{1}\otimes E_{1}^{1}+E_{2}^{2}\otimes
E_{2}^{2}+E_{1}^{1}\otimes
E_{2}^{2}+E_{2}^{2}\otimes E_{1}^{1}\Big)\\
\nonumber &+&\frac{(x_1^--x_1^+)(x_2^- -x_2^+)(x_2^-+x_1^+)}
{(x^-_1-x^+_2)(x_1^-x_2^--x_1^+x_2^+)}\frac{\eta_1\eta_2}{\tilde{\eta}_1\tilde{\eta}_2}\Big(E_{1}^{1}\otimes
E_{2}^{2}+E_{2}^{2}\otimes E_{1}^{1}-E_{1}^{2}\otimes
E_{2}^{1}-E_{2}^{1}\otimes E_{1}^{2}\Big)
\\
\nonumber &-& \Big(E_{3}^{3}\otimes
E_{3}^{3}+E_{4}^{4}\otimes E_{4}^{4}+E_{3}^{3}\otimes E_{4}^{4}+E_{4}^{4}\otimes E_{3}^{3}\Big)\nonumber\\
\nonumber & +&\frac{(x_1^--x_1^+)(x_2^- -x_2^+)(x_1^-+x_2^+)}
{(x^-_1-x^+_2)(x_1^-x_2^--x_1^+x_2^+)}\Big(E_{3}^{3}\otimes
E_{4}^{4}
+E_{4}^{4}\otimes E_{3}^{3}-E_{3}^{4}\otimes E_{4}^{3}-E_{4}^{3}\otimes E_{3}^{4}\Big)\\
\nonumber &+&\frac{x_2^--x_1^-}{x_2^+
-x_1^-}\frac{\eta_1}{\tilde{\eta}_1}\Big(E_{1}^{1}\otimes
E_{3}^{3}+E_{1}^{1}\otimes E_{4}^{4}+E_{2}^{2}\otimes
E_{3}^{3}+E_{2}^{2}\otimes
E_{4}^{4}\Big)\\
\nonumber &
+&\frac{x_1^+-x_2^+}{x_1^--x_2^+}\frac{\eta_2}{\tilde{\eta}_2}\Big(E_{3}^{3}\otimes
E_{1}^{1}+E_{4}^{4}\otimes
E_{1}^{1}+E_{3}^{3}\otimes E_{2}^{2}+E_{4}^{4}\otimes E_{2}^{2}\Big)\\
\nonumber & +&i\frac{(x_1^--x_1^+)(x_2^--x_2^+)(x_1^+-x_2^+)
}{(x_1^--x_2^+)(1-x_1^-x_2^-)\tilde{\eta}_1\tilde{\eta}_2}
\Big(E_{1}^{4}\otimes E_{2}^{3}+E_{2}^{3}\otimes E_{1}^{4}-E_{2}^{4}\otimes E_{1}^{3}-E_{1}^{3}\otimes E_{2}^{4}\Big)\\
\nonumber
&+&i\frac{x_1^-x_2^-(x_1^+-x_2^+)\eta_1\eta_2}{x_1^+x_2^+(x_1^--x_2^+)(1-x_1^-x_2^-)}
\Big(E_{3}^{2}\otimes E_{4}^{1}+E_{4}^{1}\otimes
E_{3}^{2}-E_{4}^{2}\otimes E_{3}^{1}-E_{3}^{1}\otimes
E_{4}^{2}\Big)\\ \nonumber &+&
\frac{x_1^+-x_1^-}{x_1^--x_2^+}\frac{\eta_2}{\tilde{\eta}_1}\Big(E_{1}^{3}\otimes
E_{3}^{1}+E_{1}^{4}\otimes E_{4}^{1}+E_{2}^{3}\otimes
E_{3}^{2}+E_{2}^{4}\otimes
E_{4}^{2} \Big) \\
\label{Smatrix} &+&\frac{x_2^+
-x_2^-}{x_1^--x_2^+}\frac{\eta_1}{\tilde{\eta}_2}
\Big(E_{3}^{1}\otimes E_{1}^{3}+E_{4}^{1}\otimes
E_{1}^{4}+E_{3}^{2}\otimes E_{2}^{3}+E_{4}^{2}\otimes E_{2}^{4}
\Big)\, ,
 \eea
} where $E_{i}^{j}$ with $i,j=1,\ldots, 4$ are the standard
$4\times 4$ matrix unities, see appendix A of \cite{AFZ} for
notations.

\smallskip

With the choice (\ref{etaun})  the S-matrix (\ref{Smatrix})
satisfies the Yang-Baxter equation and it is unitary for  real
$p$'s. The analytically continued S-matrix
$\tilde{S}(\tp_1,\tp_2)$ is then obtained from (\ref{Smatrix}) by
simply substituting \bea p\to 2i\, {\rm
arcsinh}\frac{1}{2g}\sqrt{1+\tp^2}\, , \eea c.f. section 2.2. One
can verify that {\it this matrix is also unitary for real
$\widetilde p$'s}: \bea
\widetilde{S}(\tp_1,\tp_2)\widetilde{S}^{\dagger}(\tp_1,\tp_2)=\mathbb{I}\,
. \eea

\smallskip

The only subtlety here is that the string S-matrix also depends on
a scalar factor, which has been omitted so far. Thus, one should
separately check that this factor remains unitary after the
analytic continuation. This will be discussed in section
\ref{unitarity}.

\smallskip

An exact relation between the S-matrix, $S^{\rm AFZ}$, found in
\cite{AFZ} and the S-matrix (\ref{Smatrix}) is given by the
following transformation\footnote{The finite-size correction to
the dispersion relation found in \cite{JanikTBA} involves the
coefficients $a_1$, $a_2$ and $a_6$ of $S^{\rm AFZ}$ (see
\cite{AFZ} for notation) which are unaffected by this
transformation. }
$$
S(p_1,p_2)=G_1(p_1)G_2(p_2)S^{\rm
AFZ}(p_1,p_2)G_1(p_1)^{-1}G_2(p_2)^{-1}\, ,
$$
where $G(p)={\rm diag}(1,1,e^{i\frac{p}{4}},e^{i\frac{p}{4}})$. It
is amusing to note that a similar transformation has been recently
introduced in \cite{MM}, but with a very different motivation.
Namely, as was shown in \cite{MM}, the graded version of
$S(p_1,p_2)$ coincides with the Shastry R-matrix \cite{Shastry}
for the one-dimensional Hubbard model \cite{RM1}-\cite{EFGKK}. In
section \ref{torus} we will give another interesting
interpretation to our choice (\ref{etaun}) which is based on the
requirement of generalized unitarity. We will also show there that
this choice of $\eta$'s makes the S-matrix (\ref{Smatrix}) and,
therefore, the Shastry R-matrix a meromorphic function on the
$z$-torus.

\smallskip

To summarize, in order to have a unified description of the
symmetry algebra of the $\AdS$ light-cone gauge-fixed string
theory and its mirror sigma-model we should make the symmetric
choice of the central charges (\ref{Ccs}), and choose the
fundamental representation of the centrally-extended $\su(2|2)$
with the parameters $a,b,c,d$  given by
\bea\begin{aligned}
& a = d = \sqrt{i gx^- - ig x^+\ov 2}=\sqrt{{H+1\ov
2}} \, , \\ & b=c= -\sqrt{{ig\ov 2x^+} - {ig\ov 2x^-}}
=-\sqrt{{H-1\ov 2}} \,.~~~~~~~~~~ \end{aligned}\label{adn}
 \eea
 Taking into
account (\ref{rot1}), (\ref{con1}) and (\ref{con2}), it is easy to
check that both the original  and the mirror
(analytically-continued) representations are unitary with respect
to their own reality conditions. Let us stress that the parameters
$a,b,c,d$ have the same dependence on $x^\pm$ in the original and
mirror theories. We simply regard $x^\pm$ as functions of $p$ in
the original model, and as functions of $\tp$ in the mirror one.

\subsection{Hopf algebra structure}
Formulas (\ref{adn}) define how the algebra generators of the
original and mirror theories  act on one-particle states of the
theory. We also need to know their action on an arbitrary
multi-particle state. The simplest way to have a unified
description of their action is to use the Hopf algebra structure
of the unitary graded associative algebra $\cA$ generated by the
even rotation generators $\bL_a{}^b\,,\ \bR_\a{}^\b$, the odd
supersymmetry generators $\bQ_\a{}^a,\,\ \bQ_a^{\dagger}{}^\a$ and
two central elements $\bH$ and $\hP$ subject to the algebra
relations (\ref{su22}) with the central elements $\bC$ and
$\bC^\dagger$ expressed through the world-sheet momentum $\hP$ by
the formula (\ref{Ccs}). We will be using the Hopf algebra
introduced in \cite{AFZ} which is basically equivalent to the Hopf
algebras discussed in \cite{Gomez:2006va}, see also \cite{sualg}
for further discussion of algebraic properties of the centrally
extended $\su(2|2)$ algebra.

\smallskip

Let us recall that the unit, $\epsilon: {\mathcal A} \rightarrow  {\bf C}$, is defined by
\begin{eqnarray}
\epsilon({\rm id})  = 1 \,,\quad  \epsilon(\bJ) = 0 \,,\quad  \epsilon(\bQ) = 0 \,, \quad \epsilon(\bQ^\dagger) = 0 \, ,
\end{eqnarray}
and the co-product is given by the following formulas\footnote{To
derive these expressions from the ones given in \cite{AFZ} one
should rescale the supersymmetry generators in \cite{AFZ} by
$e^{\pm i\hP/4}$.} \bea \nonumber \Delta({\bf J}) &=&{\bf
J}\otimes {\rm id} + {\rm id}\otimes {\bf J}\, ,\quad
\\\la{coprod} \Delta(\bQ_\a{}^a) &=&\bQ_\a{}^a\otimes e^{i\hP/4}
+ e^{-i\hP/4}\otimes \bQ_\a{}^a\,,\\\nonumber
\Delta(\bQ_a^{\dagger}{}^\a) &=&\bQ_a^{\dagger}{}^\a\otimes
e^{-i\hP/4} + e^{i\hP/4}\otimes \bQ_a^{\dagger}{}^\a\, ,\eea where
${\bf J}$ is any even generator. Here we use the graded tensor
product, that is for any algebra elements $a,b,c,d$ $$(a\otimes
b)(c\otimes d) = (-1)^{\epsilon(b)\epsilon(c)}(ac\otimes bd),$$
where $\epsilon(a)=0$ if $a$ is an even element, and
$\epsilon(a)=-1$ if $a$ is an odd element of the algebra $\cA$.

\smallskip

It is interesting to note that the antipode $S$ is trivial for
{\it any} algebra element, that is \bea S(\bJ) = -\bJ\,,\quad
S(\bQ)=-\bQ\,,\quad S(\bQ^\dagger)=-\bQ^\dagger\, . \eea This
action of the antipode arises for the symmetric choice (\ref{Ccs})
of the central elements $\bC$ and $\bC^\dagger$ only.

\smallskip

The co-product is obviously compatible with the hermiticity conditions one imposes on
the algebra generators in the $\AdS$ string theory, and this ensures that the tensor product of two unitary representations is unitary. To check if the co-product is also compatible with the hermiticity conditions one imposes on
the algebra generators of the mirror model we compute the co-product action on the supersymmetry generators $\tbQ\,, \widetilde{\bQ}^\dagger$
\begin{eqnarray}
\Delta (\widetilde{\bQ}_\a{}^a) &=& \widetilde{\bQ}_\a{}^a\otimes
\cosh \Big(\frac{\widetilde{\bH}}{4}\Big)  +   \cosh
\Big(\frac{\widetilde{\bH}}{4}\Big) \otimes \tbQ_\a{}^a\ \nonumber
\\\la{ncop}
&&+ i \epsilon^{a d} \widetilde{\bQ}_d^\dagger{}^\delta  \epsilon_{\delta \a} \otimes \sinh
\Big(\frac{\widetilde{\bH}}{4}\Big)- i \sinh
\Big(\frac{\widetilde{\bH}}{4}\Big) \otimes \epsilon^{a d} \widetilde{\bQ}_d^\dagger{}^\delta  \epsilon_{\delta \a}\,,  \\
\Delta (\widetilde{\bQ}_a^\dagger{}^\a) &=&
\widetilde{\bQ}_a^\dagger{}^\a\otimes \cosh
\Big(\frac{\widetilde{\bH}}{4}\Big)  +    \cosh
\Big(\frac{\widetilde{\bH}}{4}\Big) \otimes \tbQ_a^\dagger{}^\a\
\nonumber \\\nonumber &&- i \epsilon^{\a \delta}
\widetilde{\bQ}_\delta{}^d  \epsilon_{d a} \otimes \sinh
\Big(\frac{\widetilde{\bH}}{4}\Big) + i \sinh
\Big(\frac{\widetilde{\bH}}{4}\Big) \otimes \epsilon^{\a \delta}
\widetilde{\bQ}_\delta{}^d  \epsilon_{d a}  \,.
\end{eqnarray}
Since in the mirror theory $\tbH$ is hermitian, the co-product  is
also compatible with the hermiticity conditions of the mirror
theory. This guarantees that an $\su(2|2)$-invariant S-matrix can
be always chosen to be unitary.

\smallskip

The co-product (\ref{ncop}) can be used to find the commutation
relations of the supersymmetry generators with the
Zamolodchikov-Faddeev (ZF) operators $A(\tp)$ and
$A^{\dagger}(\tp)$ which create asymptotic states of the mirror
model. The relations can be then used to determine the
antiparticle representation, and to derive the crossing relation
following the steps in \cite{AFZ}. A simple computation gives \bea
\label{QA}
 \widetilde{\bQ}_\a{}^a A^\dagger(\tp)
&=& A^\dagger(\tp)\mathscr{Q}_\a{}^a \cosh
 \Big(\frac{\widetilde{\bH}}{4}\Big)  +   \cosh
 \Big(\frac{\widetilde{\bH}}{4}\Big) A^\dagger(\tp) \Sigma \widetilde{\bQ}_\a{}^a
\\ \nonumber  &+& i A^\dagger(\tp)
 \big(\epsilon^{a d} \overline{\mathscr{Q}}_d{}^\delta  \epsilon_{\delta
    \a}\big) \sinh \Big(\frac{\widetilde{\bH}}{4}\Big)- i A^\dagger(\tp)
 \sinh \Big(\frac{\widetilde{\bH}}{4}\Big) \Sigma \big( \epsilon^{a d}
 \widetilde{\bQ}_d^\dagger{}^\delta  \epsilon_{\delta \a}\big)\,,
  \eea
 \vspace{-0.5cm}
 \bea
\label{bQA}
 \widetilde{\bQ}_a^\dagger{}^\a A^\dagger(\tp)
&=& A^\dagger(\tp)\, \overline{\mathscr{Q}}_a{}^\a \cosh
 \Big(\frac{\widetilde{\bH}}{4}\Big)  +    \cosh
 \Big(\frac{\widetilde{\bH}}{4}\Big) A^\dagger(\tp) \Sigma \widetilde{\bQ}_a^\dagger{}^\a\\ \nonumber
&-& i  A^\dagger(\tp) \big( \epsilon^{\a \delta}
\mathscr{Q}_\delta{}^d
 \epsilon_{d a} \big)\sinh \Big(\frac{\widetilde{\bH}}{4}\Big) + i
 \sinh  \Big(\frac{\widetilde{\bH}}{4}\Big) A^\dagger(\tp) \Sigma \big(\epsilon^{\a \delta} \widetilde{\bQ}_\delta{}^d  \epsilon_{d a}\big) \,,
\eea where $\mathscr{Q}_\a{}^a$ and
$\overline{\mathscr{Q}}_a{}^\a$ are the matrices of the symmetry
algebra structure constants corresponding to the fundamental
representation (\ref{adn}) and $\Sigma = {\rm diag}(1,1,-1,-1)$.

\smallskip

As was already noted, the unitarity of the mirror S-matrix can be,
however, broken by a scalar factor. In section \ref{ellipticS} we
show that the physical unitarity of the mirror S-matrix (the
scalar factor) follows from the crossing relations.

\section{Double Wick rotation and the rapidity torus}\la{torus}
\subsection{The rapidity torus}

The universal cover of the parameter space describing the
fundamental representation of the centrally extended $\su(2|2)$
algebra is an elliptic curve \cite{Janik}. Indeed, the dispersion
formula
 \bea\la{dispn} H^2 - 4g^2\sin^2{p\ov 2} = 1\, ,
 \eea
 which originates from the relation between the central charges of the
fundamental representation, can be naturally uniformized in terms
of Jacobi elliptic functions \bea\la{pez}
p=2\,{\am\,z}\,,~~\quad~~ \sin{p\ov 2} = \sn(z,k)\,,~~\quad~~ H =
\dn(z,k)\, , \eea where we introduced the elliptic
modulus\footnote{Our convention for the elliptic modulus is the
same as accepted in the {\it Mathematica} program, e.g., ${\rm
sn}(z,k)={\rm JacobiSN}[z, k]$.  Since the modulus is kept the
same throughout the paper we will often indicate only the
$z$-dependence of Jacobi elliptic functions. }
$k=-4g^2=-\frac{\lambda}{\pi^2}<0$. The corresponding elliptic
curve (the torus) has two periods $2\omega_1$ and $2\omega_2$, the
first one is real and the second one is imaginary \bea\nonumber
2\omega_1=4{\rm K}(k)\, , ~~~~~~~~~ 2\omega_2=4i{\rm K}(1-k)-4{\rm
K}(k)\, ,
 \eea
where ${\rm K}(k)$ stands for the complete elliptic integral of
the first kind. The dispersion relation  is obviously invariant
under the shifts of $z$ by $2\omega_1$ and $2\omega_2$. The torus
parametrized by the complex variable $z$ is an analog of the
rapidity plane in two-dimensional relativistic models.

\smallskip

In this parametrization the real $z$-axis can be called the
physical one for the original string theory, because for real
values of $z$ the energy is positive and the momentum is real due
to \bea\nonumber 1 \leq\dn(z,k)\leq \sqrt{k'}\, ,~~~~~~~~z\in
{\mathbb R}\, , \eea where $k'\equiv 1 -k$ is the complementary
modulus.
\smallskip

We further note that the representation parameters $x^{\pm}$,
which are subject to the following constraint \bea\la{consxpxm}
x^+ + {1\ov x^+} - x^--{1\ov x^-}={2i\ov g}\, ,\eea are expressed
in terms of Jacobi elliptic functions as \bea \la{xpxmz}
x^{\pm}=\frac{1}{2g}\Big(\frac{\cn z}{\sn z} \pm i \Big)(1+\dn z)
\, . \eea This form of $x^\pm$ follows from the requirement that
for real values of $z$ the absolute values of $x^\pm$ are greater
than unity: $|x^\pm| > 1$ if $z\in  {\mathbb R}$. Note also that
for real values of $z$ we have $\mbox{Im}(x^+)>0$  and
$\mbox{Im}(x^-)<0$ .

\smallskip

Since both the dispersion relation and the parameters $x^\pm$ are
periodic with the period $\om_1$,  the range of  the variable
${\rm Re}\, z$ can be restricted to the interval from $-\om_1/2$
to $\om_1/2$ which corresponds to $-\pi\le p\le\pi$.

\smallskip

Postponing an extensive discussion of the bound states till
section \ref{bound}, we note here that the latter problem requires
consideration of complex values of particle momenta. According to
eq.(\ref{pez}), a rectangle $-\om_1/2\le \mbox{Re}(z)\le
\om_1/2\,;\ -\om_2/2i\le \mbox{Im}(z)\le \om_2/2i$ is mapped
one-to-one onto the complex $p$-plane. By this reason, it is
tempting to call this rectangle by the {\it physical region in the
complex $z$-plane},\footnote{ In relativistic field theories
treated in terms of the rapidity $\theta = \theta_2-\theta_1$, the
physical region is defined as a strip $0<\mbox{Im}\,\theta<\pi$
and it incorporates the bound states. Correspondingly,  the
physical region of an individual particle is ${\rm Im}\,\theta\,
\in (-\pi/2, \pi/2)$ and it covers the complex $p$-plane (with a
cut) through the relation $p=\sinh\theta$. } and, therefore, to
restrict the allowed values of the $z$-coordinates of the
particles forming a bound state by this region. An advantage of
adopting such a choice is that all the bound states would have
positive energy. We will see, however, that this is not the only
option, and there are other two regions in the complex $z$-plane
which could equally deserve the name ``{\it physical}''. As it
will become clear later on, counting the degeneracy of the bound
states drastically depends on the choice of a physical region.

\smallskip

Each solution of eq.(\ref{consxpxm}) corresponds  to a point of
the half-torus, i.e. of the rectangle\footnote{We made slightly
asymmetric choice for ${\rm Im}(z)$ to achieve better visual
clarity.} $-\om_1/2\le \mbox{Re}(z)\le \om_1/2\,;\ -3\om_2/4i\le
\mbox{Im}(z)\le 5\om_2/4i$. In what follows we will be loosely
referring to this rectangle as the torus. The torus covers the
complex $p$-plane twice. Since the space of solutions of
eq.(\ref{consxpxm}) is mapped one-to-one on the torus, the latter
could be also chosen as the physical region. Such a choice is
however problematic because half of all the states would have
negative energy, i.e. the region would contain both particles and
anti-particles, as well as bound states and anti-bound states. We
point out, however, that there exist positive energy solutions of
the bound state equations with some of the particles falling
outside of the rectangle $-\om_1/2\le \mbox{Re}(z)\le \om_1/2\,;\
-\om_2/2i\le \mbox{Im}(z)\le \om_2/2i$ that covers the complex
$p$-plane once.

\begin{figure}[t]
\begin{center}
\includegraphics*[width=0.99\textwidth]{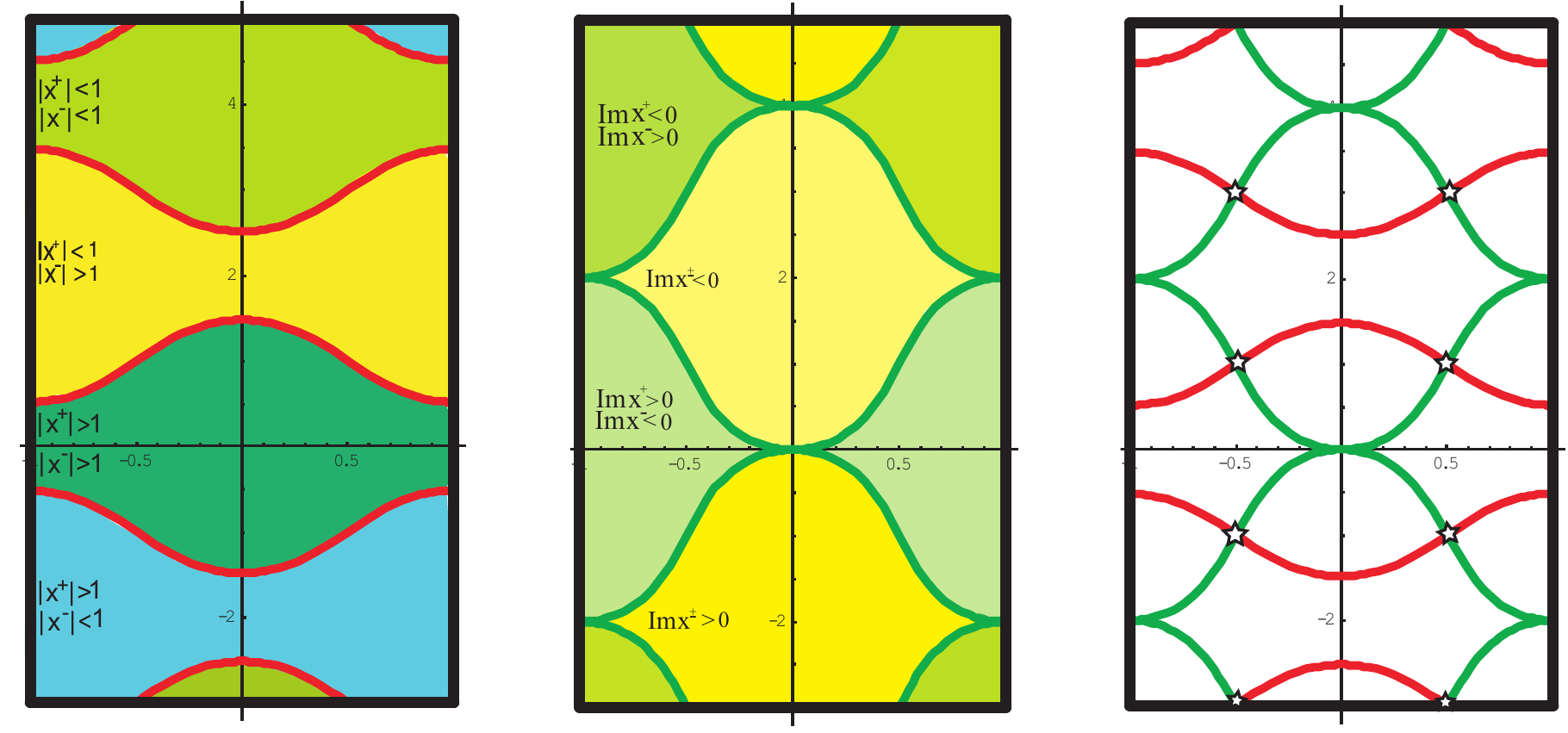}
\end{center}
\caption{On the left figure the torus is divided  by the curves
$|x^+|=1$ and $|x^-|=1$ into four non-intersecting regions. The
middle figure represents the torus divided by the curves ${\rm
Im}(x^+)=1$ and ${\rm Im}(x^-)=1$, also in four regions. The right
figure contains all the curves of interest.
 }
 \label{torus1}
\end{figure}

\smallskip

Constraint (\ref{consxpxm}) implies that if a pair $(x^+,x^-)$
satisfies it then $(1/x^+\,,\ x^-)$,  $(x^+\,,\ 1/x^-)$ and
$(1/x^+\,,\ 1/x^-)$ also do. Each of these four pairs corresponds
to a different point on the torus. Taking into account that for
any complex number $w$ if $|w|>1$ then $|1/w|<1$, and if
$\mbox{Im}(w)>0$ then $\mbox{Im}(1/w)<0$, one can divide the torus
into four non-intersecting regions in the following two natural
ways, see Fig.\ref{torus1}:
\begin{itemize}
\item
$\{|x^\pm|>1\}$, $\{|x^\pm|<1\}$, $\{|x^+|<1\,, |x^-|>1\}$ and
$\{|x^+|>1\,, |x^-|<1\}$; the division is done by the curves
$|x^\pm|=1$.
\item
$\{\mbox{Im}(x^{\pm})>0\}$, $\{\mbox{Im}(x^\pm)<0\}$,
$\{\mbox{Im}(x^+)>0\,,\ \mbox{Im}(x^-)<0\}$ and $\{
\mbox{Im}(x^+)<0\,,\ \mbox{Im}(x^-)>0 \}$; the division is done by
the curves $\mbox{Im}(x^\pm)=0$.
\end{itemize}

\begin{figure}[t]
\begin{center}
\includegraphics*[width=0.9\textwidth]{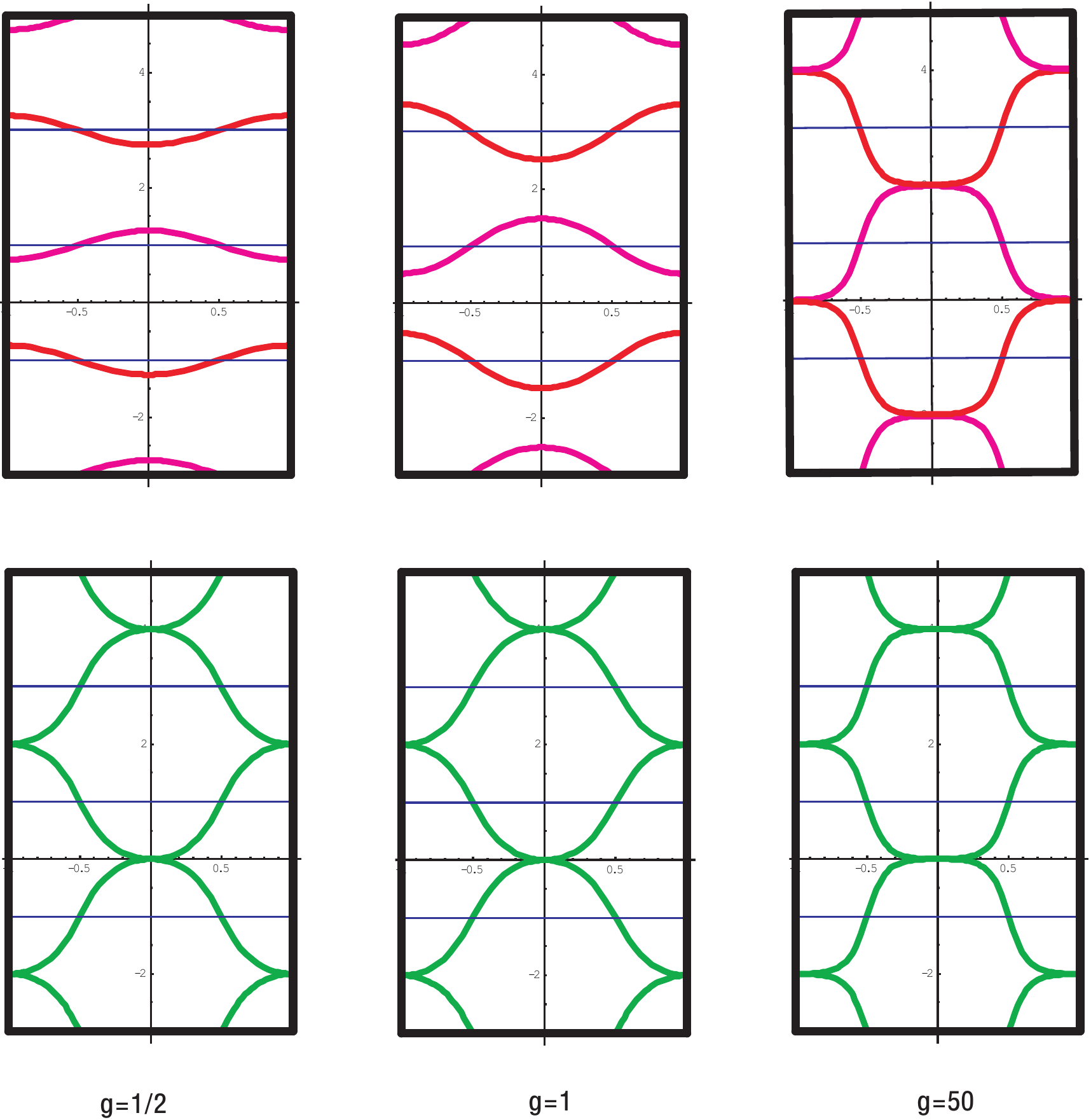}
\end{center}
\caption{Divisions of the torus by the curves $|x^\pm|=1$ (upper
figures) and by the curves ${\rm Im}\, x^{\pm}=0$ (lower figures)
for $g=1/2$, $ g=1$ and $g=50$. The red curves are $|x^-|=1$, and
the pink ones are $|x^+|=1$. The coordinates $x$ and  $y$ are the
rescaled real and imaginary parts of $z$: $x = {\rm Re}(
{2\ov\om_1}z)$, $y = {\rm Re}( {4\ov\om_2}z)$.  In the limit
$g\to\infty$ the curves $|x^{\pm}|=1$ and  ${\rm Im}\, x^{\pm}=0$
are related by the shift $z\to z+\sfrac{\omega_2}{2}$. }
\label{torus1b}
\end{figure}

 The shape of the regions depends on the value of the coupling
constant $g$, see Fig.\ref{torus1b}. Quite remarkably, in the
strong coupling limit $g\to\infty$ two divisions of the torus
produced by the red ($|x^{\pm}|=1$) and green (${\rm
Im}(x^{\pm})=0$) curves become related to each other through a
global shift by $\omega_2/2$.

\smallskip

There are eight special points on the torus where the curves
$|x^\pm|=1$ intersect with the curves $\mbox{Im}(x^\pm)=0$, see
Fig.\ref{torus1}. These points are $z = \pm {1\ov 4}\om_1 +
{2n+1\ov 4}\om_2\,,\ n = -2,-1,0,1$. It is known \cite{AF06} that
these points are the branch points of the one-loop correction
\cite{HL} to the dressing phase.  It is unclear, however, if they
remain the branch points of the exact dressing phase. One could
try to use the integral representation \cite{DHM} of the BES
dressing phase \cite{BES} to understand this issue.  In fact, all
currently available representations for the dressing phase are
defined for $|x^\pm|\ge 1$, and this is another reason to figure
out the location of the curves $|x^\pm|= 1$ on the $z$-torus.

\smallskip
Both divisions play an important role in the analysis of the bound
states of string and mirror theories. To understand the meaning of
the equations  $|x^\pm|=1$ and $\mbox{Im}(x^\pm)=0$,  it is
convenient to use another parameter $u$ which is similar to the
rapidity parameter of the Heisenberg spin chain. In terms of
$x^\pm$ it is defined as follows \bea\la{u}
 u=x^++\frac{1}{x^+}-\frac{i}{g}=x^-+\frac{1}{x^-}+\frac{i}{g}\,.
\eea By using eqs.(\ref{u}) and (\ref{xpxmz}), one can express the
rapidity $u$ as a meromorphic function on the torus \bea u = {\cn
z\dn z\ov g\, \sn z} \,. \eea It is not difficult to check that
the eight special points on the torus are mapped onto the four
points on the $u$-plane with coordinates $u=\pm2\pm {i\ov g}$,
while the points $z=\pm \om_1/2$ are mapped to $u=0$, and the
points $z=\pm \om_1/2 + \om_2/2\pm i 0$ are mapped to $u=\pm
\infty\ \pm\ i\,\infty$.

\smallskip

A special role of the points $u=\pm2\pm {i\ov g}$ can be also
understood by expressing $x^\pm$ in terms of $u$
\bea\la{xpmu}\begin{aligned}
&&x^+ = {1\ov 2}\left( u + \frac{i}{g} \pm \sqrt{\left(u - 2 + \frac{i}{g}\right)\left(u + 2 + \frac{i}{g}\right)} \right)\,,\\
&&x^- = {1\ov 2}\left( u - \frac{i}{g} \pm \sqrt{\left(u - 2 -
\frac{i}{g}\right)\left(u + 2 - \frac{i}{g}\right)} \right)\,.
\end{aligned}\eea Thus, on the $u$-plane there are four branch points  with
coordinates $u=\pm2\pm {i\ov g}$ corresponding to $x^\pm=\pm 1$
and $\mbox{Im}(x^\pm)=0$. Therefore, we can naturally choose the
cuts either connecting  the points $-2 \pm {i\ov g}$ and $2 \pm
{i\ov g}$ , or going from $\pm\infty$ to $\pm2 \pm {i\ov g}$ along
the horizontal lines. Let us determine what values of $x^\pm$
correspond to the lines $u = u_0 \pm {i\ov g}$ with $u_0$ real. We
see that \bea\nonumber &&u_0=x^++\frac{1}{x^+}\,,\quad x^+ = {1\ov
2}\left( u_0 \pm \sqrt{u_0^2-4} \right)\,,\quad  \mbox{if}\ \ u =
u_0 - {i\ov g}\,,\\\nonumber &&u_0=x^-+\frac{1}{x^-}\,,\quad x^- =
{1\ov 2}\left( u_0 \pm \sqrt{u_0^2-4} \right)\,,\quad  \mbox{if}\
\ u = u_0 + {i\ov g}\,. \eea It is clear that points $x^\pm$ and
$1/x^\pm$ of the complex $x^\pm$-plane correspond to the same
point $u$ of the $u$-plane. Then, the points of the circle
$|x^+|=1$ map to points $u$ in the interval $[-2 - {i\ov g}\,, 2 -
{i\ov g}]$ , while the points of $|x^-|=1$ correspond to $u\in [-2
+ {i\ov g}\,, 2 + {i\ov g}]$. On the other hand, the points of the
lines $\mbox{Im}(x^+)=0$ and  $\mbox{Im}(x^-)=0$ correspond to
points $u$ outside the intervals $[-2 - {i\ov g}\,, 2 - {i\ov g}]$
and $[-2 + {i\ov g}\,, 2 + {i\ov g}]$, respectively. Note also
that if one chooses a definite sign in eq.(\ref{xpmu}) then the
interval $[-2 \mp {i\ov g}\,, 2 \mp {i\ov g}]$ maps onto a half of
a unit circle in the  $x^\pm$-plane. One has to use both signs to
cover the unit circles $|x^\pm|=1$ and real lines
$\mbox{Im}(x^\pm)=0$.

\begin{figure}[t]
\begin{center}
\hskip -2cm \includegraphics*[width=0.8\textwidth]{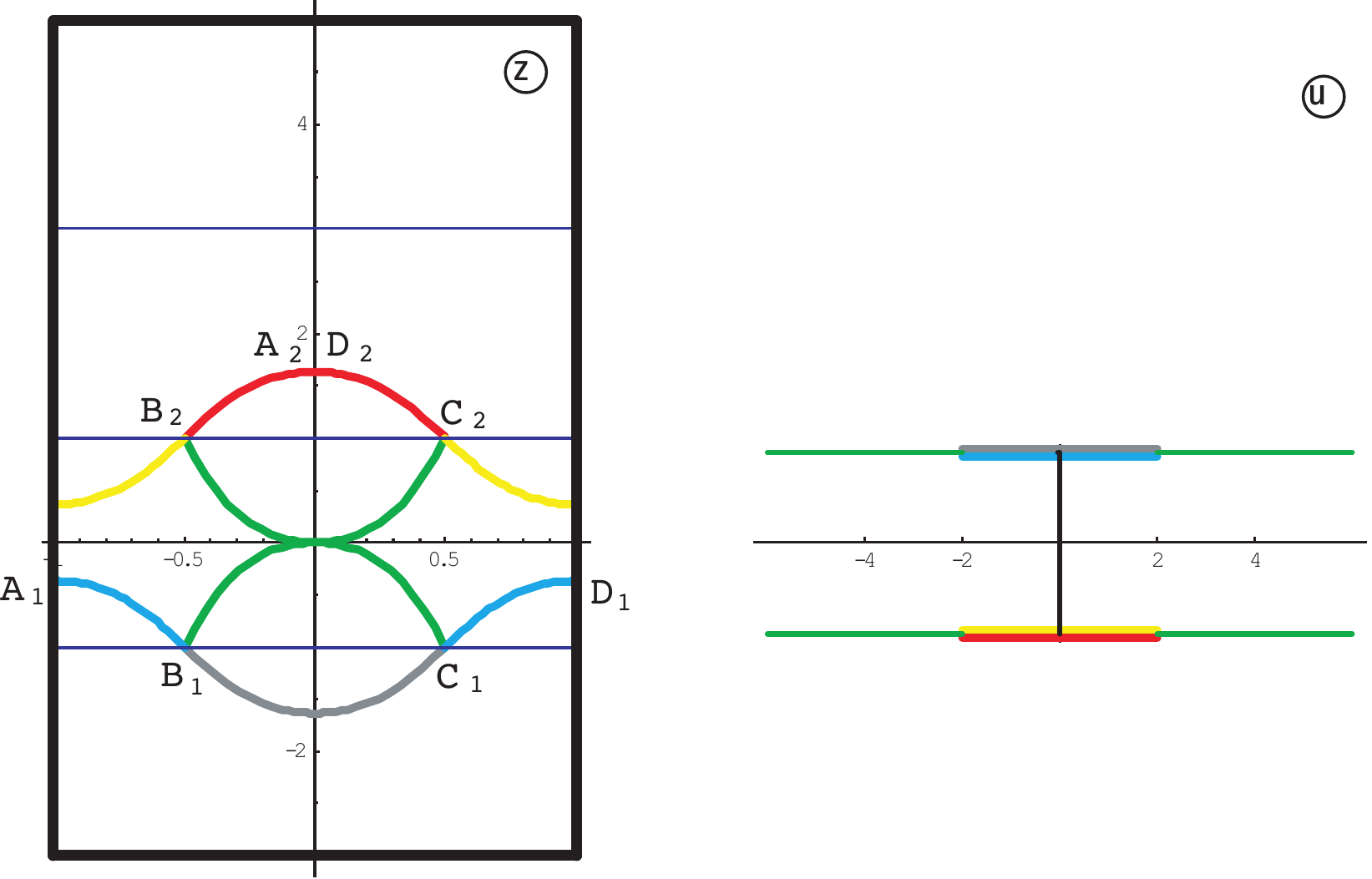}
\end{center}
\vspace{-0.5cm} \caption{On the left figure the upper and  lower
curves correspond to $|x^+|=1+0$ and $|x^-|=1+0$, respectively.
The map $z\to u(z)$ folds each of these curves onto the
corresponding cut on the $u$-plane.
 } \label{toruscuts1}
\end{figure}

\smallskip

To determine the location of the upper and lower edges of the
$u$-plane cuts $[-2 \mp {i\ov g}\,, 2 \mp {i\ov g}]$ on  the
$x^\pm$-planes, we introduce a small real parameter $\epsilon$ and
write \bea x^\pm = e^\epsilon e^{i\varphi}\,,\quad |x^\pm| =
e^\epsilon\,,\quad \mbox{Im}(x^\pm) =
e^\epsilon\sin\varphi\,,\quad u \approx 2\cos\vp \mp {i\ov g} +
2i\epsilon\sin\vp\,.~~~~ \eea We see that the upper edges $[-2 \mp
{i\ov g} + i0\,, 2 \mp {i\ov g} +i0]$ are mapped either outside
the upper halves  or inside the lower  halves of the circles
$|x^\pm|=1$, and the lower edges $[-2 \mp {i\ov g} - i0\,, 2 \mp
{i\ov g} -i0]$ are mapped either outside the lower halves  or
inside the upper halves of the circles $|x^\pm|=1$, and vice
verse: \bea && [-2 \mp {i\ov g} + i0\,, 2 \mp {i\ov g} +i0]\
\Longleftrightarrow\  \left\{
\begin{array}{c} |x^\pm| = 1 + 0\,,\
\mbox{Im}(x^\pm) >0 \\
|x^\pm| = 1 - 0\,,\ \mbox{Im}(x^\pm) <0
\end{array}\right.
\,,\\
&& [-2 \mp {i\ov g} - i0\,, 2 \mp {i\ov g} -i0]\
\Longleftrightarrow\  \left\{
\begin{array}{c} |x^\pm| = 1 + 0\,,\
\mbox{Im}(x^\pm) <0 \\
|x^\pm| = 1 - 0\,,\ \mbox{Im}(x^\pm) >0
\end{array}\right.
\,. \eea

\smallskip

As we discussed above, the $z$-torus can be divided into four
non-intersecting regions by the curves $|x^\pm|=1$. Now it is easy
to show that each of the regions is mapped one-to-one onto the
$u$-plane with the two cuts. Let us consider for definiteness the
region with $|x^\pm|>1$. Then, according to the discussion above,
the boundaries  of the region with  $|x^+| = 1 + 0\,,\
\mbox{Im}(x^+) >0$ and  $|x^+| = 1 + 0\,,\ \mbox{Im}(x^+) <0$ are
mapped onto the upper and lower edges of the cut $[-2 - {i\ov
g}\,, 2 - {i\ov g}]$  in the $u$-plane, respectively. In the same
way the boundary of the region with $|x^-| = 1$ is mapped onto the
upper and lower edges of the cut $[-2 + {i\ov g}\,, 2 + {i\ov
g}]$, see Fig.\ref{toruscuts1}.

\smallskip

Another way to understand how different copies of the $u$-plane
are glued together is to consider any of the curves $|x^\pm(z)|=1$
and shift its variable $z$ by a small positive $\epsilon$ in the
imaginary direction. For the image of the corresponding shifted
curve on the $u$-plane one obtains \bea {\rm
Im}~u(z+i\epsilon)=\mp\frac{1}{g}+\epsilon~ {\rm
Re}\left(\frac{\pa u}{\pa z}\right)+\ldots \, ,\eea where ${\rm
Re}\left(\frac{\pa u}{\pa z}\right)$ is  computed along
$|x^\pm|=1$. Further analysis shows that along any of the curves
$|x^\pm|=1$ the expression ${\rm Re}\left(\frac{\pa u}{\pa
z}\right)$ is positive for $-\frac{\om_1}{4}<{\rm Re}~z <
\frac{\om_1}{4}$ and negative otherwise. This determines how the
edges of the cuts $|x^\pm|=1$ are mapped onto the edges of the
corresponding cuts on the $u$-plane (see Fig.\ref{toruscuts1} for
an example ).

\begin{figure}[t]
\begin{center}
\includegraphics*[width=0.7\textwidth]{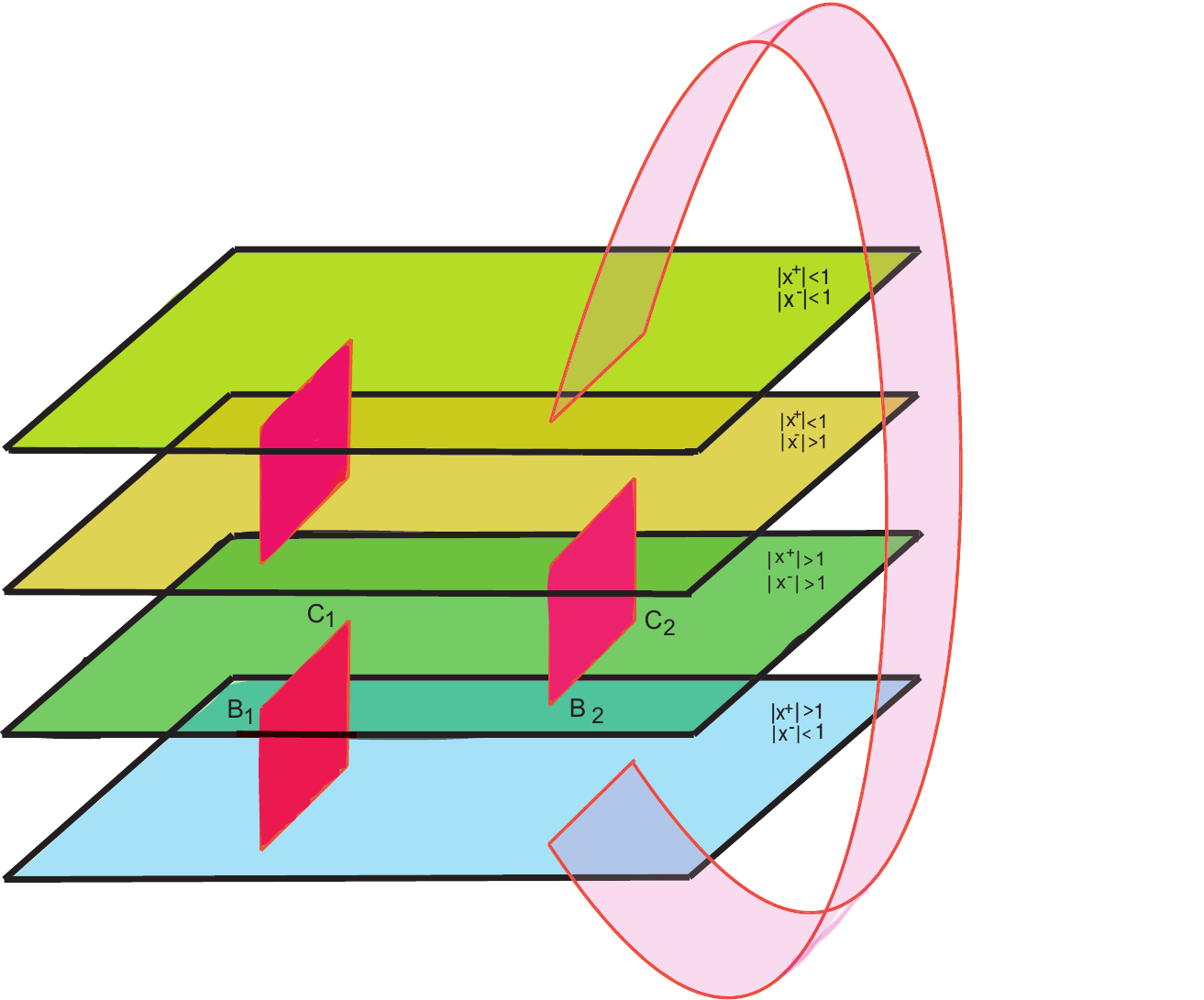}
\end{center}
\caption{Four copies of the $u$-plane (the Riemann sphere) glued
together through the cuts to produce the torus of the kinematical
variable $z$.
We indicated four branch points ${\rm \bf B}_{1,2}$ and ${\rm \bf
C}_{1,2}$ which are images of those on Fig.3.
 } \label{torus3}
\end{figure}

\smallskip

To summarize, any region confined between the curves $|x^{\pm}|=1$
is mapped under $z\to u(z)$ onto a single copy of the $u$-plane
with a point at infinity added, i.e. onto the Riemann sphere.
Extended to the whole torus, {\it this map
 defines a four-fold
covering of the Riemann sphere by the torus which has eight
ramification points}:\footnote{In agreement with the
Riemann-Hurwitz formula.} a generic point on the $u$-plane has
four images belonging to the four regions. There are two cuts on
each copy of the $u$-plane \bea \nonumber
  1)&&  [ -2 + i/g , 2 + i/g ] \\
  \nonumber
  2)&&  [ -2 - i/g , 2 - i/g ]
\eea which are images of the curves $|x^{-}|=1$ and $|x^+|=1$,
respectively.

\medskip

In the same way we can determine the images of the upper and lower
edges of the $u$-plane cuts $(-\infty,-2 \mp {i\ov g}]\,, [2 \mp
{i\ov g},\infty)$ on the $x^\pm$-planes. We again introduce a
small real parameter $\epsilon$ and write \bea x^\pm = r
e^{i\epsilon}\,,\quad |x^\pm| = |r|\,,\quad \mbox{Im}(x^\pm)
\approx r\epsilon\,,\quad u \approx r+{1\ov r} \mp {i\ov g} +
i\epsilon(r-{1\ov r})\,.~~~~ \eea We see that the upper edges
$(-\infty,-2 \mp {i\ov g}+i0]\,, [2 \mp {i\ov g}+i0,\infty)$ are
mapped either onto the upper edge of the intervals
$(-\infty,-1]\,, [1,\infty)$ or the lower edge of the interval
$[-1,1]$, and the lower edges $(-\infty,-2 \mp {i\ov g}-i0]\,, [2
\mp {i\ov g}-i0,\infty)$ are mapped either onto the lower edge of
the intervals $(-\infty,-1]\,, [1,\infty)$ or the upper edge of
the interval $[-1,1]$ of the real lines $\mbox{Im}(x^\pm)=0$, and
vice verse: \bea && (-\infty,-2 \mp {i\ov g}+i0]\cup[2 \mp {i\ov
g}+i0,\infty)\  \Longleftrightarrow\ \left\{
\begin{array}{c}  \mbox{Im}(x^\pm) = +0 \,,\ |x^\pm| >1
\\
 \mbox{Im}(x^\pm) = -0 \,,\ |x^\pm| <1
\end{array}\right.
\,,~~~~~\\
&& (-\infty,-2 \mp {i\ov g}-i0]\cup [2 \mp {i\ov g}-i0,\infty)\
\Longleftrightarrow\  \left\{
\begin{array}{c}  \mbox{Im}(x^\pm) = +0 \,,\ |x^\pm| <1
\\
 \mbox{Im}(x^\pm) = -0 \,,\ |x^\pm| >1
\end{array}\right.
\,, \eea Again, dividing the $z$-torus into four non-intersecting
regions by the curves $\mbox{Im}(x^\pm)=0$, we see that each of
the regions also maps one-to-one onto the $u$-plane with the two
cuts. This gives a different (but equivalent) four-fold covering
of the Riemann sphere by the torus.

\smallskip

When a point on the $z$-plane runs along the curve $|x^+|=1$ or
$|x^-|=1$ its image covers the corresponding interval on the
$u$-plane twice. To appreciate this fact, let us note that if $z$
is, e.g., on the curve $|x^+|=1$ then the points \bea z_{\pm}=-z
\pm \frac{\omega_1}{2}+\frac{\omega_2}{2} \eea are also on this
curve. Indeed, since $|x^+(z)|^2=x^+(z)x^-(z^*)$, we have \bea
|x^+(z_\pm)|^2=x^+\Big(-z\pm
\frac{\omega_1}{2}+\frac{\omega_2}{2}\Big)
x^-\Big(-z^*\pm\frac{\omega_1}{2}-\frac{\omega_2}{2}\Big)=\frac{1}{|x^+(z)|^2}=1\,
, \nonumber \eea where we have used the properties of Jacobi
elliptic functions under the shifts by quarter-periods. In the
same way one finds that if $z$ lies on a curve $|x^-|=1$ then the
points $z_{\pm}$ belong to another copy of $|x^-|=1$ which is
obtained from the original one by the shift by $\omega_2$.
Finally, using the properties of the Jacobi elliptic functions it
is easy to show that $u(z_{\pm})=u(z)$, i.e. the points $z$ and
$z_{\pm}$ have  one and the same image on the $u$-plane.

\smallskip

It is clear that the half of the torus and, therefore, the complex
$p$-plane is mapped onto the $u$-plane twice. The coordinate $u$
is real for real $z$, and in this case we can easily express it in
terms of $p$ \cite{BDS} \bea\la{up}
 u(p)=\frac{1}{g}\cot \frac{p}{2} \sqrt{1+4 g^2 \sin ^2\frac{p}{2}}\,
. \eea In the limit $g\to 0$ the relation (\ref{up}) turns to the
one between the rapidity and momentum variables of the Heisenberg
spin chain; the latter describes the gauge theory at the one-loop
level. This supports an idea that the physical region could be
identified with a single copy of the $u$-plane, namely the one
which maps to  the region $|x^\pm|>1$ of the $z$-torus. There are
certain advantages of such a choice which we will discuss later
on. The main disadvantage is, however, that the region $|x^\pm|>1$
is not big enough to cover the whole complex $p$-plane.

\smallskip

It is interesting to see what happens with our three candidates
for the physical region in the limits $g\to\infty$ and $g\to 0$.
In the limit $g\to\infty$ the periods of the torus have the
following behavior \bea \om_1 \to {\log g\ov g}\,,\quad \om_2\to
{i\pi\ov 2g}\qquad \makebox{if}\quad g\to\infty\,. \eea To keep
the range of $\makebox{Im}( z)$ finite, we rescale $z$ as $z\to
z/(2g)$, and the momentum as $p\to p/g$. Then the dispersion
relation (\ref{dispn}) takes the relativistic form $H^2 - p^2 =1$,
the variable $z$ plays the role of $\theta$ because $p = \sinh z$,
and we have
\begin{itemize}
\item
The torus degenerates to the strip with $-\pi<\makebox{Im}(
z)<\pi$ and $-\infty<\makebox{Re}( z)<\infty$
\item
The half-torus corresponding to the complex $p$-plane degenerates
to the strip with $-\pi/2<\makebox{Im}( z)<\pi/2$ and
$-\infty<\makebox{Re}( z)<\infty$
\item
The region $|x^\pm|>1$ corresponding to the complex $u$-plane
degenerates to the strip with $-\pi/2<\makebox{Im}( z)<\pi/2$ and
$-\infty<\makebox{Re}( z)<\infty$
\end{itemize}
We see that both the half-torus and the region $|x^\pm|>1$
degenerate to the physical strip of a relativistic field theory.

\smallskip

In the limit $g\to 0$ the periods of the torus have the following
behavior \bea\label{gtozero} \om_1 \to \pi\,,\quad \om_2\to 2i\log
g\qquad \makebox{if}\quad g\to 0\,. \eea We see that all the three
regions degenerate into the strip with $-\pi/2<\makebox{Re}(
z)<\pi/2$ and $-\infty<\makebox{Im}( z)<\infty$. The properties of
the S-matrix arising in the limit $g\to 0$ will be discussed in
appendix \ref{oneloop}.

\subsection{Double Wick rotation}

The $z$-torus can be also used to describe the mirror model. Since
we know the relation between $p=2\, {\rm am}\,z$ and the mirror
momentum $\tilde{p}$, we can express $\tilde{p}$ in terms of  $z$.
Indeed, the equality \bea \label{eqmom} 2\, {\rm am}\, z=2i\, {\rm
arcsinh} \frac{1}{2g}\sqrt{1+\tp^2} \eea implies \bea\la{tpz}
\widetilde{p}=-i\dn z\, . \eea The energy in the mirror theory
takes the form \bea \widetilde{H}=2\, {\rm
arccoth}\frac{\sqrt{k'}}{\dn z}\, . \eea

\smallskip

The formulae above show that real  values of $z$ correspond to
imaginary $\tilde{p}$. Now we would like to understand for which
values of $z$ the corresponding values of $\widetilde{p}$ are
real. One can see that if we shift the variable $z$ by $\om_2/2$, $z\to z+\om_2/2$, that is  if we write \bea
\widetilde{p}=-i\dn\Big(z+\frac{\omega_2}{2},k\Big)\equiv
\sqrt{k'}\, \frac{\sn z }{\cn z}\, , \eea then for real values of the shifted variable $z$ the
corresponding values of $\widetilde{p}$ are real as well.
 We also recognize here a close analogy with the relativistic
case -- making the double Wick rotation corresponds to the shift
by a quarter-period on the rapidity plane. The function $\cn(z,k)$
has zeroes at $z=-\frac{1}{2}\omega_1$ and $z=\frac{1}{2}\omega_1$
(and $\dn(z,k)$ has poles at $z=\frac{1}{2}(-\omega_1+\omega_2)$
and $z=\frac{1}{2}(\omega_1+\omega_2)$) which explains the
apparent absence of the periodicity in $\widetilde{p}$. Thus, when
the shifted variable $z$ runs from $-\sfrac{1}{2}\omega_1$ to
$\sfrac{1}{2}\omega_1$ the momentum $\tilde{p}$ monotonically
increases from $-\infty$ to $+\infty$ and it passes though zero
for $z=0$.

\smallskip

One further finds that the parameters $x^\pm$ are expressed in
terms of the shifted parameter $z$ of the mirror model as follows
\bea\la{xpxmzm} x^{\pm} =-i\frac{\sqrt{k'}\mp \dn z}{\sqrt{-k}\dn
z}\Big(1+i\sqrt{k'}\frac{\sn z}{\cn z}\Big)\, . \eea We can now
find how $x^{\pm}$ are expressed in terms of the mirror momentum.
Indeed, since \bea \nonumber \Big(\frac{\cn z}{\sn
z}\Big)^2=-1+\frac{k}{1-\dn^2z}\,, \eea
 we deduce from eq.(\ref{xpxmzm}) that \bea \nonumber
x^{\pm}
=\frac{1}{2g}\left(\sqrt{1+\frac{4g^2}{1+\widetilde{p}^2}}\mp
1\right)(\widetilde{p}-i)\, .\eea This, of course, agrees with the
formula (\ref{xpmtp}).

\smallskip

The variables $x^{\pm}$ of the mirror theory obey a relation
\bea\nonumber x^+x^-=\frac{\widetilde{p}-i}{\widetilde{p}+i} \eea
which implies that $|x^+x^-|=1$ for $\widetilde{p}$ real.

\smallskip

It is also not difficult to show that the dispersion relation in
the mirror theory takes the form (\ref{mirror}) \bea\nonumber
\widetilde{H}=2\, {\rm arccoth}\frac{\sqrt{k'}}{\dn z}=2\, {\rm
arccoth} \sqrt{1-\frac{k}{1+\widetilde{p}^2}}=2\, {\rm
arcsinh}\frac{1}{\sqrt{-k}}\sqrt{1+\widetilde{p}^2}\, . \eea This
completes the proof that the double-Wick rotation corresponds to a
shift of the $z$ variable by a quarter of the imaginary period of
the torus, and the real axes of the shifted  $z$ corresponds to
real values of the momentum of the mirror theory.\footnote{After
having performed the shift, one can do various physically
equivalent  transformations of the shifted $z$-variable preserving
the axes of real $z$. Particular useful examples of these
transformations are $z\to z + {\om_1\ov 2}\,,\  z\to -z +
{\om_1\ov 2}\,,\ z\to -z \pm {\om_1\ov 2}$.}

\smallskip

Finally, it is useful to express the rapidity $u$ in terms of the
shifted parameter $z$ of the mirror model and $\tp$. We have
\bea\nonumber u =
{2\cn\big(z+\frac{\omega_2}{2},k\big)\dn\big(z+\frac{\omega_2}{2},k\big)\ov
\sqrt{- k}\sn\big(z+\frac{\omega_2}{2},k\big)} =
-{2i\sqrt{k'}\dn\big(z+\frac{\omega_2}{2},k\big)\ov \sqrt{-
k}\dn\big(z,k\big)}\,. \eea Then one can check that the points
$z=\pm \om_1/2 \pm i 0$ are mapped to $u=\pm \infty\ \pm\
i\,\infty$. The coordinate $u$ is real for real $z$, and in this
case we can express it in terms of $\tp$ \bea\nonumber
 u=\frac{2\tp}{\sqrt{- k}}\sqrt{1-\frac{k}{1+\tp^2}}=\frac{\tp}{g}\sqrt{1+\frac{4g^2}{1+\tp^2}}\,
. \eea

\smallskip

Again, there are three choices of the physical region. It is the
half-torus corresponding to the complex $\tp$-plane, the whole
torus, and the region $\mbox{Im}(x^\pm)<0$ which is mapped onto
the $u$-plane. The third choice is different from the one made for
the string theory, and is motivated by the analysis of the bound
states of the mirror model.

\section{S-matrix on elliptic curve}\la{ellipticS}

\subsection{Elliptic S-matrix and its properties}
The dispersion relation (\ref{dispn}) is naturally parametrized by
the elliptic curve. Without imposing the unitarity condition for
the S-matrix, the phase $\eta$ in (\ref{phaseeta}) can be chosen
in an arbitrary way, for instance, $\eta(p)=1$. In the latter
case, the S-matrix (\ref{Smatrix}) is well defined on the elliptic
curve but it is non-unitary. It is therefore tempting to assume
that the unitary S-matrix also admits an analytic continuation
into the complex $z$-plane. To find such a continuation one has to
resolve the branch cut ambiguities arising due to the
$\eta$-factor in the S-matrix (\ref{Smatrix}):
$\eta(p)=e^{\frac{i}{4}p}\, \sqrt{ix^-(p)-ix^+(p)}$.

\smallskip

This can be done in the following way. First, we recall the
elliptic parametrization (\ref{xpxmz}) which gives \bea
\eta(p)=e^{\frac{i}{4}p}\, \sqrt{ix^-(p)-ix^+(p)}&=&
\frac{1}{\sqrt{g}}e^{\frac{i}{2}{\rm am}\, z}\,\sqrt{1+{\rm dn}\,
z}\,
 = \nonumber \\
&&~=\frac{1}{\sqrt{g}}\sqrt{(1+{\rm dn}\, z)({\rm
cn}\, z+i {\rm sn}\, z)}\, .\eea Second, by using the following
formulae  (recall $k = -4g^2$)
$$
1+{\rm dn}\, z=\frac{2\, {\rm dn}^2\, \frac{z}{2}}{1-k\, {{\rm
sn}^4\frac{z}{2}}}\, , ~~~~~~~~~~~~ {\rm cn}\, z+i\, {\rm sn}\,
z=\frac{\big({\rm cn}\, \frac{z}{2}+i \, {\rm sn}\,
\frac{z}{2}{\rm dn}\, \frac{z}{2}\big)^2}{1-k\, {{\rm
sn}^4\frac{z}{2}}}\,
$$
relating elliptic functions to those of the half argument, we can
resolve the branch cut ambiguities by means of the relation \bea
e^{\frac{i}{4}p}\, \sqrt{ix^-(p)-ix^+(p)}=
\frac{\sqrt{2}}{\sqrt{g}}\frac{{\rm dn}\, \frac{z}{2}\big({\rm
cn}\, \frac{z}{2}+i \, {\rm sn}\, \frac{z}{2}{\rm dn}\,
\frac{z}{2}\big)}{1+4g^2\, {{\rm sn}^4\frac{z}{2}}}\equiv \eta(z)
\eea valid in the region $-\frac{\om_1}{2}<{\rm Re}\,
z<\frac{\om_1}{2}$ and $-\om_2/i<{\rm Im}\, z<\om_2/i$. Further, we
notice that the non-local dependence of $\eta$'s on the momentum
of another particle enters as $e^{\frac{i}{2}p}=e^{i{\rm am}\,z}$
and, therefore, can be naturally treated as $e^{\frac{i}{2}p}={\rm
cn}\, z+ i\, {\rm sn}\, z$.

\smallskip

Thus, we define an analytic continuation of the S-matrix onto the
rapidity torus for each of the complex variables $z_1$ and $z_2$
by means of eq.(\ref{Smatrix}), where the variables $\eta_{1,2}$
and $\tilde{\eta}_{1,2}$ are given by \bea\begin{aligned}
\eta_1=\eta(z_1) ({\rm cn}\, z_2+ i\, {\rm sn}\, z_2)\,
,~~~~~~~~~\eta_2=\eta(z_2)\, , \nonumber \\
\tilde{\eta}_2=\eta(z_2)({\rm cn}\, z_1+ i\, {\rm sn}\, z_1)\,
,~~~~~~~~~ \tilde{\eta}_1=\eta(z_1) \, .\end{aligned}\eea In this
way we completely resolved the branch cut ambiguities of the
S-matrix (\ref{Smatrix}) and defined it as a meromorphic function
on the elliptic curve (for each $z$-variable). It is remarkable to
observe that such a continuation becomes possible due to
additional phase factors, $e^{\frac{i}{4}p}$, introduced in the
previous section to guarantee unitarity of the mirror theory.

\smallskip

Let us now analyze the basic properties of the elliptic S-matrix.
One can check that it satisfies the Yang-Baxter equation and the
usual unitarity requirement \bea S_{12}(z_1,z_2)
S_{21}(z_2,z_1)=\mathbb{I}\, .\la{standardunitarity}\eea Further,
it obeys the {\it generalized unitarity condition}: \bea
S_{12}(z_1^*,z_2^*)\big[ S_{12}(z_1,z_2)\big]^{\dagger}={\mathbb
I}\, . \la{generalunit}\eea Here $``\dagger"$ means hermitian
conjugation. For $z_1$ and $z_2$ real the last condition reduces
to the requirement of physical unitarity. In fact, one can see
that the elliptic S-matrix is compatible with the generalized
unitarity condition only due to our specific choice for the phase
factors discussed above. Then, unitarity and generalized unitarity
imply hermitian analyticity: $S_{21}(z_2^*,z_1^*)= \big[
S_{12}(z_1,z_2)\big]^{\dagger}\, $.

\smallskip

Let us now compute monodromies of the S-matrix (\ref{Smatrix})
over the real and imaginary periods. We find \bea \begin{aligned}
 S(z_1+2\om_1,z_2)& =\Sigma_1\,
S(z_1,z_2)\Sigma_1=\Sigma_2\, S(z_1,z_2)\Sigma_2\, , \, \\
 S(z_1+2\om_2,z_2)& =\Sigma_1\, S(z_1,z_2)\Sigma_1=\Sigma_2\,
S(z_1,z_2)\Sigma_2\, .
\end{aligned}\label{mon2w} \eea
Hence, the S-matrix exhibits the same monodromies over real and
imaginary cycles and it is a periodic function on a double torus
with periods $4\omega_1$ and $4\omega_2$. Here
$\Sigma_1=\Sigma\otimes {\mathbb I}$ and  $\Sigma_2= {\mathbb
I}\otimes\Sigma$, where $\Sigma$ is defined in section 3.4, and
the S-matrix commutes with the product $\Sigma\otimes \Sigma$.
Note that $\Sigma$ is in the center of the group ${\rm
SU}(2)\times {\rm SU}(2)$.

\smallskip Second, we establish the monodromy properties w.r.t.
shifts by half-periods. Under the shift by the real half-period we
get \bea S(z_1+\om_1,z_2)=\big(V\otimes\Sigma\big)\,
S(z_1,z_2)\big(V^{-1}\otimes \mathbb{I}\big)\, ,\eea where $V={\rm
diag}\big(e^{-\frac{i\pi}{4}},e^{-\frac{i\pi}{4}},e^{\frac{i\pi}{4}},e^{\frac{i\pi}{4}}\big)
$.

\smallskip

The shift by the imaginary half-period corresponds to the crossing
symmetry transformation \cite{Janik}. To discuss it, we multiply
the S-matrix (\ref{Smatrix}) with a scalar factor $S_0$ to produce
the string S-matrix obeying crossing symmetry
 \bea \la{strS}
 {\cal
S}(z_1,z_2)=S_0(z_1,z_2)S(z_1,z_2) \, .\eea We then find that with
a proper choice for $S_0(z_1,z_2)$ the string S-matrix exhibits
the following crossing symmetry relations \bea \label{cross1}
 {\mathscr
C}_1^{-1}{\cal S}_{12}^{t_1}(z_1,z_2){\mathscr C}_1{\cal
S}_{12}(z_1+\om_2,z_2)=\mathbb{I} \, , \la{cr1}\quad
{\mathscr C}_1{\cal S}_{12}^{t_1}(z_1,z_2){\mathscr
C}_1^{-1}{\cal S}_{12}(z_1-\om_2,z_2)=\mathbb{I} \,,~~~~~
  \eea
and also \bea \label{cross2} {\mathscr C}_1^{-1}{\cal
S}_{12}^{t_1}(z_1,z_2){\mathscr C}_1{\cal
S}_{12}(z_1,z_2-\om_2)=\mathbb{I} \, , \la{cr2}\quad {\mathscr
C}_1{\cal S}_{12}^{t_1}(z_1,z_2){\mathscr C}_1^{-1}{\cal
S}_{12}(z_1,z_2+\om_2)=\mathbb{I} \, .~~~~~
 \eea
Here $t_1$ denotes transposition in the first matrix space and
${\mathscr C}$ is a {\it constant}\footnote{This is in opposite to
\cite{AFZ}, where the charge conjugation matrix was found to
depend on the sign of the particle momentum. This dependence is,
in fact, spurious and it gets removed by a proper resolution of
the branch cut ambiguities we propose here. } charge conjugation
matrix \bea\la{cc} \mathscr{C}=\left(\begin{array}{cc} \sigma_2 &
0 \\ 0 & i\, \sigma_2
\end{array}\right)\, ,
\eea where $\sigma_2$ is the Pauli matrix. The compatibility of
eqs.(\ref{cr1}) and (\ref{cr2}) with (\ref{mon2w}) is guaranteed
by the identity ${\mathscr C}\Sigma={\mathscr C}^{-1}$ which is
equivalent to ${\mathscr C}^2=\Sigma$.

The crossing symmetry relations lead to the following equations
for the scalar factor $S_0$ \cite{Janik} \bea\la{crf}
S_0(z_1,z_2)\,S_0(z_1+\om_2,z_2)=f(z_1,z_2)\,,\quad
S_0(z_1,z_2)\,S_0(z_1,z_2-\om_2)=f(z_1,z_2)\,,~~~~~ \eea where the
function $f$ is expressed through $x^\pm$ as follows
\bea\la{ff}
f(z_1,z_2)=\frac{\Big(\frac{1}{x_1^-}- x_2^- \Big)(x_1^--x_2^+)}
{\Big(\frac{1}{x_1^+}-x_2^-\Big)(x_1^+ -x_2^+)}\, . \eea One can
easily check that the function $f(z_1,z_2)$ obeys the following
properties
$$
f(z_2,z_1)f(z_1+\om_2,z_2)=1=f(z_2,z_1)f(z_1,z_2+\om_2)\, , \quad
f(z_1+\om_2,z_2+\om_2)=f(z_1,z_2)\, ,
$$
which are, however, incompatible with the assumption that the scalar
factor is an analytical function of $z_1,z_2$.

\smallskip

Another important property of the string S-matrix (\ref{strS}) is
that it remains invariant under the simultaneous shift of $z_1$
and $z_2$ by $\omega_2$: \bea\la{Sm12}
\S(z_1+\om_2,z_2+\om_2)=\S(z_1,z_2)\, . \eea This follows  from
the fact that both the S-matrix (\ref{Smatrix}) and the scalar
factor $S_0$ are invariant under the shift. This property together
with the crossing relations (\ref{cross1}), (\ref{cross2}) implies
\bea\nonumber \S^{t_1,t_2}(z_1,z_2)= {\mathscr C}_1{\mathscr
C}_2\S(z_1,z_2){\mathscr C}_1^{-1}{\mathscr C}_2^{-1}={\mathscr
C}_1^{-1}{\mathscr C}_2^{-1} \S(z_1,z_2){\mathscr C}_1{\mathscr
C}_2\, , \eea where $t_1$ and $t_2$ mean the transposition in the
first and in the second matrix spaces, respectively.

\smallskip

Assuming that the above-mentioned properties of the S-matrix
(\ref{Smatrix}) are shared by ${\cal S}$, we can now see that the
string S-matrix allows one to define consistently an elliptic
analog of the ZF algebra, i.e. \bea\label{ZF1}
\begin{aligned}
A_1(z_1)A_2(z_2)&={\cal S}_{12}(z_1,z_2)A_2(z_2)A_1(z_1)\, , \\
A_1^{\dagger}(z_1)A_2^\dagger(z_2)&=A_2^\dagger(z_2)A_1^\dagger(z_1){\cal
S}_{12}(z_1,z_2)\, ,
\end{aligned}
\eea where the creation and annihilation ZF operators are now
functions of the complex variable $z$. In addition, away from the
line $z_1=z_2$ we can impose the following relation between the
creation and annihilation operators \bea \label{ZF2}
A_1(z_1)A_2^\dagger(z_2)= A_2^\dagger(z_2){\cal
S}_{21}(z_2,z_1)A_1(z_1)\, . \eea As usual, the absence of cubic
and higher relations for the ZF operators is guaranteed by the
Yang-Baxter equation for ${\cal S}$. Furthermore, the validity of
relations (\ref{ZF1}), (\ref{ZF2}) for all values of $z_1$ and
$z_2$ is due to unitarity condition (\ref{standardunitarity}).

\smallskip

Transposing the second equation in (\ref{ZF1}) in the first matrix
space we get \bea\nonumber
(A_1^{\dagger}(z_1))^{t_1}A_2^\dagger(z_2)&=A_2^\dagger(z_2){\cal
S}_{12}^{t_1}(z_1,z_2)(A_1^{\dagger}(z_1))^{t_1}\, , \eea On the
other hand, shifting in eq.(\ref{ZF2}) the variable $z_1$ by the
imaginary half-period we obtain \bea\nonumber
A_1(z_1+\om_2)A_2^\dagger(z_2)= A_2^\dagger(z_2){\cal
S}_{12}(z_1+\om_2,z_2)^{-1}A_1(z_1+\om_2)\, . \eea Since the
string S-matrix satisfies the crossing relation  we see that the
algebra structure is compatible with the following identification
\bea A(z+\om_2)=\mathscr{C}^{-1}A^{\dagger}(z)^t\, , \la{A1}\quad
A^{\dagger}(z-\om_2)=-A(z)^t\mathscr{C}\, .  \eea Analogously, we
establish \bea A(z-\om_2)=\mathscr{C}A^{\dagger}(z)^t\, ,
\la{A2}\quad ~~~~A^{\dagger}(z+\om_2)=-A(z)^t\mathscr{C}^{-1}\, .
\eea These relations together with the monodromy properties
(\ref{mon2w}) of the S-matrix  further imply \bea\nonumber
&&A(z+2\om_1)=\Sigma A(z)\, , \quad
A^{\dagger}(z+2\om_1)=A^{\dagger}(z) \Sigma\, ,\\
&&\nonumber  A(z+2\om_2)=\Sigma A(z)\, , \quad
A^{\dagger}(z+2\om_2)=A^{\dagger}(z) \Sigma\, .\eea This means
that the bosonic operators are unchanged under the shift around
the torus while fermionic ones acquire the minus sign. Thus, the
monodromy properties of the \mbox{S-matrix} imply the spin
structure $(-,-)$ for the fermionic ZF operators.

\smallskip

Finally, the generalized unitarity condition (\ref{generalunit})
allows one to impose  the following hermiticity conditions on the
ZF operators: \bea \la{hc} \begin{aligned} &~[A^i(z)]^{\dagger} =
A_i^{\dagger}(z^*)\,
~~~~~~~~~{\rm for}~~~~~0<|{\rm Im}\, z|<\frac{\om _2}{2i}; \\
&~[A^i(z)]^{\dagger} = -A_i^{\dagger}(z^*)\, ~~~~~~~{\rm for}~~~
\frac{\om_2}{2i}<|{\rm Im}\, z| <{\om_2\ov i} \, . \end{aligned} \eea
The hermiticity condition for the ZF creation and annihilation operators in the anti-particle region $\om_2/{2i}<|{\rm Im}\, z| <\om_2/i$  is compatible with the hermiticity condition for the ZF operators in the particle region $0<|{\rm Im}\, z| <\om_2/2i$ and the identifications (\ref{A1}) and (\ref{A2}).

\subsection{Unitarity of the scalar factor in mirror theory}\la{unitarity}
It is clear from the discussion above that the S-matrix of the
mirror theory is obtained from the string S-matrix just by the
shift of the $z$-variables by $\om_2/2$ \bea\la{mS} \widetilde\S (
z_1,  z_2) = \S( z_1+{\om_2\ov 2},  z_2+{\om_2\ov 2})\,. \eea The
momentum of the mirror theory is expressed in terms of the
variable $z$ by eq.(\ref{tpz}) and is real for real values of $z$,
and the generalized unitarity of the mirror S-matrix in terms of
the shifted coordinates $z$ takes the usual form \bea\la{umS}
\left[\widetilde\S (z_1, z_2)\right]^\dagger \widetilde\S (z_1^* ,
z_2^* )= \mathbb{I}\, . \eea This just follows from the
generalized unitarity of the string S-matrix and relation
(\ref{Sm12}) which is a consequence of the crossing equations
\bea\nonumber \left[\widetilde\S_{12} (z_1, z_2)\right]^\dagger =
\S_{21}( z_2^*-{\om_2\ov 2}, z_1^*-{\om_2\ov 2}) = \S_{21}(
z_2^*+{\om_2\ov 2}, z_1^*+{\om_2\ov 2}) =\widetilde\S_{21} (z_2^*,
z_1^*)\,. \eea In fact, since both the S-matrix (\ref{Smatrix})
and the scalar factor $S_0$ satisfy the generalized unitarity
condition and relation (\ref{Sm12}), the same holds for the mirror
theory.

\smallskip

It is of interest to understand how the dressing factor of the mirror theory transforms under the complex conjugation. To this end  we
recall that in the $a=0$ light-cone gauge\footnote{It is easy to
check that the additional $a$-dependent factor   does not break
any of the properties of the S-matrix.} the scalar factor of the
string S-matrix can be written in the form \cite{AF06}
\bea
\la{scf} S_0(z_1,z_2)^2 =s(z_1,z_2) \,\s(z_1,z_2)\, , \quad s(z_1,z_2) =
\frac{x^-_1-x^+_2}{x^+_1-x^-_2}\, {1-{1\ov x_1^+ x_2^-}\ov 1-{1\ov
x_1^- x_2^+}} \,.~~~~~~
\eea Here the gauge-independent dressing
factor $\s(z_1,z_2)$ has the following structure
\cite{AFS} \bea \la{dph} {1\ov i}\ln\s(z_1,z_2)\equiv \theta(z_1,z_2) =
\sum_{r=2}^\infty \sum_{s=r+1}^\infty c_{r,s}(g)\Big[
q_r(z_1)q_{s}(z_2) - q_r(z_2)q_{s}(z_1) \Big]\, ,
\eea where
$q_r(z)=\frac{i}{r-1}\big[(x^+)^{1-r}-(x^-)^{1-r}\big]$ are
the local conserved charges. At any order of the perturbative
expansion in powers of $1/g$ the sums in $r$ and $s$ define the convergent series for
$|x^{\pm}_1|>1$ and $|x^{\pm}_2|>1$. {\it Thus, the S-matrix is by
construction well-defined only in the region where $|x^{\pm}|>1$ and it
should be analytically continued for other values of $x^{\pm}$. }

\smallskip

The string theory dressing factor satisfies the generalized unitarity condition that follows from the fact  that under the complex conjugation the variables $x^\pm$ transform as
$\left[ x^\pm(z)\right]^\dagger = x^\mp(z^*)$. In the mirror theory the variables $x^\pm$ depend on the shifted coordinate $z$ and, as a result, satisfy the following complex conjugation rule
\bea\nonumber
\left[ x^\pm(z+{\om_2\ov 2})\right]^\dagger = {1\ov x^\mp(z^*+{\om_2\ov 2})}\,.
\eea
By using this rule one can easily check that the factor $s(z_1,z_2)$ in (\ref{scf}) transforms under the complex conjugation as follows
\bea\la{ccs}
\left[s(z_1^*, z_2^*)\right]^\dagger s(z_1 , z_2 )=\left( {x^-_1 x^+_2\ov x^+_1 x^-_2}\right)^2\, ,
\eea
where $x^\pm_i = x^\pm(z_i+{\om_2\ov 2})$. Taking into account that the scalar factor $S_0$ of the mirror theory satisfies the generalized unitarity condition, we find the complex conjugation rule for the dressing factor of the mirror theory
\bea\la{scs}
\left[\s(z_1^*, z_2^*)\right]^\dagger \s(z_1 , z_2 )=\left( {x^+_1 x^-_2\ov x^-_1 x^+_2}\right)^2\, .
\eea
In particular, for real values of $z$'s corresponding to real $\widetilde p$'s the dressing factor of the mirror theory is not unitary.

\smallskip

It is interesting
to note that the scalar factor can be split into a product of two
factors satisfying the generalized unitarity condition in both string and mirror theory
 \bea\la{sp1} S_0(z_1,z_2)^2 =
\frac{x^-_1-x^+_2}{x^+_1-x^-_2}\, {x_1^+ x_2^- -1\ov
x_1^- x_2^+ - 1}\,\times \,  \frac{x_1^-x_2^+}{x_1^+x_2^-} \,\s(z_1,z_2) \,. \eea
Another interesting splitting is given by
 \bea\la{sp2} S_0(z_1,z_2)^2 =
\frac{u_1-u_2 - {2i\ov g}}{u_1-u_2 + {2i\ov g}}\,\times \,  \left({1-{1\ov x_1^+ x_2^-}\ov 1-{1\ov x_1^- x_2^+}}\right)^2 \s(z_1,z_2) \,. \eea
This splitting is
useful for analyzing the bound state spectrum of the mirror
model.

\smallskip

Knowing the series representation  for the dressing phase in the
original theory \cite{BHL}, it is interesting to understand what
is precisely the source of its unitarity breakdown in the mirror
theory.

\smallskip

To clarify this issue,  we  recall that the dressing phase can be
conveniently written in terms of a single function $\chi(x_1,x_2)$
\cite{AF06} \bea \theta (z_1,z_2) &=&
\chi(x_1^+,x_2^+)-\chi(x_1^+,x_2^-)-\chi(x_1^-,x_2^+)+\chi(x_1^-,x_2^-)- \nonumber \\
&-&\chi(x_2^+,x_1^+)+\chi(x_2^-,x_1^+)+\chi(x_2^+,x_1^-)-\chi(x_2^-,x_1^-)\,
,\nonumber \eea which admits the following strong coupling
expansion \bea\nonumber
\chi(x_1,x_2)=g\sum_{n=0}^{\infty}\chi^{(n)}(x_1,x_2)\,
\Big(\frac{g}{2}\Big)^{-n}\, . \eea Here \bea\nonumber
\chi^{(0)}(x_1,x_2)=-\frac{1}{x_2}-\frac{x_1x_2-1}{x_2}\log\frac{x_1
x_2-1}{x_1 x_2} \eea is the leading AFS factor \cite{AFS}. The
next-to-leading contribution is \cite{HL}: \bea \label{hl}
\chi^{(1)}(x_1,x_2)=&-&\frac{1}{2\pi}{\rm
Li}_2\frac{\sqrt{x_1}-1/\sqrt{x_2}}{\sqrt{x_1}-\sqrt{x_2}}
-\frac{1}{2\pi}{\rm
Li}_2\frac{\sqrt{x_1}+1/\sqrt{x_2}}{\sqrt{x_1}+\sqrt{x_2}} \nonumber \\
&+&\frac{1}{2\pi}{\rm
Li}_2\frac{\sqrt{x_1}+1/\sqrt{x_2}}{\sqrt{x_1}-\sqrt{x_2}}
+\frac{1}{2\pi}{\rm
Li}_2\frac{\sqrt{x_1}-1/\sqrt{x_2}}{\sqrt{x_1}+\sqrt{x_2}}\, .
 \eea
All higher terms are rational functions of $x_1,x_2$ \cite{BHL}.
As we will now show, the unitarity breakdown of the dressing phase
is due to the leading AFS contribution only, the
Hern\'andez-L\'opez term (\ref{hl}), as well as all higher order
terms do not influence the unitarity condition.

To simplify the notations in what follows we only consider the
case of real $z$'s in the mirror theory. It is easy to see that
the complex conjugate of the function $\chi^{(0)}$ is given by
\bea\nonumber
\big[\chi^{(0)}(x_1^{\pm},x_2^{\pm})\big]^*&=&-\chi^{(0)}(x_2^{\mp},x_1^{\mp})
-\frac{i\pi+1}{x_1^{\mp}}+(i\pi-1)x_2^{\mp}
-\Big(\frac{1}{x_1^{\mp}}-x_2^{\mp}\Big)\log x_1^{\mp}x_2^{\mp}\,
. \eea Using this formula for computing the leading value
$\theta_{\rm AFS}$, we find that the contribution of
non-logarithmic terms cancels out and we get \bea \theta^{*}_{\rm
AFS} =\theta_{\rm AFS}
&+&g\frac{x_1^--x_2^-}{x_1^-x_2^-}(1-x_1^-x_2^-)\log
x_1^-x_2^-+g\frac{x_1^+-x_2^+}{x_1^+x_2^+}(1-x_1^+x_2^+)\log
x_1^+x_2^+ \nonumber \\  \nonumber
&-&g\frac{x_1^--x_2^+}{x_1^-x_2^+}(1-x_1^-x_2^+)\log x_1^-x_2^+
+g\frac{x_2^--x_1^+}{x_1^+x_2^-}(1-x_1^+x_2^-)\log x_1^+x_2^- \, .
\eea Using identity (\ref{consxpxm}), it is easy to show that all
logarithmic terms are neatly combined to produce the following
answer \bea \label{AFSc} \theta^{*}_{\rm AFS} =\theta_{\rm
AFS}+i\log\Big(\frac{x_1^+x_2^-}{x_1^-x_2^+}\Big)^2\, , \eea which
coincides with the  logarithmic form of eq.(\ref{scs}).

\smallskip

Since we have shown that the shift of the phase under the complex
conjugation occurs due to the leading contribution, all the higher
order terms in the expansion of $\theta$ must be real functions.
To convince oneself that this is indeed the case, we consider the
next-to-leading term in the strong coupling expansion of $\theta$.
As was shown in \cite{BHL}, this term admits the following
representation \bea \theta_{\rm
HL}=\psi(q_1^+-q_2^+)-\psi(q_1^+-q_2^-)-\psi(q_1^--q_2^+)+\psi(q_1^--q_2^-)\,
. \eea Here the function $\psi(q)$ is \bea
\psi(q)=\frac{1}{2\pi}{\rm Li}_2(1-e^{iq})-\frac{1}{2\pi}{\rm
Li}_2(1-e^{iq+i\pi})-\frac{i}{2}\log(1-e^{iq+i\pi})+\frac{\pi}{8}\,
, \eea
where the variables $q^{\pm}$ are related to $x^{\pm}$ through
\bea e^{iq^{\pm}}=\frac{x^{\pm}+1}{x^{\pm}-1}\, . \eea Taking into
account the conjugation rule in the mirror theory,
eq.(\ref{con2}), we obtain \bea (q^{\pm})^*=-q^{\mp}-\pi\, . \eea
Since $\theta_{\rm HL}$ depends on the difference of two $q's$,
the shift by $\pi$ arising upon the complex conjugation will
cancel out. Thus, taking the complex conjugate we obtain \bea
\theta_{\rm
HL}^*=\bar{\psi}(q_1^--q_2^-)-\bar{\psi}(q_1^--q_2^+)-\bar{\psi}(q_1^+-q_2^-)+\bar{\psi}(q_1^+-q_2^+)\,
, \eea where the function $\bar{\psi}(q)$ is defined as \bea
\bar{\psi}(q)= \frac{1}{2\pi}{\rm
Li}_2(1-e^{iq})-\frac{1}{2\pi}{\rm
Li}_2(1-e^{iq-i\pi})-\frac{i}{2}\log(1-e^{iq-i\pi})+\frac{\pi}{8}\,
. \eea Taking into account the following transformation property
of the dilogarithm function
$$
{\rm Li}_2(1-e^{iq-i\pi})={\rm Li}_2(1-e^{iq+i\pi-2\pi i})={\rm
Li}_2(1-e^{iq+i\pi})+2\pi i \log(1-e^{iq})\, ,
$$
we find that
$$
\bar{\psi}(q)=\psi(q)+\pi\, .
$$
Since the shift by $\pi$ in the previous formula does not
contribute to $\theta^*_{\rm HL}$, we conclude that $\theta^*_{\rm
HL}=\theta_{\rm HL}$. Finally, by working out several higher order
terms $\chi^{(k)}$, one can easily check that they always lead to
the real functions $\theta$, in accord with eqs.(\ref{scs}) and
(\ref{AFSc}).

\smallskip

Thus, we have shown that under the double Wick rotation the scalar
factor remains unitary, while the dressing factor does not; the
non-unitarity of the dressing factor is  only due to the leading
contribution $\theta_{\rm AFS}$, which is another distinguished
property of $\theta_{\rm AFS}$. Concluding this section, we note
that it would be interesting to understand whether the BES factor
\cite{BES} exhibits the same kind of non-unitarity behavior in the
mirror theory.


\section{Bethe ansatz equations}

In this section we discuss the nested Bethe equations for the
light-cone string theory on $\AdS$ and its mirror model. These
equations  based on the $\su(2|2)\oplus\su(2|2)$-invariant string
S-matrix \cite{AFZ} were recently  derived by using the algebraic
\cite{MM} and the coordinate \cite{Le} Bethe ansatz approaches. In
the sector with even winding number, i.e. with the total momentum
$P$ satisfying $e^{iP/2}=1$, the set of equations found in these
papers coincides with the one previously obtained in \cite{BS,B}
by using the spin chain description of the gauge theory. It
appears, however, that in the sector with odd winding number,
where $e^{iP/2}=-1$, the Bethe equations by \cite{MM,Le} differ
from the ones derived from the gauge theory. The origin of this
disagreement can be traced back to the fact that in the light-cone
gauge the fermions of the string sigma model are {\it
anti-periodic in the odd winding number sector} \cite{AAF1,We},
and this changes the periodicity conditions for wave functions
which one imposes to get the Bethe equations. Indeed, in the
light-cone gauge one of the fields, an angle $\phi$ which
parametrizes the five-sphere,  appears to be unphysical and it is
solved in terms of (transversal) physical fields. In particular,
the equation of motion for $\phi$ implies
$$
\phi(2\pi)-\phi(0)=P\, .
$$
Since $\phi$ enters into parametrization of the five-sphere via
$e^{i\phi}$, the closed string periodicity condition for physical
fields gives rise to the winding sectors
$$
\phi(2\pi)-\phi(0)=2\pi m\, ,
$$
where $m$ is an integer. Now, we recall that fermions of the
original string sigma-models are charged w.r.t. the U(1) isometry
acting on $\phi$ as $\phi\to \phi+{\rm const}$. Also, the
Wess-Zumino term in the sigma-model action contains $e^{i\phi}$,
i.e. it is non-local in terms of physical fields. To uncharge the
fermions under the U(1) isometry, as well as to make the
Wess-Zumino term local, one has to redefine the fermions as
$$
\psi\to e^{\sfrac{i}{2} \phi}\psi \, .
$$
Thus, the redefined fermions acquire the periodicity properties
which do depend on the winding sector
$$
\psi(0)= e^{\sfrac{i}{2}P}\psi(2\pi)=e^{i\pi m}\psi(2\pi)\, ,
$$
i.e. they are periodic in the even winding sector and they are
ant-periodic in the odd winding sector \cite{AAF1,We}.

\smallskip

As a result, the Bethe equations obtained in \cite{MM,Le}
correctly describe the light-cone string theory in the sector with
periodic fermions only. Changing the boundary conditions for
fermions to anti-periodic, one derives a new set of Bethe
equations which does agree with the gauge theory one for physical
states satisfying \mbox{$e^{iP/2}=-1$}.

\subsection{BAE  for a model with the $\su(2|2)$-invariant S-matrix}
The asymptotic states of both the original and the mirror theory
are constructed by applying the ZF operators
$A^{\dagger}_{M\dot{M}}$ to the vacuum state. The indices $M$ and
$\dot{M}$ are associated  with two factors of the
centrally-extended $\su(2|2)\oplus\su(2|2)$ algebra; the latter
being the symmetry algebra of the light-cone string theory
\cite{B,AFPZ}. For our present purpose it is convenient to think
about the ZF operator as being  a product of two (anti)commuting
operators each transforming in a fundamental representation of
$\su(2|2)$: $A^{\dagger}_{M\dot{M}}\sim
A^{\dagger}_{M}A^{\dagger}_{\dot{M}}$. Since the string S-matrix
is a tensor product of two $\su(2|2)$-invariant S-matrices, the
Bethe equations for the string model are, in a sense, the square
of the Bethe equations for a model with the $\su(2|2)$-invariant
S-matrix. We start with discussing the Bethe equations for such a
model.

\smallskip

The multi-particle wave function which satisfies the Bethe
periodicity conditions  can be written as a superposition of the
asymptotic states (see appendix \ref{bae1} for details)
 \bea\label{A}
|\Psi\rangle=\sum \Psi^{M_1\cdots M_{K^{\mathrm{I}}}}
 A_{M_1}^{\dagger}(p_1)\cdots
A_{M_{K^{\rm I }}}^{\dagger}(p_{K^I})|0\rangle\, , \eea where $K^{\rm I}$ is a
number of particles in the asymptotic state and $p_i$ are their momenta.
Denote by $N(M)$ the number of particles of type $M$ (that is number of indices of type $M$) occurring in
the wave function (\ref{A}). Obviously,
$$
K^{\mathrm{I}}=N(1)+N(2)+N(3)+N(4)\,.
$$
Since the scattering is elastic, the number of particles $K^{\rm
I}$ is a conserved quantity.

\smallskip

The form of the Bethe equations derived through the nesting
procedure of the coordinate Bethe ansatz depends on the choice of
the initial reference state. Due to the $\su(2)^2$ bosonic
symmetry there are two inequivalent choices for a model with the
$\su(2|2)$-invariant S-matrix. This is obviously related to the
two forms of the Dynkin graph for $\su(2|2)$.

\smallskip

First, one can choose  a ``bosonic'' reference state which is
created by acting with $K^{\rm I}$ bosonic operators
$A_{1}^\dagger$ on the vacuum:
$$
A_{1}^\dagger(p_1)\ldots A_{1}^\dagger(p_{K^{\rm
I}})|0\rangle\, .
$$
Then, we define \bea\nonumber
K^{\mathrm{II}}_+=2N(2)+N(3)+N(4)\,,\quad
K^{\mathrm{III}}=N(2)+N(4)\,.  \eea It appears that in the
scattering process not only $K^{\mathrm{I}}$ but also these
numbers are conserved \cite{B}. By this reason, the values of
 $K^{\mathrm{II}}_+$ and
$K^{\mathrm{III}}$ are the {\it same} for
any term in the sum (\ref{A}). In particular,  $K^{\mathrm{II}}_+$
plays the role of the fermionic number, because in the background of the $A_1^\dagger$-particles $A_2^{\dagger}$
counts for two fermions. The number $K^{\mathrm{III}}$ has a
similar meaning.

\smallskip

Then the asymptotic Bethe
equations based on the $\su(2|2)$-invariant
S-matrix  for a sigma model on a circle of length $R$ and with (anti)-periodic
fermions can be written in the form \cite{B, MM, Le}
\begin{eqnarray}
 e^{i p_{k} R}&=& \prod_ {\textstyle\atopfrac{l=1}{l\neq k}}
^{K^{\mathrm{I}}}S_{0}(p_{k},p_{l})\frac{x_{k}^{+}-x_{l}^{-}}{x_{k}^{-}-x_{l}^{+}}
\sqrt{\frac{x_l^+ x_k^-}{x_l^- x_k^+}}
\prod_{l=1}^{K^{\mathrm{II}}_+}\frac{{x_{k}^{-}-y_{l}}}{x_{k}^{+}-y_{l}}
\sqrt{\frac{x_k^+}{x_k^-}} \nonumber \\
(-1)^\epsilon &=&\prod_{l=1}^{K^{\mathrm{I}}}\frac{y_{k}-x^{+}_{l}}{y_{k}-x^{-}_{l}}\sqrt{\frac{x_l^-}{x_l^+}}
\prod_{l=1}^{K^{\mathrm{III}}}\frac{v_{k}-w_{l}+\frac{i}{g}}{v_{k}-w_{l}-\frac{i}{g}} \la{BEsu22a}\\
\nonumber
1&=&\prod_{l=1}^{K^{\mathrm{II}}_+}\frac{w_{k}-v_{l}-\frac{i}{g}}{w_{k}-v_{l}+\frac{i}{g}}
\prod_ {\textstyle\atopfrac{l=1}{l\neq
k}}^{K^{\mathrm{III}}}\frac{w_{k}-w_{l}+\frac{2i}{g}}{w_{k}-w_{l}-\frac{2i}{g}}.
\end{eqnarray}
Here  $\epsilon = 0$ for a sector with periodic fermions and
$\epsilon = 1$ for a sector with anti-periodic fermions, $x_k^\pm$
depend on the momentum $p_k$ of the model, $y_l$ and $w_l$ are
auxiliary roots of the second and third levels, respectively, and
$v=y+\frac{1}{y}$.

\smallskip

On the other hand, if one chooses  a ``fermionic'' reference state
created by $K^{\rm I}$ fermionic operators $A_{3}^\dagger$:
$$
A_{3}^\dagger(p_1)\ldots A_{3}^\dagger(p_{K^{\rm
I}})|0\rangle\, ,
$$
then, one should define \bea\nonumber
K^{\mathrm{II}}_-=2N(4)+N(1)+N(2)\,,\quad
K^{\mathrm{III}}=N(2)+N(4)\,, \eea because these numbers are also
conserved in the scattering process. Then,  $K^{\mathrm{II}}_-$
plays the role of the bosonic number, because in the background of
the $A_3^\dagger$-particles $A_4^{\dagger}$ counts for two bosons.

\smallskip

Then the corresponding Bethe equations take the following form
\begin{eqnarray}
e^{i p_{k} R}&=&(-1)^\epsilon\, \prod_ {\textstyle\atopfrac{l=1}{l\neq k}}
^{K^{\mathrm{I}}}S_{0}(p_{k},p_{l})
\prod_{l=1}^{K^{\mathrm{II}}_-} \frac{{x_{k}^{+}-y_{l}}}{x_{k}^{-}-y_{l}}
\sqrt{\frac{x_k^-}{x_k^+}} \nonumber \\
(-1)^\epsilon &=&\prod_{l=1}^{K^{\mathrm{I}}}\frac{y_{k}-x^{+}_{l}}{y_{k}-x^{-}_{l}}\sqrt{\frac{x_l^-}{x_l^+}}
\prod_{l=1}^{K^{\mathrm{III}}}\frac{v_{k}-w_{l}+\frac{i}{g}}{v_{k}-w_{l}-\frac{i}{g}} \la{BEsu22b}\\
\nonumber
1&=&\prod_{l=1}^{K^{\mathrm{II}}_-}\frac{w_{k}-v_{l}-\frac{i}{g}}{w_{k}-v_{l}+\frac{i}{g}}
\prod_ {\textstyle\atopfrac{l=1}{l\neq
k}}^{K^{\mathrm{III}}}\frac{w_{k}-w_{l}+\frac{2i}{g}}{w_{k}-w_{l}-\frac{2i}{g}}.
\end{eqnarray}
Equations (\ref{BEsu22b}) can be derived either by using the
nesting procedure of the coordinate Bethe ansatz (see appendix
\ref{bae2} for an example) or by applying the duality
transformation discussed in \cite{BS} to eqs.(\ref{BEsu22a}).
Comparing the two sets of Bethe equations (\ref{BEsu22a}) and
(\ref{BEsu22b}), we see that only the first lines in two sets are
different. Let us stress, however, that in general
$K^{\mathrm{II}}_-\neq K^{\mathrm{II}}_+$. We further note that
the bosonic reference state corresponds to, say, $\eta_1=1$ and
the fermionic one  to $\eta_1=-1$, where $\eta_1$ and $\eta_2$ are
the gradings introduced in \cite{BS}.

\subsection{BAE based on the $\su(2|2)\oplus\su(2|2)$-invariant string
S-matrix} The  Bethe equations based on the
$\su(2|2)\oplus\su(2|2)$-invariant string S-matrix for both string
and mirror models can be now easily written by
 taking a ``product" of two copies of the Bethe equations for the $\su(2|2)$-invariant model.
 Since any of the two sets, (\ref{BEsu22a}) and (\ref{BEsu22b}),
 can be used there are four different forms of the asymptotic Bethe equations  based on the $\su(2|2)\oplus\su(2|2)$-invariant S-matrix \cite{BS}. The corresponding bosonic reference states of the coordinate Bethe ansatz are of the form
$$
A_{1\dot{1}}^\dagger(z_1)\ldots A_{1\dot{1}}^\dagger(z_{K^{\rm
I}})|0\rangle\,,\quad \eta_1=\eta_2=1\,;\quad A_{3\dot{3}}^\dagger(z_1)\ldots A_{3\dot{3}}^\dagger(z_{K^{\rm
I}})|0\rangle\,,\quad \eta_1=\eta_2=-1\,,
$$
and fermionic reference states are
$$
A_{1\dot{3}}^\dagger(z_1)\ldots A_{1\dot{3}}^\dagger(z_{K^{\rm
I}})|0\rangle\,,\quad \eta_1=-\eta_2=1\,;\quad A_{3\dot{1}}^\dagger(z_1)\ldots A_{3\dot{1}}^\dagger(z_{K^{\rm
I}})|0\rangle\,,\quad \eta_1=-\eta_2=-1\, ,
$$
where for the original theory the $z$-variables lie on the real
line, while for the mirror theory they have ${\rm Im}\,  z=\om_2/2i$,
and we also indicated the corresponding gradings.

\smallskip

To discuss the bound states of the light-cone string  sigma model,
it is convenient to choose as the reference state  the one created
by the bosonic operators $A_{1\dot{1}}^\dagger$. These reference
states are dual to gauge theory operators from the $\su(2)$
sector. Then the corresponding Bethe equations based on the
$\su(2|2)\oplus\su(2|2)$-invariant string S-matrix can be written
in the form \cite{BS, B, MM, Le}
\begin{eqnarray}\la{BEn}
 e^{i p_{k}J}&=& \prod_ {\textstyle\atopfrac{l=1}{l\neq k}}
^{K^{\mathrm{I}}}\left[S_{0}(p_{k},p_{l})\frac{x_{k}^{+}-x_{l}^{-}}{x_{k}^{-}-x_{l}^{+}}
\sqrt{\frac{x_l^+ x_k^-}{x_l^- x_k^+}}\, \right]^2
\prod_{\a=1}^{2}\prod_{l=1}^{K^{\mathrm{II}}_{(\a)}}\frac{{x_{k}^{-}-y_{l}^{(\a)}}}{x_{k}^{+}-y_{l}^{(\a)}}
\sqrt{\frac{x_k^+}{x_k^-}} \nonumber \\
(-1)^\epsilon&=&\prod_{l=1}^{K^{\mathrm{I}}}\frac{y_{k}^{(\a)}-x^{+}_{l}}{y_{k}^{(\a)}-x^{-}_{l}}\sqrt{\frac{x_l^-}{x_l^+}}
\prod_{l=1}^{K^{\mathrm{III}}_{(\a)}}\frac{v_{k}^{(\a)}-w_{l}^{(\a)}+\frac{i}{g}}{v_{k}^{(\a)}-w_{l}^{(\a)}-\frac{i}{g}} \label{BE}\\
\nonumber
1&=&\prod_{l=1}^{K^{\mathrm{II}}_{(\a)}}\frac{w_{k}^{(\a)}-v_{l}^{(\a)}-\frac{i}{g}}{w_{k}^{(\a)}-v_{l}^{(\a)}+\frac{i}{g}}
\prod_ {\textstyle\atopfrac{l=1}{l\neq
k}}^{K^{\mathrm{III}}_{(\a)}}\frac{w_{k}^{(\a)}-w_{l}^{(\a)}+\frac{2i}{g}}{w_{k}^{(\a)}-w_{l}^{(\a)}-\frac{2i}{g}}\,.
\end{eqnarray}
Here we take into account that the string sigma model in the $a=0$
light-cone gauge  is defined on a circle of length $J$,  $\a =1,2$
reflects the two copies of $\su(2|2)$ and $y_l^{(\a)}$ and
$w_l^{(\a)}$ are auxiliary roots of the second and third levels,
respectively, and   $v=y+\frac{1}{y}$.

For the reader's convenience
we point out that the excitation numbers in the set of Bethe equations are related to the ones used in
\cite{BS} as follows
$$(K_{(1)}^{\rm III}\,,\ K_{(1)}^{\rm II}\,,\ K^{\rm I}\,,\ K_{(2)}^{\rm II}\,,\ K_{(2)}^{\rm III}) = (K_2,K_1+K_3,K_4,K_5+K_7,K_6)\,,$$
and the Dynkin labels $[q_1,p,q_2]$ of $\su(4)$ and $[s_1,s_2]$ of
$\su(2)\oplus\su(2)\subset \su(2,2)$ are expressed in terms of the
excitation numbers by the following formulas\footnote{Let us note
in passing that in recent papers \cite{St, st2} the anomalous
dimension of the operator $\Tr{\cal F}^L$ was computed by using
the asymptotic Bethe ansatz with an understanding that in the
large $L$ limit one may trust the corresponding result to an
arbitrary loop order. One can notice, however, that the excitation
pattern of Bethe roots for the operator is $(K_{(1)}^{\rm III}\,,\
K_{(1)}^{\rm II}\,,\ K^{\rm I}\,,\ K_{(2)}^{\rm II}\,,\
K_{(2)}^{\rm III}) = (0,2L-3,2L-2,2L-4,L-2)$ with $J={3\ov 2}$,
and, therefore, one would expect the breakdown of the asymptotic
ansatz due to the finite size effects already at two loops. It may
happen that the asymptotic ansatz could still be used to determine
the leading $L$ behavior of the anomalous dimension of $\Tr{\cal
F}^L$ if the finite-size corrections are subleading at large $L$,
but this is currently unknown. } \bea
\begin{aligned} & q_1=K^{\rm I}-K_{(1)}^{\rm
II}\, ,  &~~ s_1= K_{(1)}^{\rm II}-2K_{(1)}^{\rm III}  \, ,\\
& p=J-K^{\rm I }+\frac{1}{2}(K_{(1)}^{\rm II}+K_{(2)}^{\rm II})\,
, &~~
s_2= K_{(2)}^{\rm II}-2K_{(2)}^{\rm III} \, ,\\
& q_2=K^{\rm I}-K_{(2)}^{\rm II}\, .
\end{aligned}
\eea

\medskip

To analyze the bound states of the mirror theory, it is more
convenient, however, to choose as an initial reference state the
one created by the operators $A_{3\dot 3}^\dagger\ $. The reason
is that the operators $A_{3\dot 3}^\dagger$ create states from the
$\sl(2)$ sector, and, as we have seen, it is this sector which
gives rise to mirror magnons. Analogously, there are $M$-particle
bound states made only out of the $A_{3\dot 3}^\dagger$-type
particles.
\smallskip

 If we choose in the mirror theory the above-described reference
 state
then the corresponding Bethe equations take the form
\begin{eqnarray}\la{BEn2}
e^{i\tp_{k} R}&=& \prod_ {\textstyle\atopfrac{l=1}{l\neq k}}
^{K^{\mathrm{I}}}\left[S_{0}(\tp_{k},\tp_{l})\right]^2
\prod_{\a=1}^{2}\prod_{l=1}^{K^{\mathrm{II}}_{(\a)}}\frac{{x_{k}^{+}-y_{l}^{(\a)}}}{x_{k}^{-}-y_{l}^{(\a)}}
\sqrt{\frac{x_k^-}{x_k^+}} \nonumber \\
-1&=&\prod_{l=1}^{K^{\mathrm{I}}}\frac{y_{k}^{(\a)}-x^{+}_{l}}{y_{k}^{(\a)}-x^{-}_{l}}\sqrt{\frac{x_l^-}{x_l^+}}
\prod_{l=1}^{K^{\mathrm{III}}_{(\a)}}\frac{v_{k}^{(\a)}-w_{l}^{(\a)}+\frac{i}{g}}{v_{k}^{(\a)}-w_{l}^{(\a)}-\frac{i}{g}} \\
\nonumber
1&=&\prod_{l=1}^{K^{\mathrm{II}}_{(\a)}}\frac{w_{k}^{(\a)}-v_{l}^{(\a)}-\frac{i}{g}}{w_{k}^{(\a)}-v_{l}^{(\a)}+\frac{i}{g}}
\prod_ {\textstyle\atopfrac{l=1}{l\neq
k}}^{K^{\mathrm{III}}_{(\a)}}\frac{w_{k}^{(\a)}-w_{l}^{(\a)}+\frac{2i}{g}}{w_{k}^{(\a)}-w_{l}^{(\a)}-\frac{2i}{g}}.
\end{eqnarray}
Note that in the mirror model we do not have $(-1)^\epsilon$ in
the middle equation because the fermions are always
anti-periodic\footnote{We are grateful to R. Janik and M. Martins
for drawing our attention to this point.} with respect to
$\tilde\sigma$. In terms of excitation numbers, the Dynkin labels
read now as follows \bea
\begin{aligned} & q_1=K^{\rm II}_{(1)}-K_{(1)}^{\rm
III}\, ,  &~~ s_1= K^{\rm I}-K_{(1)}^{\rm II}  \, ,\\
& p=J-\frac{1}{2}(K_{(1)}^{\rm II}+K_{(2)}^{\rm II})+K_{(1)}^{\rm
III}+K_{(2)}^{\rm III}\, , &~~
s_2= K^{\rm I}-K_{(2)}^{\rm II} \, ,\\
& q_2=K^{\rm II}_{(2)}-K_{(2)}^{\rm III}\, .
\end{aligned}
\eea

\section{Bound states of the $\AdS$ gauge-fixed model}
\la{bound} Bound states arise as poles of the multi-particle
S-matrix corresponding to complex values of the particle momenta,
see e.g. \cite{Faddeev:1996iy}. In the thermodynamic limit they
are described by string-like solutions known as ``Bethe strings".
In this section we discuss in detail the bound states of the
string sigma-model. They have been already analyzed in
\cite{D,CDO}. The main outcome of this analysis is that the
$M$-particle bound states comprise into short (BPS) multiplets of
the centrally extended $\su(2|2)\oplus \su(2|2)$ symmetry algebra.
Although the S-matrix exhibits additional simple and double poles
beyond those corresponding to the BPS multiplets, these
singularities do not lead however to the appearance of new bound
states \cite{DHM}. In other words, the only bound states in the
theory are the BPS ones. As we will see, they exist for all values
of the (real) bound state  momentum $-\pi\leq p\leq \pi$, but have
a rather intricate structure. Moreover, depending on the choice of
the physical region for a given value of the bound state momentum
there could be 1, 2 or $2^{M-1}$ $M$-particle bound states sharing
the same set of global conserved charges: $Q_r = \sum_{i=1}^M
q_r(z_i)$. It is unclear to us whether this indicates that the
actual physical region is the one that contains only a single
$M$-particle bound state (it is the one with $|x^\pm|>1$) or it is
a sign of a hidden symmetry of the model responsible for the
degeneracy of the spectrum.

\subsection{Two-particle bound states}

Let us consider a bound state made of two excitations from the
$\su(2)$ sector of the string sigma-model. In terms of the ZF
creation operators we can think about this state as
$$
A_{1\dot 1}^\dagger(p_1)A_{1\dot 1}^\dagger(p_2)|0\rangle\,
,
$$
where the particle momenta $p_1$ and $p_2$ are complex. We
find it convenient to parametrize the momenta as follows \bea p_1
= \frac{p}{2} -i q\,, ~~\quad~~ p_2 = \frac{p}{2} + i q\,,
~~\quad~~ {\rm Re}\, q
> 0\,,
\eea where  $p$ is the real total momentum of the bound state.
When $q$ is real then $p_1$ and $p_2$ are complex conjugate to
each other and the energy of the corresponding bound state being
the sum of the (complex) energies of individual particles is
obviously real. Interestingly, as we will show below, there
necessarily exists a branch of BPS bound states which corresponds
to complex values of $q$ with ${\rm Re}\, q
> 0$. Such solutions can be reinterpreted as solutions
parametrized by a new real variable $q$: $q\to {\rm Re}\, q$ and
for which the real parts of $p_1$ and $p_2$ are not anymore equal
to each other. Of course, one has to check that the energy of
these solutions is real.

\smallskip

The first equation in the set of the Bethe equations \cite{AFS}
takes the form \bea\la{BEsu2} e^{i(p/2+{\rm Im}\, q) L}e^{ {\rm Re}\, q \,
L}= e^{iP}\prod_ {l=2} ^{K^{\mathrm{I}}}
\frac{x_{1}^{+}-x_{l}^{-}}{x_{1}^{-}-x_{l}^{+}}{1-{1\ov x_1^+
x_l^-}\ov 1-{1\ov x_1^- x_l^+}}\,\s_{1l}\,
, \eea where $P =p_1+p_2+\cdots +p_{K^{\mathrm{I}}}$ and $L=J +
K^{\mathrm{I}}$ with $J$ being one of the global charges
corresponding to the isometries of the five-sphere.

\smallskip

We see that for large $L$ the l.h.s. is exponentially divergent.
 Then, there should exist a root $p_2$ such that  for
${\rm Re}\, q>0$ we have\footnote{We assume here and in what
follows that the dressing factor $\sigma_{12}$ is non-singular on
solutions of the bound state equation. } \bea \label{bseqs} \left(
x_1^- - x^+_2\right) \Big(1-{1\ov x_1^- x_2^+}\Big)\, \sim\, e^{-
{\rm Re}\, q\,  L}\,. \eea In the infinite $L$ limit
eq.(\ref{bseqs}) becomes \bea \label{bseqs2} \left( x_1^- -
x^+_2\right) \Big(1-{1\ov x_1^- x_2^+}\Big)= 0\, ,\eea which is
equivalent to \bea\label{cond}
 x_1^- - x^+_2=0\quad ~~{\rm or}~~\quad
 1-{1\ov x_1^- x_2^+}=0\,.
\eea The first equation  \bea \label{bps}
 x_1^- - x^+_2=0
\eea implies that the central charges corresponding to the
two-particle bound state saturate the BPS condition \cite{D}
\bea\label{bpsen}
H^2=2^2+4g^2\sin^2\frac{p}{2}\, . \eea On the contrary, solutions
of the second equation in (\ref{cond}) do not saturate the BPS
bound, and as was argued in \cite{DHM}, this pole of the S-matrix does not correspond to a bound state.

\smallskip

It is easy to see that equation (\ref{bps}) is equivalent to
vanishing the following fourth order polynomial\footnote{For
any $p$ there are two solutions for $x^-$ and, therefore, for
$x^+=e^{ip}x^-$. The fourth order polynomial is universal and it
does not depend on which solution for $x^-$ we take. } in the
variable $t=\cos\frac{p}{2}$
\bea
4e^q(t-e^q)(1-e^qt)+g^2(t^2-1)(1-2e^q t+e^{2q})^2=0\, .
\label{4thpol}
\eea
The equation has four solutions which can be cast to the following simple form
\bea\la{gsbse4}
e^q=\frac{ \left(\sqrt{1+g^2 \sin ^2\frac{p}{2}}\, \pm 1\right)
   \left( \cos \frac{p}{2}\,\sqrt{1+g^2 \sin ^2\frac{p}{2}}\pm \sqrt{\cos
   ^2\frac{p}{2}-g^2 \sin ^4\frac{p}{2}}\right)}{g^2\sin^2\frac{p}{2}}\,,~~~~~
\eea
where any choice of the $\pm$ sign is possible.

\smallskip

Analysis of eq.(\ref{gsbse4}) immediately  shows that solutions
corresponding to real values of $q$ exist if and only if the total
momentum $p$ does not exceed a critical value $p_{\rm cr}$
determined by \bea\la{pcr} \sin^2\frac{p_{\rm cr}}{2}=\,
 {1\ov 2g^2}\left(\sqrt{1+4g^2}
-1\right)\, .
 \eea
For any given $p$ obeying $|p|\, < \, p_{\rm cr}$ equation
(\ref{4thpol}) has four real roots for $q$, two of them are
positive and the other two are negative. According to our
assumption ${\rm Re}\, q>0$, only positive roots are acceptable.\footnote{The solutions with negative $q$ correspond to bound states of anti-particles with negative energy.}
They are given by the formula
\bea\la{gsbse}
e^{q_{\pm}}=\frac{ \left(\sqrt{1+g^2 \sin ^2\frac{p}{2}}\, + 1\right)
   \left( \cos \frac{p}{2}\,\sqrt{1+g^2 \sin ^2\frac{p}{2}}\pm \sqrt{\cos
   ^2\frac{p}{2}-g^2 \sin ^4\frac{p}{2}}\right)}{g^2\sin^2\frac{p}{2}}\, .~~~~~
\eea
Various expansions of eq.(\ref{gsbse}) for small
and large values of $g$ can be found in Appendix \ref{expan}.

 It turns out that from the two positive roots only the
smaller one, $q_-$, falls inside the region confined by the curves
$|x^{\pm}|=1$. We therefore arrive at the conclusion that inside
the region $|x^{\pm}|>1$ there is a unique solution with real $p$
and $q$, and it exists if and only if \footnote{The energy of the
corresponding bound state is $ E<\sqrt{ 2\sqrt{1+4g^2} +2}\,. $}
\bea |p|\, < \, p_{\rm cr}\,,~~\quad~~ 0\le q\, <\, \log {{2g}
+\sqrt{ 2\sqrt{1+4g^2} -2} \ov\sqrt{1+4g^2} -1}\, . \eea The
second solution with $q=q_+$ lies outside the region with
$|x^{\pm}|>1$ but inside the region with
$-\om_2/2i<\makebox{Im}(z)<\om_2/2i$; the latter maps onto the
complex  $p$-plane, see section \ref{torus}. Both solutions have
the same values of all global conserved charges $Q_r = q_r(z_1) +
q_r(z_2)=\frac{i}{r-1}\big[(x^+_1)^{1-r}-(x^-_2)^{1-r}\big]$
because $x^+_1$ and $x^-_2$ are the same on both solutions.

\smallskip

We see that if we choose the physical region to be the one with
$|x^\pm|>1$ then there is a unique bound state with $|p|<p_{\rm
cr}$. This region, however, does not cover the whole complex
$p$-plane. One the other hand, if the physical region is the half
of the torus corresponding to the $p$-plane, then there are two
solutions with the same energy and other conserved charges.
Finally, if one considers solutions on the $z$-torus then there
are four solutions but only two of them have positive energy.

\medskip
Continuing above the critical value, $|p|>p_{\rm cr}$, two
solutions (\ref{gsbse}) acquire imaginary parts and become
complex-conjugate to each other, or, equivalently, the real parts
of $p_1$ and $p_2$ become different. Thus, we see that {\it the
BPS bound states naturally split into two families depending on
whether the total momentum is below (the first family) or above
(the second family) the critical value $p_{\rm cr}$.}

\smallskip

The two complex conjugate roots give two
different solutions beyond criticality:
 \bea p_1^{\pm}=\frac{p}{2}\pm {\rm Im}\,
q-i\,{\rm Re}\, q\, ,~~~~~~~~ p_2^{\pm}=\frac{p}{2}\mp
{\rm Im}\, q+i\,{\rm Re}\, q\, ,~~~~~~{\rm Re}\, q>0\, .
\eea We can choose either $(p_1^+,p_2^+)$ or $(p_1^-,p_2^-)$ as a
possible solution of the BPS condition (\ref{bps}). Note that the second solution is the complex conjugate of the first one.  A remarkable
fact to be proven below is that both solutions lie precisely on
the boundary of the  region defined by the curves
$|x^{\pm}|=1$.

\smallskip

Now if we adopt the physical region (sheet) to be $|x^\pm|>1$ with
the boundary $|x^\pm|=1$, then it should contain only one solution
from the second BPS family. Indeed, we do not expect the doubling
of the number of BPS bound states moving beyond the critical
point. The second solution can be then naturally interpreted as
lying on the boundary of another (unphysical) sheet joint to the
physical one through the cut. It is unclear however what is the
precise origin for such an asymmetry. A possible explanation would
be the absence of parity invariance of the string sigma-model but
a concrete implication of this fact is unknown to us.

\begin{figure}[t]
\begin{center}
\includegraphics*[width=1.0\textwidth]{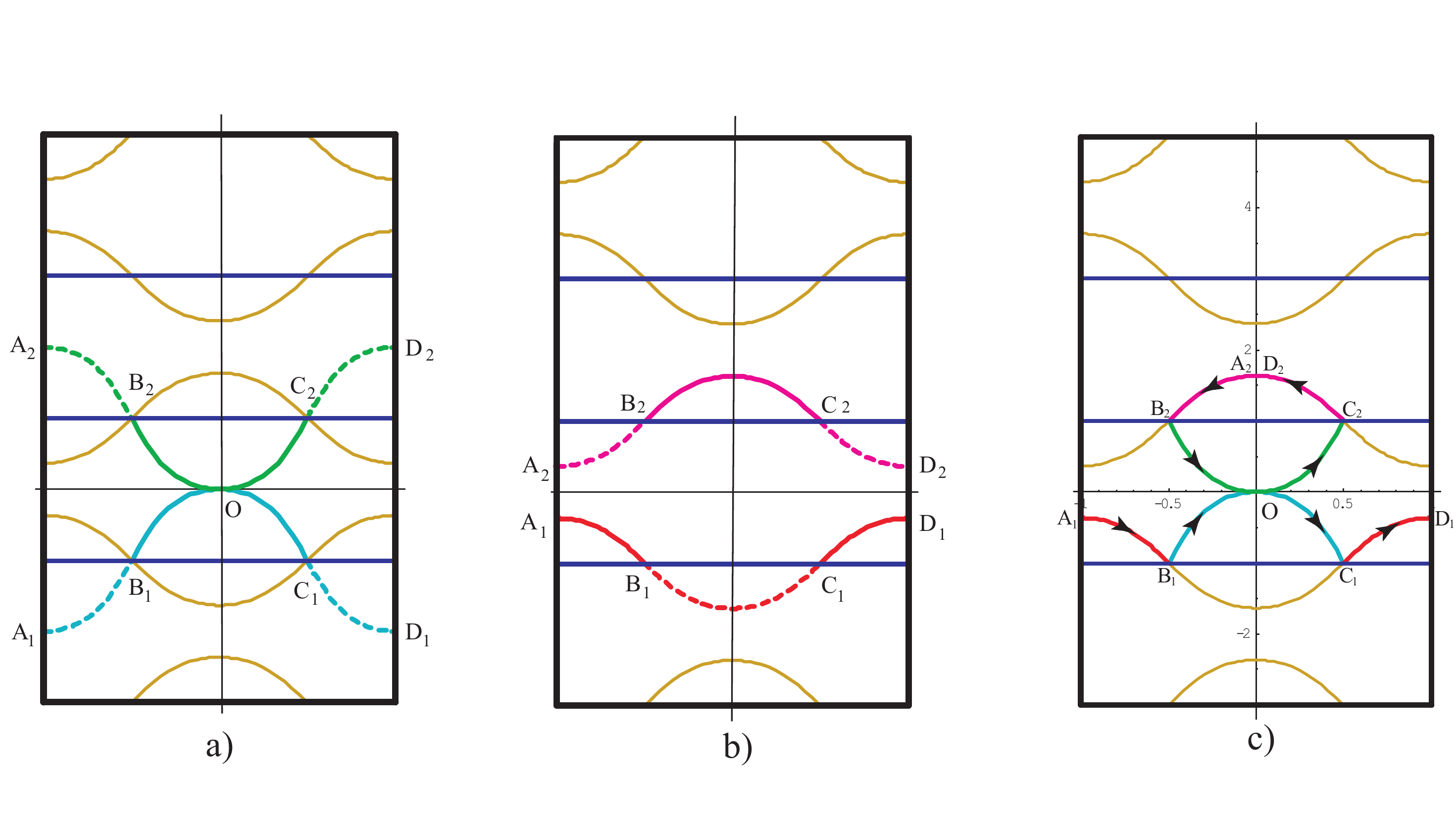}
\end{center}
\vspace{-0.5cm} \caption{ Two-particle bound states of string
theory. Figure a) describes the first BPS family corresponding to
$p<p_{\rm cr}$. The green curves are ${\rm Im}(x^-)=0$ for ${\rm
Im}(z) <0$ and ${\rm Im}(x^+)=0$ for ${\rm Im}(z)
>0$. For any $p<p_{\rm cr}$ there are two solutions: the first one is
represented by the continuous curves ${\rm\bf B}_1{\bf O}{\rm\bf
C}_1 $ (1st particle) and ${\rm\bf B}_2{\bf O}{\rm\bf C}_2 $ (2nd
particle), the second one corresponds to the dashed curves
${\rm\bf A}_1{\rm\bf B}_1 \cup {\rm\bf C}_1{\rm\bf D}_1$ (1st
particle) and ${\rm\bf A}_2{\rm\bf B}_2 \cup {\rm\bf C}_2{\rm\bf
D}_2$ (2nd particle). Figure b) describes the second BPS family
corresponding to $p>p_{\rm cr}$. Again, for any  $p>p_{\rm cr}$
there are two solutions ${\rm\bf B}_2{\rm\bf C}_2 \cup {\rm\bf
A}_1{\rm\bf B}_1\cup {\rm\bf C}_1{\rm\bf D}_1$ and ${\rm\bf
B}_1{\rm\bf C}_1 \cup {\rm\bf A}_2{\rm\bf B}_2\cup {\rm\bf
C}_2{\rm\bf D}_2$. Figure c) represents one of the four
possibilities to smoothly connect solutions from the first and the
second BPS families. Here the variable $z_1$ of the 1st particle
is on the curve ${\rm\bf A}_1{\rm\bf B}_1{\bf O}{\rm\bf
C}_1{\rm\bf D}_1$. When $z_1$ runs along the curve from ${\rm \bf
A}_1$ to ${\rm \bf D}_1$ the real part of the momentum $ {\rm
Re}(p_1)$ increases monotonically from $-\pi$ to $\pi$. At the
same time, the variable $z_2$ corresponding to the 2nd particle
encloses the curve ${\rm\bf A}_2{\rm\bf B}_2{\bf O}{\rm\bf
C}_2{\rm\bf D}_2$. } \label{torus2}
\end{figure}

\smallskip

To visualize the singularities of the string S-matrix and also to
verify that energy is real for the second BPS family, it is
instructive to analyze eqs.(\ref{cond}) in terms of the
generalized rapidity variables $z_1$ and $z_2$ associated to the
first and the second particles, respectively. It is not hard to
see that the first family of the BPS states corresponds to
imposing the reality condition $z_2^*=z_1$. In this case,
eqs.(\ref{cond}) are equivalent to \bea {\rm Im}(x_1^-)=0\quad
{\rm or}\quad |x_1^-|=1\,  , \eea where the first equation defines
the first BPS family. Solving the bound state equation for $z_1$,
one gets a curve in the torus. The part of the curve that  lies
inside the region $|x^\pm|>1$ is represented in Fig.\ref{torus2}
by the green curve ${\rm \bf B}_1{\rm \bf O}{\rm \bf C}_1$, and
the corresponding momentum $p_1$ has $\makebox{Im}(p_1)=-q_-$. The
variable $z_2=z_1^*$ of the second particle runs along another
(conjugate) green curve ${\rm \bf B}_2{\rm \bf O}{\rm \bf C}_2$,
which can be also viewed as describing solutions of the equation
${\rm Im}(x^+_2)=0$ for $z_2$. The dashed curves on
Fig.\ref{torus2}a, which are outside the region $|x^\pm|>1$,
represent solutions of the equations ${\rm Im}(x_1^-)={\rm
Im}(x_2^+)=0$ for $z_1,z_2$ corresponding to the momentum $p_1$
with $\makebox{Im}(p_1)=-q_+$.

\smallskip

To describe the second family of the BPS states corresponding to
the complex values of $q$ one has to take \bea \label{ncrule}
z_2=-z_1^*+\frac{\om_1}{2}+\frac{\om_2}{2}\, . \eea In this case
\bea
x^+(z_2)=x^+\Big(-z_1^*+\frac{\om_1}{2}+\frac{\om_2}{2}\Big)=\frac{1}{x^+(z_1^*)}=\frac{1}{[x^-(z_1)]^*}\,
, \eea where we have used the properties of Jacobi elliptic
functions under the shifts by quarter-periods. Hence, due to the
BPS equation $x^-_1=x^+_2$, the points $z_1$ and $z_2$ lie on the
curves $|x^-|=1$ and $|x^+|=1$, respectively.

\smallskip

As was discussed above, there are two different ways to choose the
second BPS family which is equivalent to deciding what is the
physical sheet. Consider the point $z_1$ corresponding to the
first particle, Fig.\ref{torus2}c. When it moves along the curve
${\rm \bf B}_1 {\rm \bf O}{\rm \bf C}_1$ corresponding to the
first BPS family and reaches, e.g., the point ${\rm \bf C}_1$ then
there are two possibilities  to continue its path along the curve
$|x^-|=1$: either one moves along ${\rm \bf C}_1{\rm \bf D}_1$ or
along ${\rm \bf C}_1{\rm \bf B}_1$. In the case when $z_1$ moves
along the curve ${\rm \bf C}_1{\rm \bf D}_1$, the second point
$z_2$ follows the path ${\rm \bf C}_2{\rm \bf D}_2$. In the
opposite situation, when $z_1$ moves along ${\rm \bf C}_1{\rm \bf
B}_1$, the point $z_2$ follows ${\rm \bf C}_2{\rm \bf B}_2$.
Similar discussion applies to continuing the first BPS family
beyond ${\rm \bf B}_1$. Obviously, for the second family $z_1$ and
$z_2$ are not complex conjugate anymore, rather they obey the
relation (\ref{ncrule}). The bound state energy
$H=ig(x_2^--x_1^+)-2$ is however real, as one can also check by
using the shift/reflection properties of the elliptic functions.

\smallskip

Our discussion reveals that there are four special points on the
$z$-plane \bea z_{\rm cr}=\pm \frac{\om_1}{4} \pm
\frac{\om_2}{4}\,  \eea where both equations ${\rm Im} (x_1^-)=0$
and $|x_1^-|=1$ or ${\rm Im} (x_2^+)=0$ and $|x_2^+|=1$ are
simultaneously satisfied. These are the critical points where two
BPS families meet.

\smallskip

The most transparent description of the bound states is achieved
in terms of the rapidity variable $u$ introduced in section
\ref{torus}, rather than in terms of momentum $p$ or the variable
$z$. Indeed, in terms of $u$ eq.(\ref{bseqs}) becomes \bea
\label{bseqs2b} \left( x_1^- - x^+_2\right) \Big(1-{1\ov x_1^-
x_2^+}\Big)= u_1-u_2 - {2i\ov g} =0\, , \eea i.e. the rapidity
variables $u_1$ and $u_2$ of the first and the second particle lie
on a straight line running parallel to the imaginary axis.
Moreover, for the first BPS family the variables $u_{1,2}$ are
subject to the following conjugation rule $u_1^*=u_2$ which,
together with eq.(\ref{bseqs2b}) allows one to conclude that
\bea\la{Bss} u_{1,2}=u_0\pm \frac{i}{g}\, , ~~~~~~~~u_0\in
{\mathbb R}\,. \eea This is a typical pattern of ``Bethe string".
One can further see that for the first BPS family corresponding to
$p\leq p_{\rm cr}$ the variable $u_0$ is restricted to satisfy
 \bea |u_0|  \ge 2\,,~~~\quad~~ u_{1, {\rm cr}}
=\pm 2 + { i\ov g} \, , \eea where $u_{1, {\rm cr}}$ is a critical
value of rapidity $u_1$ for which the first BPS family ceased to
exist. Under the map  to the $u$-plane the four critical points
$z_{\rm cr}$ are mapped to the four branch points on the
$u$-plane (see Fig.\ref{toruscuts1} in section \ref{torus}) \bea u_{ {\rm cr}} =\pm 2 \pm { i\ov g}\, .
\eea

\smallskip

Let us now turn to the second BPS family. First, by using
eq.(\ref{u}) and the properties of the elliptic functions, we
derive  \bea
u_1^*=x^-\Big(-z_2+\frac{\om_1}{2}+\frac{\om_2}{2}\Big)+\frac{1}{x^-\Big(-z_2+\frac{\om_1}{2}
+\frac{\om_2}{2}\Big)}+\frac{i}{g}=u_2\, .
 \eea
We see that for both families of BPS states the conjugation rule
for $u$'s is the one and the same. By this reason, a solution to
the BPS condition is always represented by the Bethe string
(\ref{Bss}). However, one finds that for the second family a
solution exists for $|u_0|\leq 2$ only. Thus, on the $u$-plane
both families of BPS states admit a uniform description in terms
of the Bethe string with $u_0$ running over the whole real line.

\subsection{Multi-particle bound states}

The consideration of the two-particle bound states can be easily
generalized to the $M$-particle case. The corresponding set of
bound state equations reads \cite{D} \bea \la{bsem}
x^-_j-x^+_{j+1}=0,~~~~~j=1,\ldots, M\, . \eea The total momentum
of a state satisfying these equations is given by \bea\nonumber
e^{ip}= {x^+_1\ov x^-_1}{x^+_2\ov x^-_2}\cdots {x^+_M\ov x^-_M} =
{x^+_1\ov x^-_M}\,, \eea and the energy of the state is
\bea\la{enM} H_M= \sum_{i=1}^M\left(-1 - igx^+_i + igx^-_i\right)
= -M - igx^+_1 + igx^-_M\, . \eea Both the energy and momentum
depend on the values of $x^+_1$ and $x^-_M$ only. Since the energy
is real, $x^-_M$ must be the complex conjugate of $x^+_1$:
$(x^-_M)^*=x^+_1$. In fact, a simple but important observation is
that any global conserved charge of a state obeying (\ref{bsem})
depends only on $x^+_1$ and $x^-_M$:
$$Q_r = \sum_{i=1}^M q_r(z_i) =\sum_{i=1}^M\frac{i}{r-1}\big[(x^+_i)^{1-r}-(x^-_i)^{1-r}\big]=\frac{i}{r-1}\big[(x^+_1)^{1-r}-(x^-_M)^{1-r}\big]\,. $$

\smallskip

Another important consequence of eqs.(\ref{bsem}) is that the
coordinates $x^+_1$ and $x^-_M$ satisfy the following quadratic
constraint \bea\la{constraintM} x^+_1 + {1\ov x^+_1} - x^-_M-{1\ov
x^-_M}={2M\ov g}i\, . \eea This is the same constraint as the one
satisfied by $x^\pm$ (\ref{consxpxm}) with $g\to g/M$, and we get
immediately the dependence of $x^+_1$ and $x^-_M$ on the total
real momentum $p$ \footnote{In general for a given momentum $p$
there are two solutions of the constraint (\ref{constraintM}), and
there could be any sign in front of the square root in
(\ref{xpxmp}). The positive sign guarantees the positivity of the
energy.} \bea\la{xpxmp}\begin{aligned} x^+&=\frac{e^{i{p\ov
2}}}{2g\sin{p\ov 2}}\Big({M+
\sqrt{M^2+4g^2\sin^2\frac{p}{2}}}\Big) \, ,\\
x^-&=\frac{e^{-i{p\ov 2}}}{2g\sin{p\ov 2}}\Big({M+
\sqrt{M^2+4g^2\sin^2\frac{p}{2}}}\Big) \, , \end{aligned}\eea and,
using (\ref{enM}),  the BPS energy formula \bea\nonumber
H^2_M=M^2+4g^2\sin^2\frac{p}{2} \, . \eea Moreover, we see that
the set of global conserved charges $Q_r $ is the same for any
solution of (\ref{bsem}) with a given total momentum $p$.

\smallskip

It is also easy to see that the number of different solutions with
a real momentum $p$ and positive energy is equal to $2^{M-1}$
because for a given $x^+$ there are two different $x^-$ solving
the constraint (\ref{consxpxm}), see the diagram below for $M=4$
\bea\nonumber x^+_1\longrightarrow\left\{\begin{array}{c}
x^-_1=x^+_2\longrightarrow\left\{\begin{array}{c}
x^-_2=x^+_3\longrightarrow\left\{\begin{array}{c}
x^-_3=x^+_4\\x^-_3=x^+_4\end{array}  \right.
\\x^-_2=x^+_3\longrightarrow\left\{\begin{array}{c} x^-_3=x^+_4\\x^-_3=x^+_4\end{array}  \right.\end{array}  \right.\\
x^-_1=x^+_2\longrightarrow\left\{\begin{array}{c}
 x^-_2=x^+_3\longrightarrow\left\{\begin{array}{c} x^-_3=x^+_4\\x^-_3=x^+_4\end{array}  \right.\\x^-_2=x^+_3\longrightarrow\left\{\begin{array}{c} x^-_3=x^+_4\\x^-_3=x^+_4\end{array}  \right.\end{array}  \right.\end{array}  \right\}\longrightarrow x^-_4
\eea
To have all these solutions one would have to allow the parameters $z_i$ of the particles to be anywhere on the $z$-torus, in particular, some of them would be in the anti-particle region with $|x^\pm|<1$.

\smallskip

However, if we require that all the constituent particles of the
bound state belong to the region $|x^\pm|>1$ then we are left with
a unique solution because for a given $x^+$ only one solution for
$x^-$ satisfies the condition $|x^-|\ge1$. For $M$ even it is also
necessary to specify what parts of the boundaries $|x^\pm|=1$
belong to the region because if  the momentum of a bound state
exceeds a critical, $g-$ and $M$-dependent, value then there are
several solutions of the bound state equations with $|x^-_{M/2}| =
|x^+_{M/2+1}|=1$.

\smallskip

Finally, if the parameters $z_i$ of the particles belong to the half of the torus corresponding to the complex $p$-plane, then one can show that for any $M$ there are two solutions of the bound state equations.

\smallskip

Just as for the case of two-particle bound states, the simplest
description of $M$-particle bound states is provided by the
$u$-plane where a solution is given by the Bethe string \bea
u_j=u_0+(M-2j+1)\frac{i}{g}\, ,~~~~~~~j=1,\ldots, M\, .
 \eea
We can choose one and the same map of the $u$-plane with the cuts
described in section \ref{torus} onto the region of the $z$-torus
with $|x^\pm|>1$ for all the particles. It is then obvious that
for a given momentum $p$ there is just a single $M$-particle bound
state that falls inside the physical region. Its structural
description however becomes rather involved.

\subsection{Finite-size corrections to the bound states}
It is of interest to analyze finite-size corrections to the energy
of the BPS bound states, and to see what restrictions on the
dressing factor could be derived from the condition that the
energy corrections are real. To this end, we consider two-particle
states in the $\su(2)$ sector described by the following two
equations, see (\ref{BEsu2}) \bea \la{basu2}\left( {x^+_1\ov
x^-_1}\right)^J= \Sigma_{12}\,
\frac{x_{1}^{+}-x_{2}^{-}}{x_{1}^{-}-x_{2}^{+}}{1-{1\ov x_1^+
x_2^-}\ov 1-{1\ov x_1^- x_2^+}}\,,\quad \left( {x^+_1x^+_2\ov
x^-_1x^-_2}\right)^J =1\, , \eea where $\Sigma_{12} =
{x^-_1x^+_2\ov x^+_1x^-_2}\sigma_{12}$ is the unitary factor that
appeared in the splitting (\ref{sp1}) of the scalar factor, and
$J$ is one of the global charges corresponding to the isometries
of the five-sphere. The variables $x^\pm_i$ also satisfy the
constraint (\ref{consxpxm}). These equations are supposed to be
valid asymptotically  for large values of $J$, and have to be
modified for finite $J$.

\smallskip

We will analyze these equations for large values of $J$ in the vicinity of a bound state satisfying the bound state equation $x^-_1=x^+_2$ and having a fixed total momentum
$p = {2\pi m\ov J}$ where $m$ is an integer. The quantization condition for the total momentum follows from the second equation in (\ref{basu2}).

\smallskip
Let ${\bf x}^\pm_i$ denote the values of ${x}^\pm_i$ satisfying
the bound state equation and the second equation in (\ref{basu2}).
Then, $\left({{\bf  x}^-_1\ov {\bf  x}^+_1}\right)^J \sim e^{-q
J}$ exponentially decreases at large $J$, and we can look for a
solution of the form \bea\nonumber x^\pm_i={\bf x}^\pm_i \left( 1
+ \left({{\bf  x}^-_1\ov {\bf  x}^+_1}\right)^J y^\pm_i\right)\,.
\eea Expanding the equations (\ref{basu2}) and the  constraint
(\ref{consxpxm}) in powers of $y^\pm_i$, we find a system of
linear equations for $y^\pm_i$. The solution of the system is
given below \bea\nonumber y^-_1&=& \frac{4\Sigma_{12} (i g ({\bf
x}^-_2-{\bf x}^+_2)+{\bf x}^-_2
   {\bf x}^+_2) \left(2 i {\bf x}^+_1 {\bf x}^+_2+g \left({\bf x}^+_1
   \left(({\bf x}^+_2)^2+1\right)-2 {\bf x}^+_2\right)\right)}{g^2 (g
   {\bf x}^-_2-(g+2 i {\bf x}^-_2) {\bf x}^+_1) \left(({\bf x}^+_2)^2-1\right)^2}\\\nonumber
y^+_1&=& \frac{4\Sigma_{12} {\bf x}^+_1 (i g ({\bf x}^-_2-{\bf x}^+_2)+{\bf x}^-_2
   {\bf x}^+_2)}{g (g {\bf x}^-_2-(g+2 i {\bf x}^-_2) {\bf x}^+_1)
   \left(({\bf x}^+_2)^2-1\right)}\\\nonumber
y^-_2&=&    - \frac{4 i\Sigma_{12} {\bf x}^-_2 (g
   ({\bf x}^+_1-{\bf x}^+_2)+i {\bf x}^+_1 {\bf x}^+_2)}{g ((g+2 i
   {\bf x}^-_2) {\bf x}^+_1-g {\bf x}^-_2) \left(({\bf x}^+_2)^2-1\right)}\\\nonumber
y^+_2&=&  \frac{4\Sigma_{12}(g
   ({\bf x}^+_1-{\bf x}^+_2)+i {\bf x}^+_1 {\bf x}^+_2) \left(g {\bf x}^-_2
  ( {\bf x}^+_2)^2-2 (g+i {\bf x}^-_2) {\bf x}^+_2+g {\bf x}^-_2\right)}{g^2 (-i g
   ({\bf x}^-_2-{\bf x}^+_1)-2 {\bf x}^-_2 {\bf x}^+_1)
   \left(({\bf x}^+_2)^2-1\right)^2}\,,
\eea
where $\Sigma_{12}$ is evaluated on the solution to the bound state equation. The leading correction to the energy of the state is easily found by expanding
\bea
E = E_1 + E_2\,,\quad E_i = 1 + {i g\ov x^+_i} -  {i g\ov x^-_i} = -1 - {i g x^+_i} + {i g x^-_i}\,.
\eea
By using $E_i = 1 + {i g\ov x^+_i} -  {i g\ov x^-_i}$, we obtain
\bea\la{de1}
\delta E =  \left(\frac{{\bf x}^-_1}{{\bf x}^+_1}\right)^{J}\Sigma_{12}\frac{4 i
   ({\bf x}^-_2 (2 {\bf x}^+_1-{\bf x}^+_2)-{\bf x}^+_1 {\bf x}^+_2)}{(g
   ({\bf x}^-_2-{\bf x}^+_1)-2 i {\bf x}^-_2 {\bf x}^+_1)
   \left(({\bf x}^+_2)^2-1\right)}\,.
\eea
On the other hand by using $E_i =  -1 - {i g x^+_i} + {i g x^-_i}$, we get
\bea\la{de2}
\delta E =  \left(\frac{{\bf x}^-_1}{{\bf x}^+_1}\right)^{J}\Sigma_{12} \frac{4 i {\bf x}^-_2 {\bf x}^+_1 {\bf x}^+_2 ({\bf x}^-_2+{\bf x}^+_1-2
   {\bf x}^+_2)}{(g({\bf x}^-_2-{\bf x}^+_1)-2 i {\bf x}^-_2 {\bf x}^+_1)
   \left(({\bf x}^+_2)^2-1\right)}\,.
\eea Even though the expressions look different they coincide on
solutions to the bound state equation. In what follows we will be
using the simpler eq.(\ref{de2}). Note also that the perturbation
theory breaks down at $p=p_{\rm cr}$. Due to the quantization
condition for the momentum $p$ it may happen only at special
values of the coupling constant $g$ depending on $m/J$.

\smallskip

It is clear that the energy correction cannot be real for any choice of the dressing factor $\Sigma_{12}$.
The imaginary part of the correction depends on the branch of the bound state under consideration.

\smallskip

In the first case with $\makebox{Im}({\bf x}^-_1)=\makebox{Im}({\bf x}^+_2)=0$ and the total momentum smaller than the critical value (\ref{pcr}),  the parameters ${\bf x}^\pm_i$ satisfy the complex conjugation rule $({\bf x}^\pm_1)^*={\bf x}^\mp_2$, and we get
\bea\nonumber
\delta E - \delta E^* = \left( \left(\frac{{\bf x}^-_1}{{\bf x}^+_1}\right)^{J}\Sigma_{12} - \left(\frac{{\bf x}^+_2}{{\bf x}^-_2}\right)^{J}\Sigma_{12}^*\right) \frac{4 i {\bf x}^-_2 {\bf x}^+_1 {\bf x}^+_2 ({\bf x}^-_2+{\bf x}^+_1-2
   {\bf x}^+_2)}{(g({\bf x}^-_2-{\bf x}^+_1)-2 i {\bf x}^-_2 {\bf x}^+_1)
   \left(({\bf x}^+_2)^2-1\right)}\,.
\eea
Taking into account that $ \left(\frac{{\bf x}^-_1}{{\bf x}^+_1}\right)^{J} =\left(\frac{{\bf x}^+_2}{{\bf x}^-_2}\right)^{J}$, we conclude that in this case  the correction is real only if the dressing factor is real $
\Sigma_{12} =\Sigma_{12}^*\,.
$
This property of the dressing factor can be easily shown by using the representation (\ref{dph}) for the dressing phase.

\smallskip

In the second case with $|{\bf x}^-_1|=|{\bf x}^+_2|=1$ and the
total momentum exceeding the critical value (\ref{pcr}), the
parameters ${\bf x}^\pm_i$ satisfy the complex conjugation rule
$({\bf x}^+_1)^*={\bf x}^-_2$, $({\bf x}^-_1)^*=1/{\bf x}^+_2$,
and we obtain \bea\nonumber &&\delta E - \delta E^* = \nonumber
\\ \nonumber &&=\frac{4 {\bf
x}^-_2 {\bf x}^+_1 \left(\Sigma_{12}^*
  ( {\bf x}^+_2{\bf x}^-_2)^{-J} (2-({\bf x}^-_2+{\bf x}^+_1) {\bf x}^+_2)-\Sigma_{12}
  ( {\bf x}^+_1)^{-J}( {\bf x}^+_2)^{
   J+1} ({\bf x}^-_2+{\bf x}^+_1-2 {\bf x}^+_2)\right)}{(i g ({\bf x}^-_2-{\bf x}^+_1)+2 {\bf x}^-_2 {\bf x}^+_1)
   \left(({\bf x}^+_2)^2-1\right)}\,.
\eea We see that the imaginary part of the correction would vanish
only if \bea \nonumber \Sigma_{12}^* = \Sigma_{12}( {\bf
x}^+_2)^{2J+1}\frac{ ({\bf x}^-_2+{\bf x}^+_1-2 {\bf
x}^+_2)}{(2-({\bf x}^-_2+{\bf x}^+_1) {\bf x}^+_2)}\,. \eea Since
the last equation depends on $J$ and on a particular bound state
solution, it cannot be satisfied for any choice of the dressing
factor. The complex energy of the state would mean that the
Hamiltonian of the model is not hermitian for finite $J$.

\smallskip

One might conclude from this result that the S-matrix poles with
\mbox{$|{\bf x}^-_1|=|{\bf x}^+_2|=1$} are spurious and do not
correspond to bound states, and, therefore, should be omitted.
That would mean, however, that for any value of the total momentum
the bound states satisfying the equations $\makebox{Im}({\bf
x}^-_1)=\makebox{Im}({\bf x}^+_2)=0$ would disappear as soon as
the coupling constant $g$ reaches a critical (momentum-dependent)
value. This seems to contradict to the statement that the bound
states are BPS. We believe that such a conclusion might be
erroneous and the result above indicates, in fact,  that the
asymptotic Bethe ansatz cannot be used to analyze the finite-size
corrections to the energy of bound states with the total momentum
exceeding the critical value (\ref{pcr}).

\smallskip

To show that this is indeed the case, let us recall that, as was
shown in \cite{magnon}, at large values of the string tension $g$
and the charge  $J$  the dispersion relation receives finite-size
corrections of the order $e^{-J/(g\sin{p/2})}$. On the other hand,
the energy correction we computed above is of the order $e^{-q J}$
where $q$ is the imaginary part of the momentum $p_2$. It depends
on the total momentum $p$ and the string tension $g$. By using
eq.(\ref{bexa1}), it is not difficult to determine the large $g$
dependence of the momenta $p_1$ and $p_2$ of a bound state
\bea\la{bex} p_1 = {\cos{p\ov 2}\ov 2 g^2 \sin^3{p\ov 2}} - {i\ov
g \sin{p\ov 2}} +{\cal O}({1\ov g^3})\,,\quad p_2 = p - {\cos{p\ov
2}\ov 2 g^2 \sin^3{p\ov 2}} + {i\ov g \sin{p\ov 2}} +{\cal
O}({1\ov g^3})\, .~~~ \eea The second solution of
eq.(\ref{4thpol}) (with $q>0$) is related to (\ref{bex}) as $p_1
\to p_2^*\,,\ p_2\to p_1^*$ that is one exchanges the real parts
of momenta $p_i$. A surprising result of the computation is that
$q$ is equal to ${1\ov g \sin{p\ov 2}}$, and, therefore, $e^{-q
J}$ is exactly equal to the magnitude of the  finite-size
correction to the dispersion relation. That means that computing
the finite $J$ correction to the energy of such a bound state one
has to take into account the necessary modifications of the
asymptotic Bethe ansatz. As a result of these modifications, one
should be able to get a real finite-size correction to the energy
of a bound state carrying momentum exceeding the critical value.
In fact, this would be a non-trivial check of  finite $J$ ``Bethe"
equations.

\smallskip

The analysis performed above raises the question if one can use
the asymptotic Bethe ansatz to compute the corrections to the
energy of the bound states with momenta smaller than the critical
value. At large $g$ we can again compare the value of $q$ with
${1\ov g \sin{p\ov 2}}$. If $q$ is less than ${1\ov g \sin{p\ov
2}}$ then the energy correction (\ref{de2}) is bigger than the
correction due to  finite $J$  modifications of the asymptotic
Bethe ansatz, and we can trust  (\ref{de2}). Since $p_{\rm
cr}=2/\sqrt g$ at large values of $g$ one should consider a bound
state with momentum $p$ of the order $1/\sqrt g$. The large $g$
dependence of the momenta $p_1$ and $p_2$ of a bound state is
easily found by using eq.(\ref{bexa2})
\bea\la{bex2} p_1 = {p\ov 2} - 2 i
{1 \pm \sqrt{1 - {p^4g^2\ov 16 }}\ov gp} + \cdots  \,,\quad p_2 =
{p\ov 2} + 2 i {1 \pm \sqrt{1 - {p^4g^2\ov 16 }}\ov gp} + \cdots \,,~~~~~
\eea
leading  for $p < p_{\rm cr}$ to the following two real
solutions for $q$ \bea q_\pm = 2 {1 \pm \sqrt{1 - {p^4g^2\ov 16
}}\ov gp}\, . \eea Comparing these values with ${1\ov g \sin{p\ov
2}}\approx {2\ov g p}$, we see that $q_- < {2\ov g p}$ and $q_+ >
{2\ov g p}$. Thus, the asymptotic Bethe ansatz can be used to
analyze finite $J$ corrections to the energy of a bound state with
momentum smaller than $p_{\rm cr}$  for the bound state with $q_-$
only.

\smallskip

Actually, the fact that the energy correction (\ref{de2}) to the
bound state with $q_+$ is smaller than the corrections due to finite $J$
modifications of the asymptotic Bethe ansatz
  raises a question if these solutions correspond to the actual bound states.
  It may happen that  finite $J$ Bethe equations would not have any solution
  that would reduce to the solution  with $q_+$ in the limit $J\to\infty$.

\smallskip

A similar analysis can also be performed for small values of $g$.
Then we expect that the finite $J$ effects (in gauge theory they are due to the wrapping interactions) become important at order $g^{2J}$, and therefore we could trust the asymptotic Bethe ansatz and the energy correction (\ref{de2}) only if $q < -2\log g$.

\smallskip

The leading small $g$ dependence of $q$ of the  bound state  solutions with the momentum smaller than $p_{\rm cr}$  is given by eqs.(\ref{bexa3p}), (\ref{bexa3m})
\bea\la{bexa3}
q_+ = -2\log g + \cdots \,,\quad q_- = -\log\cos{p\ov 2} + \cdots \,.~~~~~
\eea
We see immediately that again  only the solution with the smaller imaginary part of the momentum $q_-$ satisfies the necessary condition.  The energy correction to the state with $q_+$ is of order $g^{2J}$ that is exactly the order of wrapping interactions, and the asymptotic Bethe ansatz again breaks down for the state.

\smallskip

Finally,  the leading small $g$ dependence of $q$ of the  bound state solutions with the momentum exceeding $p_{\rm cr}$ is given by (\ref{bexa4})
\bea\la{bex5}
q_\pm = -\log {g\ov 2} \pm i\a  +\cdots\,,~~~~~~~~~~~
\eea
where $\a$ is related to the momentum $p$ as follows
$p = \pi - 2 g \cos \a\,.$

\smallskip

We see that the real part of $q_\pm$ is smaller than $-2\log g$,
and therefore one could conclude that one might use the asymptotic
Bethe ansatz for the states in this regime. This, however, leads
to the problem of the complex energy of these states discussed
above. As before the only resolution of the problem we see is the
breakdown of the asymptotic Bethe ansatz. This would imply,
however, that for these states the wrapping interactions become
important already at the order $g^J$. The fact that in gauge
theory these states are not dual to gauge-invariant operators does
not seem to be important for this conclusion. One could for
example scatter a bound state carrying momentum $p=\pi$ which
always exceeds the critical momentum $p_{\rm cr}$ with an
elementary one carrying  momentum $-\pi$ so that the total
momentum would be zero, and such a state would be dual to a
gauge-invariant operator. We would still expect the finite $J$
corrections to this state to be of the order $g^J$. Another
puzzling property of the states with $p>p_{\rm cr}$ is that in the
limit $g\to 0$ the states are pushed away from the spectrum
because $p_{\rm cr}=\pi$, and cannot be seen in the perturbative
gauge theory.

\section{Bound states of the mirror model}\la{boundmir}


Let us now consider  in a similar fashion
bound states of the mirror model.  In this case one should consider
mirror particles of type $A_{3\dot 3}^\dagger$.

\smallskip

We begin our consideration with two-particle bound states, and
 let the complex momenta of the two particles be
\bea\nonumber \tp_1 = {p\ov 2} -i q\,, \quad \tp_2 = {p\ov 2} + i
q\,,\quad \makebox{Re}\, q > 0\,,
 \eea
where $p$ is the total momentum of the mirror bound state.

\smallskip

The first equation in (\ref{BEn2})
takes the form
\bea\la{bemt}
e^{ip R/2}e^{ q R}= \s_{12}\,
\frac{x_{1}^{-}-x_{2}^{+}}{x_{1}^{+}-x_{2}^{-}}
\frac{1-\frac{1}{x_1^+ x_2^-}}{1-\frac{1}{x_1^-x_2^+}}\,,
\eea
where we set all auxiliary roots to 0.
Assuming that the dressing factor does not vanish, we conclude that for $ \makebox{Re}\, q>0$ and in the limit $R\to\infty$  the following bound state equation should hold
\bea
\label{bseq}
x_1^+ - x^-_2=0\,.
\eea
The second factor in the denominator of the Bethe equation (\ref{bemt}) may also vanish
but the energy of the corresponding state does not satisfy the BPS condition. We expect that, just as a similar factor in the string theory, the pole due to this factor does not correspond to a bound state.

\smallskip

By using eqs. (\ref{xpmtp}) which express $x^\pm$ as functions of $\tp$, we find that
eq.(\ref{bseq}) is equivalent to
\bea
-4g^2q^2+t^3-2q^2t\left(2-t\right)+q^4t=0\, ,
\eea
where $t\equiv  1 + {p^2\ov 4}$. This equation gives the following
two solutions with a positive real part of $q$:
\bea\la{solbse}
q=\sqrt{1+{g^2\ov t}}\, \pm\, \sqrt{1-t+{g^2\ov t}} = \sqrt{1+{4g^2\ov 4+p^2}}\, \pm\, \sqrt{-{p^2\ov 4}+{4g^2\ov 4+p^2}}\, .
\eea
Solutions for $q$ are real provided the expression under the square root is nonnegative, and this implies the following restriction on the total momentum of the bound state
\bea\la{pcrm}
 |p|\leq p_{\rm cr}\equiv \sqrt{2}\sqrt{-1+\sqrt{1+4g^2}}\, .
\eea
For an exact inequality we have two positive solutions $q_\pm$, and
 when the bound on the
momentum is saturated the solution is obviously unique\footnote{The  energy of the bound state is
$
\widetilde{{\cal E}}_{\rm cr}=2\, {\rm
arcsinh}\frac{\sqrt 2}{g}\sqrt{1+1\sqrt{1+4g^2}}\,.
$}
\bea
q_-<q_{\rm cr}<q_+\,,\quad   q_{\rm cr}=\frac{1}{\sqrt{2}}\sqrt{1+\sqrt{1+4g^2}}\, .
\eea
It is interesting to notice that the dependence of $q_\pm$ on the momentum of the bound state is smoother at $p=0$ than the one for string theory bound states. We see from eq.(\ref{solbse}) that  $q_-$ reaches its minimum, and $q_+$ reaches its maximum at $p=0$
\bea
q_-^{\rm min} =\sqrt{1
+g^2}\, -\, g\,,\quad q_+^{\rm max} =\sqrt{1
+g^2}\, +\, g\,,\quad p=0\,.
\eea
In string theory the corresponding values are $0$ and $\infty$.

\smallskip

To find what curves in the $z$-torus correspond to the two solutions with real $q$ we  take into account that in this case $\tp_1{}^* = \tp_2$, and the reality condition for  $x^\pm$ in the mirror
theory is $\left(x_1^\pm\right)^* = 1/x_2^\mp$. Thus the bound state equation (\ref{bseq}) reduces to  the following equivalent conditions
$$|x_1^+|=1\quad\Longleftrightarrow\quad |x_2^-|=1\,,
$$
being represented by the two curves in the $z$-torus that bound the yellow region with $|x^{+}|<1\,,\ |x^{-}|>1$ in Figure \ref{torus1}. Note that the curves are symmetric about the horizontal line passing through the point $z = {\om_2\ov 2}$. Let us recall that hermitian conjugation in the mirror theory is defined with respect to this line, see section \ref{torus}.
It is not difficult to check that the parts of the curves $|x_1^+|=1\,,\  |x_2^-|=1$ that are
inside the region  $\makebox{Im}(x^{\pm})<0$ correspond to the smaller root $q_-$ of eq. (\ref{bseq}). The other parts of the curves correspond to
the
second solution with $q=q_+$, see Figure \ref{torus1}. Just as it was for string theory bound states, both solutions
have the same values of all global conserved charges $Q_r =
q_r(z_1) +
q_r(z_2)=\frac{i}{r-1}\big[-(x^-_1)^{1-r}+(x^+_2)^{1-r}\big]$.

\smallskip

We see that if we want to have only one bound state with $|p|<p_{\rm
cr}$ in a physical region, then we should
 choose the physical region to be the one with  $\makebox{Im}(x^{\pm})<0$ but not the one bounded by the curves
$|x^\pm|=1$ as it is for string theory.  We will see in a moment that the region $\makebox{Im}(x^{\pm})\le 0$ also contains bound states with $|p|>p_{\rm cr}$ described by the solutions with complex $q$.

\medskip
Above the critical value, $|p|>p_{\rm cr}$, the two solutions
(\ref{solbse}) acquire imaginary parts and become complex
conjugate to each other. It is convenient to denote the
corresponding solutions as follows
\bea\la{solbse1}
q_{\pm}=
\sqrt{1+{4g^2\ov 4+p^2}}\, \pm\, i\,\frac{p}{2}\,\sqrt{1-{16g^2\ov
p^2(4+p^2)}}\, .
\eea
We see that the real part of $q_\pm$ is a
decreasing function of $p$, and its minimum value is 1. On the
contrary the imaginary part of $q_\pm$ is an increasing function
of $p$ and it behaves as $\pm p/2$ at large values of $p$. As a
result, the two complex momenta
 \bea\nonumber \tp_1{}^{\pm}=\frac{p}{2}\pm {\rm Im}\,
q-i\,{\rm Re}\, q\, ,~~~~~~~~ \tp_2{}^{\pm}=\frac{p}{2}\mp
{\rm Im}\, q+i\,{\rm Re}\, q\, ,~~~~~~{\rm Re}\, q>0\, ,
\eea
have the following large $p$ behavior
\bea\nonumber
 \tp_1{}^{+}=p-i\, ,\quad\tp_2{}^{+}=i\ ; \qquad  \tp_1{}^{-}=-i\, ,\quad\tp_2{}^{-}= p+i\, .
\eea

\smallskip

\begin{figure}[t]
\begin{center}
\includegraphics*[width=1.0\textwidth]{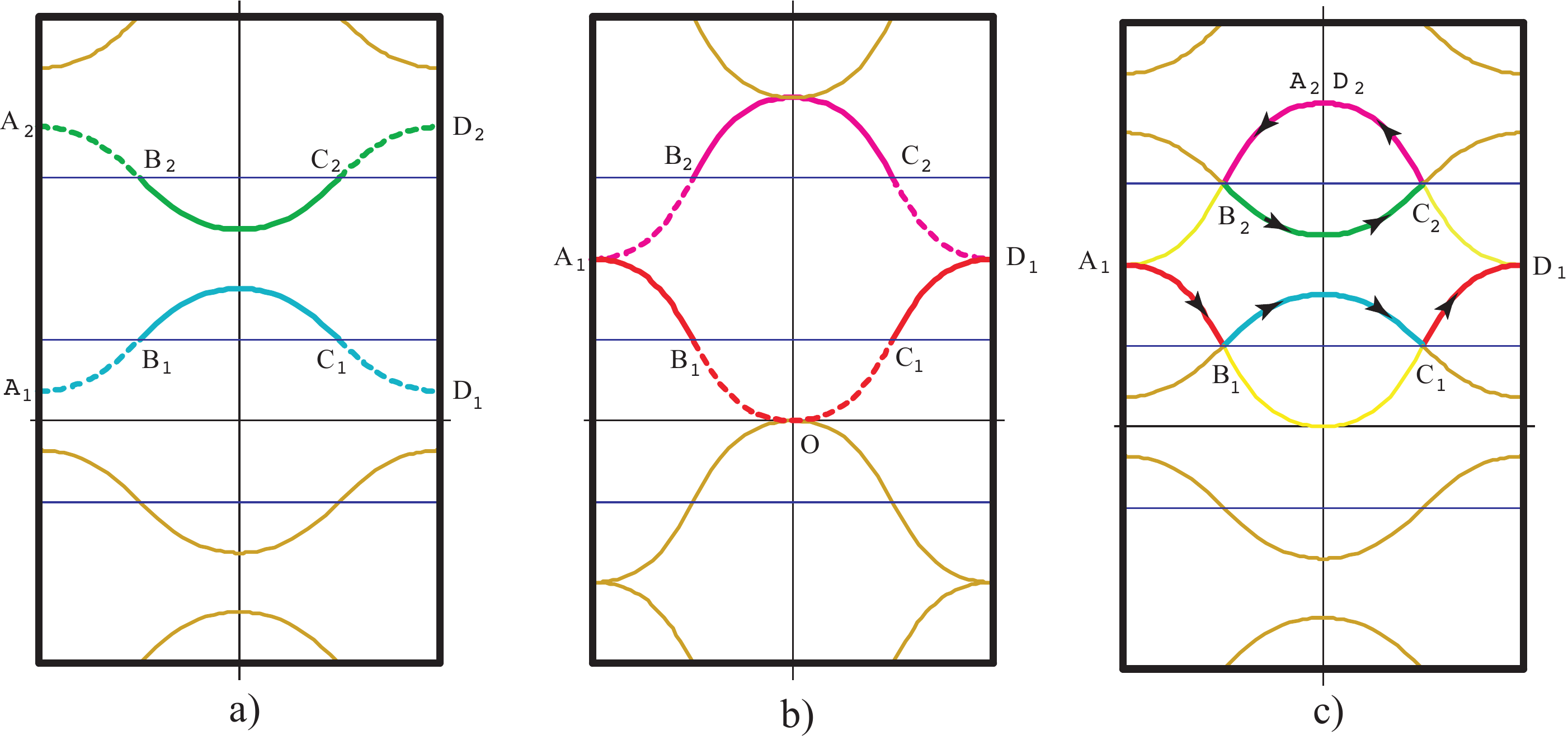}
\end{center}
\caption{Bound states of the mirror theory. Figure a) represents
the first BPS family: for any $p$ with $|p|<p_{\rm cr}$ there are
two solutions corresponding to $q_-$ (the curves ${\rm\bf
B}_1{\rm\bf C}_1 $ for the 1st particle and ${\rm\bf B}_2{\rm\bf
C}_2 $ for the 2nd one, respectively) and to $q_+$ (the curves
${\rm\bf A}_1{\rm\bf B}_1 \cup {\rm\bf C}_1{\rm\bf D}_1$ for the
1st particle and ${\rm\bf A}_2{\rm\bf B}_2 \cup {\rm\bf
C}_2{\rm\bf D}_2$ for the 2nd one, respectively). Figure b)
represents the second BPS family which is also doubly degenerate:
it is given by either ${\rm\bf A}_1{\rm\bf B}_1 \cup {\rm\bf
C}_1{\rm\bf D}_1 \cup {\rm\bf B}_2{\rm\bf D}_2$ or by ${\rm\bf
B}_1{\rm\bf C}_1 \cup {\rm\bf A}_1{\rm\bf B}_2 \cup {\rm\bf
D}_1{\rm\bf C}_2$. Figure c) corresponds to one of the four
possibilities to connect the first and the second BPS family: when
the variable $z_1$ of the 1st particle runs along the curve
${\rm\bf A}_1{\rm\bf B}_1{\rm\bf C}_1{\rm\bf D}_1$ the real part
of its momentum increases from $-\infty$ to $+\infty$. At the same
time, the variable $z_2$ of the 2nd particle encloses the curve
${\rm\bf A}_2{\rm\bf B}_2{\rm\bf C}_2{\rm\bf D}_2$. }
\label{torus4}
\end{figure}

 A remarkable
fact  is that both solutions lie precisely on the boundary of the
region $\makebox{Im}(x^{\pm})\le 0$. To see this we notice that,
just as it was for string theory bound states, the coordinates
$z_1$ and $z_2$ of the solutions with the complex values of $q$
are related by  eq.(\ref{ncrule}) \bea \label{ncrulem}
z_2=-z_1^*+\frac{\om_1}{2}+\frac{\om_2}{2}\, . \eea Then, one can
easily show that \bea\nonumber
x^-(z_2)=x^-\Big(-z_1^*+\frac{\om_1}{2}+\frac{\om_2}{2}\Big)=x^-(z_1^*)=[x^+(z_1)]^*\,
, \eea and, therefore, the bound state equation $x^+_1=x^-_2$  is
equivalent to  $\makebox{Im}(x^{+}(z_1))=
\makebox{Im}(x^{-}(z_2))=0$. We plot the corresponding curves in
Figure \ref{torus4}.

\smallskip

Thus, we have shown that these solutions lie on the boundary of
the  region $\makebox{Im}(x^{\pm})\le 0$, and, therefore,  the
region contains bound states with any value of the total momentum
and could be considered as the physical one for the mirror model.
It is also necessary to specify what part of the boundary of the
region $\makebox{Im}(x^{\pm})\le 0$ belongs to the physical
region, and this can be done by choosing properly the cuts in the
$u$-plane where the bound state equation reduces to \bea\nonumber
x_1^+ - x_2^-=0\quad\Longrightarrow\quad u_2-u_1=\frac{2i}{g}\,.
\eea As was discussed in section \ref{torus}, eqs.
$|x_1^+|=|x_2^-|=1$ describing a bound state with the momentum not
exceeding the critical value $p_{\rm cr}$ and with a real $q$ give
a Bethe string solution with the real part of $u$ lying in the
interval $[-2,2]$ \bea\nonumber u_{1,2}=u_0\mp \frac{i}{g}\, ,
~~~~~~~~-2\le u_0\le 2\,. \eea On the other hand, values of $u_0$
lying outside  the interval $[-2,2]$ correspond to solutions of
eqs. $\makebox{Im}(x^{+}_1)= \makebox{Im}(x^{-}_2)=0$. The
momentum $\tp=\tp(u)$ is a multi-valued function of $u$, and one
should choose a proper branch of the function to get the right
values  of the momenta $\tp_1\,,\ \tp_2$ of the bound state. This
fixes the cuts in the $u$-plane which run from $\pm\infty$ to $\pm
2\mp \frac{i}{g}$, and also the boundaries of the  region
$\makebox{Im}(x^{\pm})\le 0$ in the $z$-plane which is mapped onto
the $u$-plane with these cuts.

\medskip

The discussion of  bound states of $M$ particles of type $A_{3\dot
3}^\dagger$ basically repeats the one in section \ref{bound}. One
finds a system of equations \bea\la{bems} x^+_j-x^-_{j+1}=0\, ,
~~~~j=1,\ldots,M-1\, . \eea In terms of the variable $u$ the Bethe
string solution reads as \bea u_j=u_0-(M-2j+1)\frac{i}{g}\, ,
~~~~~~~j=1,\ldots, M\, , \eea and has the energy \bea {\cal
E}=\log\frac{x^-_1}{x^+_M} =2\, {\rm arcsinh}\,
\frac{1}{2g}\sqrt{M^2+\tp^2}\, ,
 \eea
where $\tp=\tp_1+\ldots+\tp_M$ is a total (real) momentum of the bound
state.

\smallskip

Depending on a choice of the physical region, the system (\ref{bems}) could
have one, two or $2^{M-1}$ solutions.  All solutions have the same global conserved charged.
They behave, however, differently for very large but finite values of $R$, and the solutions
which are not in the region $\makebox{Im}(x^{\pm})< 0$  show various signs of pathological behavior.
In particular, they might have complex finite $R$ correction to the energy, or the correction would
exceed the correction due to finite $R$ modifications of the Bethe equations thus making
the asymptotic Bethe ansatz inapplicable.


\section*{Acknowledgements}
We thank Marija Zamaklar for collaboration at the early stage of
the project. We are grateful to N. Beisert, N. Dorey, P. Dorey, D.
Hofman, R. Janik, M. de Leeuw, J. Maldacena, R. Roiban  and M.
Staudacher for valuable discussions.  The work of G.~A. was
supported in part by the RFBI grant N05-01-00758, by the grant
NSh-672.2006.1, by NWO grant 047017015 and by the INTAS contract
03-51-6346. The work of S.F. was supported in part by the Science
Foundation Ireland under Grant No. 07/RFP/PHYF104 and by a
one-month Max-Planck-Institut f\"ur Gravitationsphysik
Albert-Einstein-Institut grant. The work of G.~A. and S.~F.~was
supported in part by the EU-RTN network {\it Constituents,
Fundamental Forces and Symmetries of the Universe}
(MRTN-CT-2004-512194).

\section{ Appendices}
\subsection{Gauge-fixed Lagrangian.}\la{lagr}
The Lagrangian density of the gauge-fixed sigma-model in the
generalized $a$-gauge \cite{We,FPZ} can be written in the
following form \cite{KMRZ}
\bea
\label{Lgf} {\mathscr L} = - \frac{\sqrt{G_{\varphi \varphi}
G_{tt}}}{(1
     -a)^2 G_{\varphi \varphi} - a^2 G_{tt}}\sqrt{\cal W} +
\frac{a~G_{tt} + (1-a)~G_{\varphi \varphi}}{(1-a)^2 G_{\varphi
\varphi} - a^2
  G_{tt}}\, ,
\eea
where
\begin{multline}
{\cal W}\equiv 1 - \frac{(1-a)^2
  G_{\varphi \varphi} - a^2 G_{tt}}{2}
\Big[ \Big(1 + \frac{1}{G_{\varphi \varphi}
G_{tt}}\Big)\partial_{\a}
X \cdot \partial^{\a} X \\[1ex]
- \Big(1 - \frac{1}{G_{\varphi \varphi}
  G_{tt}}\Big)\left( \dot{X} \cdot \dot{X} + X' \cdot X'\right)\Big]\\[1ex]
 + \frac{((1-a)^2
  G_{\varphi \varphi} - a^2 G_{tt})^2}{2 G_{\varphi \varphi} G_{tt}}
\left( (\partial_{\a} X \cdot \partial^{\a} X)^2 -  (\partial_{\a}
X \cdot
\partial_{\b} X)^2 \right) \, . \nonumber
\end{multline}
Here $X=(y^i,z^i)$, where  $y^i$, $i=1,\ldots, 4$ are four fields
parametrizing five-sphere, while $z^i$ are fields parametrizing
four directions in ${\rm AdS}_5$. The fields $X$ in the Lagrangian
above are contracted with the help of the metric \bea
ds^2=-G_{tt}dt^2+G_{zz}dz^2+G_{\varphi\varphi}d\varphi^2+G_{yy}dy^2\,
. \nonumber \eea Here
\begin{eqnarray}
&&G_{tt} = \left(\frac{1 + z^2}{1-z^2}\right)^2 \, , \quad G_{zz}
=
  \frac{1}{\big(1 - z^2\big)^2}\, , \quad  G_{\varphi \varphi} = \left(\frac{1 -
  y^2}{1+y^2}\right)^2 \, , \quad G_{yy} = \frac{1}{\big(1 + y^2\big)^2} \, ,\nonumber
\end{eqnarray}
where we had used the notation $z^2 \equiv z^i z^i$ and $y^2
\equiv y^i y^i $.

\subsection{One-loop S-matrix}\la{oneloop}
Here we describe the properties of the ``one-loop" S-matrix which
is obtained from the S-matrix (\ref{Smatrix}) upon taking the
limit $g\to 0$. We will work in the elliptic parametrization
discussed in section 4.1. According to eq.(\ref{gtozero}), in this
limit Jacobi elliptic functions degenerate into the corresponding
trigonometric ones and we find the following trigonometric
S-matrix:

{\footnotesize
 \bea
 \nonumber
 S(z_1,z_2)=e^{-i(z_1-z_2)}\frac{\cot z_1-\cot z_2+2 i}{\cot z_1-\cot z_2-2
 i}& &
\Big(E_{1}^{1}\otimes E_{1}^{1}+E_{2}^{2}\otimes
E_{2}^{2}+E_{1}^{1}\otimes
E_{2}^{2}+E_{2}^{2}\otimes E_{1}^{1}\Big)\\
\nonumber - e^{-i(z_1-z_2)}\frac{2i } {\cot z_1-\cot z_2-2
i}&&\Big(E_{1}^{1}\otimes E_{2}^{2}+E_{2}^{2}\otimes
E_{1}^{1}-E_{1}^{2}\otimes E_{2}^{1}-E_{2}^{1}\otimes
E_{1}^{2}\Big)
\\
\nonumber  - &&\Big(E_{3}^{3}\otimes E_{3}^{3}+E_{4}^{4}\otimes
E_{4}^{4}+E_{3}^{3}\otimes
E_{4}^{4}+E_{4}^{4}\otimes E_{3}^{3}\Big) \\
\nonumber  -\frac{2i} {\cot z_1-\cot z_2-2 i} &&
\Big(E_{3}^{3}\otimes E_{4}^{4} +E_{4}^{4}\otimes
E_{3}^{3}-E_{3}^{4}\otimes E_{4}^{3}-E_{4}^{3}\otimes
E_{3}^{4}\Big)\\
\nonumber + e^{iz_2}\frac{\cot z_1-\cot z_2}{\cot z_1-\cot z_2-2
i}&&\Big(E_{1}^{1}\otimes E_{3}^{3}+E_{1}^{1}\otimes
E_{4}^{4}+E_{2}^{2}\otimes E_{3}^{3}+E_{2}^{2}\otimes
E_{4}^{4}\Big)\\
\nonumber  + e^{-iz_1}\frac{\cot z_1-\cot z_2}{\cot z_1-\cot z_2-2
i}&&\Big(E_{3}^{3}\otimes E_{1}^{1}+E_{4}^{4}\otimes
E_{1}^{1}+E_{3}^{3}\otimes E_{2}^{2}+E_{4}^{4}\otimes E_{2}^{2}\Big)\\
\nonumber +  e^{-i(z_1-z_2)}\frac{2i } {\cot z_1-\cot z_2-2
i}&&\Big(E_{1}^{3}\otimes E_{3}^{1}+E_{1}^{4}\otimes
E_{4}^{1}+E_{2}^{3}\otimes E_{3}^{2}+E_{2}^{4}\otimes E_{4}^{2}
\Big) \\ \nonumber
 +  e^{-i(z_1-z_2)}\frac{2i } {\cot z_1-\cot
z_2-2 i} &&\Big(E_{3}^{1}\otimes E_{1}^{3}+E_{4}^{1}\otimes
E_{1}^{4}+E_{3}^{2}\otimes E_{2}^{3}+E_{4}^{2}\otimes E_{2}^{4}
\Big)\, . \\ \label{S1loop}
\eea } The relations between the $z$-variable, momentum and the
rescaled rapidity $u\to gu$  transform in the limit $g\to 0$ into
\bea p=2z\, , ~~~~u=\cot z=\cot\frac{p}{2}\, . \eea Surprisingly
enough, this S-matrix cannot be written in the difference form,
i.e. as a function of one variable being the difference of a
properly introduced spectral parameter. By construction, this
S-matrix satisfies the usual Yang-Baxter equation \bea\label{YB}
S_{23}(z_2,z_3)S_{13}(z_1,z_3)S_{12}(z_1,z_2)=S_{12}(z_1,z_2)S_{13}(z_1,z_3)
S_{23}(z_2,z_3) \, ,\eea as one can also verify by direct
calculation. On the other hand, at one-loop there is another
``canonical" S-matrix which is a linear combination of the graded
identity and the usual permutation:
 \bea S^{\rm
can}_{12}=\frac{u_1-u_2}{u_1-u_2-2 i}I^g_{12}
+\frac{2i}{u_1-u_2-2i}P_{12} \, . \label{Schain}\eea This S-matrix
satisfies the same Yang-Baxter equation (\ref{YB}).

\smallskip

The results of \cite{AFZ} imply that the two one-loop S-matrices,
(\ref{S1loop}) and (\ref{Schain}) are related through the
following transformation \bea S^{\rm
can}(z_1,z_2)=U_2(z_1)\Big[V_1(z_1)V_2(z_2)S_{12}(z_1,z_2)V_1^{-1}(z_1)V_2^{
-1}(z_2)\Big]U_1^{-1}(z_2)\, , \eea where we have introduced the
diagonal matrices \bea U(z)&=&{\rm
diag}(e^{iz},e^{iz},1,1)\, ,\\
 V(z)&=&{\rm
diag}(e^{i\frac{z}{4}},e^{i\frac{z}{4}},e^{-i\frac{z}{4}},e^{-i\frac{z}{4}})
\, . \eea The transformation by $V$ is just a gauge transformation
which always preserves the Yang-Baxter equation. On the other
hand, transformation by $U$ is a twist, that generically
transforms the usual Yang-Baxter equation into the twisted one and
vice versa \cite{AFZ}. Indeed, $S^{\rm can}_{12}$ is nothing else
but the one-loop limit of the spin chain S-matrix \cite{B}; the
latter obeys the twisted Yang-Baxter equation \cite{AFZ}. Note
also that the twist $U$ does not belong to the symmetry group
${\rm SU }(2)\times {\rm SU}(2)$ of the string S-matrix.

\smallskip

To understand why at one loop the Yang-Baxter equation is
preserved under the twisting, we first write the Yang-Baxter
equation for $S^{\rm can}$ by using the relation\footnote{The
gauge transformation by the matrix $V$ decouples from the
Yang-Baxter equation.} (\ref{Schain}) \bea \nonumber
&&U_3(z_2)S_{23}U_2^{-1}(z_3)U_3(z_1)S_{13}U_1^{-1}(z_3)U_2(z_1)S_{12}U_1^{-
1}(z_2)=\\
&&~~~~~~~~=U_2(z_1)S_{12}U_1^{-1}(z_2)U_3(z_1)S_{13}U_1^{-1}(z_3)U_3(z_2)S_{
23}U_2^{-1}(z_3)\, , \eea which can be reshuffled as follows
\bea\nonumber
&&U_3(z_2)S_{23}U_2(z_1)U_3(z_1)S_{13}U_1^{-1}(z_3)U_2^{-1}(z_3)S_{12}U_1(z_
2)=\\
&&~~~~~~~~=U_2(z_1)U_3(z_1)S_{12}U_1^{-1}(z_2)S_{13}U_3(z_2)S_{23}U_1^{-1}(z
_3)U_2^{-1}(z_3)\, . \eea It is clear now that we will get the
usual Yang-Baxter equation for $S$ provided it obeys the following
relation \bea\label{1loopU} [S,U\otimes U]=0\, , \eea where $U$ is
an {\it arbitrary diagonal matrix}. One can easily verify that
both S-matrices, (\ref{S1loop}) and (\ref{Schain}), do indeed
satisfy this relation. At higher orders in $g$ the relation
(\ref{1loopU}) does not hold anymore. The corresponding
``all-loop" S-matrix (\ref{Smatrix}) satisfies only a weaker
condition \bea\label{allloopS} [S,G\otimes G]=0\, , ~~~~~~~~G\in
{\rm SU }(2)\times {\rm SU}(2)\, ,\eea which is nothing else but
the invariance condition for the string S-matrix. As a
consequence, the Yang-Baxter equation is preserved by the twist
transformation only at the one-loop order.

\smallskip

As a final remark, we note that it would be interesting to
understand how the derivation of the Hirota difference equations
for the canonical S-matrix \cite{Kazakov:2007fy} could be extended
to the ``twisted'' S-matrix (\ref{S1loop}). This might shed some
light on construction the fusion procedure for the all-loop
S-matrix (\ref{Smatrix}).

\subsection{BAE with nonperiodic fermions}\la{bae}
\subsubsection{Bethe wave function and the periodicity
conditions}\label{bae1} In any asymptotic domain $\Q$ with
$x_{\Q_1} \ll x_{\Q_2} \ll\cdots\ll x_{\Q_{N}}$ where $N\equiv
K^I$ and $\Q_1,\ldots ,\Q_N$ is a permutation of $1,2,\ldots, N$,
the wave function of $N$ particles with flavors $i_1,i_2,\ldots
,i_N$  can be written  as a superposition of plane waves with
momenta $p_1>p_2>\cdots >p_{N}$ \bea\la{wf} \Psi_{i_1\cdots
i_N}^\Q(x_1,\ldots ,x_{N}) =\sum_{\P}\, \cA^{\P|\Q}_{i_1\cdots
i_N}\, e^{i\, p_{\P}\cdot x_{\Q}}  \,, \eea where the sum runs
over all permutations of the momenta $p_i$. The scalar product
$p_{\P}\cdot x_{\Q}$ is defined as $p_{\P}\cdot x_{\Q} \equiv
\sum_{k=1}^N p_{_{\P_k}}x_{_{\Q_k}}$, and for any two permutations
$\P$ and $\Q$ it satisfies $p_{\P}\cdot x_{\Q}= p_{\P\Q^{-1}}\cdot
x_{\I}= \sum_{k=1}^N p_{_{(\P\Q^{-1})_k}}x_k$ where $\I$ is the
trivial permutation.

\smallskip

The amplitude $\cA^{\P|\Q}_{i_1\cdots i_N}$ is related to the probability of finding
the particle with the flavor $i_k$ (the $i_k$-th particle in what follows) carrying
the momentum $p_{_{(\P\Q^{-1})_k}}$ at the position $x_k$. That means that
the index $i_k$ is attached to the coordinate $x_k$. As a result the wave function (\ref{wf}) should
satisfy the following symmetry condition for any two indices $k,m$
\bea\nonumber
&&\Psi_{i_1\cdots i_k \cdots i_m\cdots i_N}^\Q(x_1,..., x_k,..., x_m,...,x_{N})=\nonumber \\
&&~~~~~~~~~~~~~~~~~~~=(-)^{\epsilon_{i_k}\epsilon_{i_m}}
\Psi_{i_1\cdots i_m \cdots i_k\cdots i_N}^{\P_{km}\Q}(x_1,...,
x_m,..., x_k,... ,x_{N})\, ,
 \la{wsym}\eea where $\P_{km}$ is the permutation of $k$ and $m$,
and $\epsilon_{i}=0$ if the $i$-th particle is boson and
$\epsilon_{i}=1$ if the $i$-th particle is fermion, that is one
takes the minus sign if both the $i_k$-th and $i_m$-th particles
are fermions, and the plus sign otherwise.

\smallskip

In  any two domains $\Q$ and $\overline{\Q}$  the amplitudes $\cA^{\P|\Q}_{i_1\cdots i_N}$ and $\cA^{\overline{\P}|\overline{\Q}}_{i_1\cdots i_N}$ of the same plane wave (that is $p_{\P}\cdot x_{\Q}  = p_{\overline{\P}}\cdot x_{\overline{\Q}} $) are related through the S-matrix. The relation can be easily found by representing the amplitudes as the following products of the ZF operators
\bea\la{am}
\cA^{\P|\Q}_{i_{1}\cdots i_{N}} \sim \pm A_{i_{\Q_1}}^\dagger(p_{_{\P_1}})\cdots A_{i_{\Q_N}}^\dagger(p_{_{\P_N}})\,,
\eea
and then by using the ZF algebra to relate the amplitudes in  the domains $\Q$ and $\overline{\Q}$. The $+/-$ sign in this formula is related to the even/odd number of permutations of fermions by the permutation $\Q$.
To understand the origin of this formula let us recall that the indices $i_k$ are attached to the coordinates $x_k$ which explains the order of $A_{i_{\Q_k}}^\dagger$. The dependence of $A_{i_{\Q_k}}^\dagger$ of the momentum follows from the coupling $p_{\P_k}x_{\Q_k}$ in the exponential of the wave function (\ref{wf}).

\smallskip

To proceed  it is convenient to use matrix notations. We introduce the simple permutation
$P_{12}=E^i_j\otimes E^j_i$ which permutes the spaces $V_1$ and $V_2$ but does not touch the momenta $p_i$ so that $S_{21}= P_{12}S(p_2,p_1)P_{12}$ , the graded permutation $P_{12}^g=(-1)^{\epsilon_i\epsilon_j}E^i_j\otimes E^j_i$,  and the graded two-particle S-matrix $S_{12}^g$ which can be written in the form $S_{12}^g=I_{12}^gS_{12}$
where $I_{12}^g = (-1)^{\epsilon_i\epsilon_j} E^i_i\otimes E^j_j$ is the graded identity.
We also define
$S_{21}^g= P_{12}S(p_2,p_1)P_{12}I_{12}^g = P_{12}S(p_2,p_1)P_{12}^g$ so that the unitarity condition $S_{12}^gS_{21}^g=I$ is fulfilled.

\smallskip

Then we multiply the wave function (\ref{wf}) and (\ref{am}) by
the tensor product of $N$ rows  $ E^{i_1}\otimes
E^{i_2}\otimes\cdots\otimes E^{i_N}\equiv \left(E^1E^2\cdots
E^N\right)^{i_1i_2\cdots i_N}$, and (\ref{wf}) takes the form
\bea\la{wfNp} \Psi^\Q(x_1,\ldots ,x_{N}) =\sum_{\P}\,
\cA^{\P|\Q}\, e^{i\, p_{\P}\cdot x_{\Q}}  \,, \eea where
\bea\nonumber \cA^{\P|\Q} \sim
A_{{\Q_1}}^\dagger(p_{_{\P_1}})\cdots
A_{{\Q_N}}^\dagger(p_{_{\P_N}}) I^g_{\Q}\,, \eea and the index
$\Q_k$ refers to the location of the row $E^{{\Q_k}}$, and
$I^g_{\Q}$ is the product of graded identities which can be found
by representing the permutation $\Q$ as a product of $Y$ simple
permutations $\P_{km}$: $Q = \P_{k_1m_1}\cdots\P_{k_Ym_Y} $, and
then $I^g_{\Q}=I^g_{k_1m_1}\cdots I^g_{k_Ym_Y} $.

\smallskip

Now the ZF algebra can be used to express the amplitudes $\cA^{\P|\Q} $ with $\Q_0\equiv \P^{-1}\Q$ fixed in terms of the amplitude $\cA^{\I |\Q_0}$. In particular the amplitudes $\cA^{\P|\P}$ are expressed in terms of the incoming amplitude
$\cA^{\I|\I} \sim A_{{1}}^\dagger(p_1)\cdots A_{{N}}^\dagger(p_N)$. The corresponding terms in the wave function can be used to derive the periodicity conditions.

\smallskip

To find the relations, it is convenient to represent
\bea\nonumber\begin{aligned}
&\cA^{\P|\Q} \sim A_{{\Q_1}}^\dagger(p_{_{\P_1}})\cdots A_{{\Q_N}}^\dagger(p_{_{\P_N}}) I^g_{\Q} =A_{{\P_1}}^\dagger(p_{_{\P_1}})\cdots A_{{\P_N}}^\dagger(p_{_{\P_N}})(\Q\P^{-1})_{1\cdots N} I^g_{\Q} \,,\\
&\cA^{\I |\Q_0} \sim A_1^\dagger(p_1)\cdots A_N^\dagger(p_N)\cdot
(\Q_0)_{1\cdots N}  I^g_{\Q_0}= A_1^\dagger(p_1)\cdots
A_N^\dagger(p_N) ( \P^{-1}\Q)_{1\cdots N} I^g_{\Q_0}\,,~~~~~~~~
\end{aligned}\eea where $ (\Q\P^{-1})_{1\cdots N}$ is the permutation matrix
that acting on the tensor product $E^{\P_1}\otimes\cdots\otimes
E^{\P_N}$ produces $E^{\Q_1}\otimes\cdots\otimes E^{\Q_N}$. Now we
use the ZF algebra to find the relation \bea \nonumber
&&\cA^{\P|\Q} = A_1^\dagger\cdots A_N^\dagger\cdot S_{\P_1\cdots\P_N}(p_{_{\P_1}},\ldots,p_{_{\P_N}})(\Q\P^{-1})_{1\cdots N} I^g_{\Q} =\\
\nonumber &&~~~~~~~~~~~~~~~~~~~\cA^{\I |\Q_0} \, I^g_{\Q_0}
(\Q^{-1}\P)_{1\cdots
N}S_{\P_1\cdots\P_N}(p_{_{\P_1}},\ldots,p_{_{\P_N}})(\Q\P^{-1})_{1\cdots
N} I^g_{\Q}\,,~~~~~~ \eea where
$S_{\P_1\cdots\P_N}(p_{_{\P_1}},\ldots,p_{_{\P_N}})$ is the
multi-particle S-matrix.

\smallskip

In particular, we find that \bea\nonumber \cA^{\P|\P} = \cA^{\I
|\I} \, S_{\P_1\cdots\P_N}(p_{_{\P_1}},\ldots,p_{_{\P_N}})
I^g_{\P}\equiv \cA^{\I |\I} \,
S^g_{\P_1\cdots\P_N}(p_{_{\P_1}},\ldots,p_{_{\P_N}}) \,,~~~~~~
\eea where $S^g_{\P_1\cdots\P_N}(p_{_{\P_1}},\ldots,p_{_{\P_N}})$
is the graded multi-particle S-matrix. Note that it is not a
product of two-particle graded S-matrices.

 \smallskip

This formula can be used to find the set of periodicity
conditions. We write the part of the wave function with the plane
wave $e^{ip_kx_k}$ \bea\nonumber &&\Psi(x_1,\ldots ,x_{N})
=\sum_{\P}\, \cA^{\P|\P}\, e^{i\, p_{\P}\cdot
x_{\P}}\theta(x_{\P_1}<\ldots < x_{\P_N}) \\\la{wfNp2}
&&~~~~~~~~=\cA^{\I |\I} \,  \sum_{\P}\,
S^g_{\P_1\cdots\P_N}(p_{_{\P_1}},\ldots,p_{_{\P_N}})\, e^{i\,
p_{\P}\cdot x_{\P}}\theta(x_{\P_1}<\ldots < x_{\P_N})  \,.~~~~~~~
\eea The periodicity conditions read \bea\nonumber \Psi(x_1,\ldots
,x_k=0,\ldots , x_{N}) = \Psi(x_1,\ldots ,x_k=L,\ldots ,
x_{N})W_k\,, \eea where the diagonal matrix $W$ is equal to the
identity matrix if the fermions are periodic, and it is
$W=(-1)^{\epsilon_i} E_i^i$ if the fermions are anti-periodic. For
the $\su(2|2)$ case we have $W=\Sigma =\makebox{diag}(1,1,-1,-1)$
for anti-periodic fermions.

\smallskip

By using eq.(\ref{wfNp2}), we get \bea\nonumber &&\Psi(x_1,\ldots
,x_k=0,\ldots , x_{N}) =\\\nonumber &&~~~~~~~~~ \cA^{\I |\I} \,
\sum_{\P:\P_1=p_k}\,
S^g_{k\P_2\cdots\P_N}(p_k,p_{_{\P_2}},\ldots,p_{_{\P_N}})\, e^{i\,
p_{\P}\cdot x_{\P}}\theta(x_{\P_2}<\ldots <
x_{\P_N})\,,~~~~~~~~~\\\nonumber &&\Psi(x_1,\ldots ,x_k=L,\ldots ,
x_{N}) =\\\nonumber &&e^{ip_kL} \cA^{\I |\I} \,
\sum_{\P:\P_N=p_k}\,
S^g_{\P_1\cdots\P_{N-1}k}(p_{_{\P_1}},\ldots,p_{_{\P_{N-1}}},p_k)W_k\,
e^{i\, p_{\P}\cdot x_{\P}}\theta(x_{\P_1}<\ldots <
x_{\P_{N-1}})\,,~~~~~~~~~ \eea Comparing the terms, we obtain
\bea\la{pc1} \cA^{\I |\I}
\left(S^g_{k\P_2\cdots\P_N}(p_k,p_{_{\P_2}},\ldots,p_{_{\P_N}}) -
e^{ip_kL}S^g_{\P_2\cdots\P_{N}k}(p_{_{\P_2}},\ldots,p_{_{\P_{N}}},p_k)W_k\right)=0\,.~~~
\eea To compute the S-matrices, we use their definitions
\bea\nonumber A_k^\dagger(p_k)
A_{{\P_2}}^\dagger(p_{_{\P_2}})\cdots
A_{{\P_N}}^\dagger(p_{_{\P_N}}) &=& A_1^\dagger\cdots
A_N^\dagger\cdot
S_{k\P_2\cdots\P_N}(p_k,p_{_{\P_2}},\ldots,p_{_{\P_N}})\, ,\\
\nonumber
 A_{{\P_2}}^\dagger(p_{_{\P_2}})\cdots A_{{\P_N}}^\dagger(p_{_{\P_N}})A_k^\dagger(p_k) &=&
A_1^\dagger\cdots A_N^\dagger\cdot
S_{\P_2\cdots\P_Nk}(p_{_{\P_2}},\ldots,p_{_{\P_N}},p_k)\,.~~~~~~
\eea Then we use the ZF algebra to order the product  $
A_{{\P_2}}^\dagger(p_{_{\P_2}})\cdots
A_{{\P_N}}^\dagger(p_{_{\P_N}})$ \bea \nonumber
 A_{{\P_2}}^\dagger(p_{_{\P_2}})\cdots A_{{\P_N}}^\dagger(p_{_{\P_N}}) = A_1^\dagger\cdots A_{k-1}^\dagger A_{k+1}^\dagger\cdots A_N^\dagger\cdot S_{\P_2\cdots\P_N}(p_{_{\P_2}},\ldots,p_{_{\P_N}})\,,
\eea and finally we get the multi-particle S-matrices \bea
\nonumber A_k^\dagger A_{{\P_2}}^\dagger(p_{_{\P_2}})\cdots
A_{{\P_N}}^\dagger(p_{_{\P_N}}) &=&
A_1^\dagger\cdots A_N^\dagger\cdot S_{k,k-1}S_{k,k-2}\cdots S_{k1} \cdot S_{\P_2\cdots\P_N}\\
\nonumber A_{{\P_2}}^\dagger(p_{_{\P_2}})\cdots
A_{{\P_N}}^\dagger(p_{_{\P_N}})A_k^\dagger &=& A_1^\dagger\cdots
A_N^\dagger\cdot S_{k+1,k}S_{k+2,k}\cdots S_{Nk} \cdot
S_{\P_2\cdots\P_N}\,.~~~~~~ \eea Thus, for $S_{\P_2\cdots\P_N}=1$
eq.(\ref{pc1}) takes the form \bea\la{pc2} \cA^{\I |\I}
\left(S_{k,k-1}\cdots S_{k1} I_{k,k-1}^g\cdots I_{k1}^g -
e^{ip_kL}S_{k+1,k}\cdots S_{Nk} I_{k+1,k}^g\cdots I_{Nk}^g
W_k\right)=0\, ~~~ \eea or, equivalently, \bea\la{pc3} \cA^{\I
|\I} \left( e^{ip_kL}-S_{k,k-1}\cdots S_{k1} I_{k,k-1}^g\cdots
I_{k1}^g W_k  I_{kN}^g\cdots I_{k,k+1}^gS_{kN}\cdots
S_{k,k+1}\right)=0\,.~~~ \eea It is possible to show that the same
equations follow if $S_{\P_2\cdots\P_N}\neq 1$ which uses the
identity $S_{km}I_{kn}^gI_{mn}^g= I_{kn}^gI_{mn}^g S_{km}$, and
also that the terms in the wave function with the plane wave
$e^{i\, p_{\P}\cdot x_{\Q}}$ lead to the same equations.

\smallskip

The consistency condition for the system of equations (\ref{pc3})
requires that  the matrices \bea \nonumber T_k\equiv
S_{k,k-1}\cdots S_{k1} I_{k,k-1}^g\cdots I_{k1}^g  W_k
I_{kN}^g\cdots I_{k,k+1}^gS_{kN}\cdots S_{k,k+1}~~~~~~~ \eea
mutually commute. Naturally,  we
 expect that the matrices $T_k$ should be related to
the monodromy matrix \bea\la{monT} T(p_A) =- \Str_A \, W_A
S^f_{AN}(p_A,p_N)S^f_{A,N-1}(p_A,p_{N-1})\cdots
S^f_{A1}(p_A,p_1)\,, \eea where $S^f_{jk}$ is the fermionic
$R$-operator defined, e.g.,  in eq.(102) of \cite{kor}. The
authors of \cite{kor} use index notations to define the operator.
It is more convenient, however, to use the matrix notations and
the usual convention for $S_{jk}$  to work with the operator. One
can check that it can be written in the following form
\bea\la{ferS} S^f_{jk}(p_j,p_k) =\left\{ \begin{array}{c}
I^g_{j\cdots N} I^g_{k\cdots N}\,I^g_{jk} S_{jk}(p_j,p_k)
\,I^g_{j\cdots N} I^g_{k\cdots N}\quad~~ \makebox{if}\ \ j<k\, ;\\
I^g_{j\cdots N} I^g_{k\cdots N}\,S_{jk}(p_j,p_k)I^g_{jk}\,
I^g_{j\cdots N} I^g_{k\cdots N}\quad~~  \makebox{if}\ \ j>k\, .
\end{array}\right.
\eea Here $I_{jk}^g$ is the graded identity and \bea\nonumber
I^g_{j\cdots N}\equiv I^g_{j,j+1}I^g_{j,j+2}\cdots I^g_{jN}\,.
\eea To prove the formula, one should use the following
representation for the graded projection operators $\widetilde
{E}_{j\a}^\b$ eq.(28) of \cite{kor} \bea\nonumber \widetilde
{E}_{j\a}^\b = I^g_{j\cdots N}\, E_{j\a}^\b\, I^g_{j\cdots N} \,.
\eea There are two natural choice for the index $A$ in
(\ref{monT}), that is $A=0$ or $A=N+1$. The choice leading to
$T_k$ appears to be $A=N+1>k$. Then we get \bea\nonumber
S^f_{Ak}(p_A,p_k) = I^g_{k\cdots N}\,S_{Ak}(p_A,p_k)I^g_{Ak}
\,I^g_{k\cdots N}\,. \eea Now we compute the following product
 \bea \nonumber  && S^f_{Ak}(p_A,p_k)S^f_{A,k-1}(p_A,p_{k-1}) =\\\nonumber
 &&
I^g_{k\cdots N}S_{Ak} \,I^g_{Ak} \,I^g_{k\cdots N} I^g_{k-1\cdots
N}\,S_{A,k-1} \,I^g_{A,k-1}\,I^g_{k-1\cdots N} =\\\nonumber &&
I^g_{k\cdots N} I^g_{k-1\cdots N}I^g_{k-1,k} \,S_{Ak}
\,I^g_{Ak}I^g_{k-1,k} S_{A,k-1} \,I^g_{A,k-1} \,I^g_{k\cdots
N}\,I^g_{k-1\cdots N} =\\ && I^g_{k\cdots N} I^g_{k-1\cdots
N}I^g_{k-1,k} \,S_{Ak} S_{A,k-1}\,I^g_{Ak} \,I^g_{A,k-1}
\,I^g_{k-1,k} \,I^g_{k\cdots N}\,I^g_{k-1\cdots N} \,,
\la{ss2}\eea where we used the identity \bea\nonumber
S_{A,k-1}(p_A,p_k) \, I^g_{k-1,k}\,I^g_{Ak} =
I^g_{k-1,k}\,I^g_{Ak} \, S_{A,k-1}(p_A,p_k) \,. \eea The following
generalization of the formula (\ref{ss2}) can be proven by using
the mathematical induction \bea\la{ssn}
&&S^f_{Ak}(p_A,p_k)S^f_{A,k-1}(p_A,p_{k-1})
\cdots S^f_{A,k-n}(p_A,p_{k-n}) =\nonumber \\
&&~~~~~~~=
I^g_{k\cdots N}\cdots  I^g_{k-n\cdots N}\, I^g_{k-1,k} I^g_{k-2\cdots k}\cdots I^g_{k-n\cdots k}\times\\
&&~~~~~~~\times S_{Ak}\cdots S_{A,k-n}\,I^g_{Ak} \cdots
I^g_{A,k-n} \,I^g_{k\cdots N} \cdots  I^g_{k-n\cdots N}\,
I^g_{k-1,k} I^g_{k-2\cdots k}\cdots I^g_{k-n\cdots k}\,.
\nonumber\eea To get the monodromy matrix, we  set $k=N$ and
$n=N-1$ in this formula, and using  the identity \bea\nonumber
I^g_{N-1\cdots N}\cdots  I^g_{1\cdots N}\, I^g_{N-1,N}
I^g_{N-2\cdots N}\cdots I^g_{1\cdots N} = I\,, \eea we find the
following drastic simplification \bea\nonumber &&T(p_A) =- \Str_A
\, W_A\, S_{AN}\cdots S_{A1}\,I^g_{A N}\cdots I^g_{A1} \,. \eea
Now we choose $p_A=p_k$ and use the fact that $S_{Ak}(p_k,p_k) =
-P_{Ak}$. Recalling that our goal is to show that $T(p_k)=T_k$, we
have \bea\nonumber &&T(p_k) = \Str_A \, W_A\, S_{AN}\cdots
S_{A,k+1}\,P_{Ak}\, S_{A,k-1}\cdots S_{A1}\,I^g_{A N}\cdots
I^g_{A1}\\\nonumber &&=\Str_A \, P_{Ak}\,W_k\, S_{kN}\cdots
S_{k,k+1}\cdot S_{A,k-1}\cdots S_{A1}\,I^g_{A N}\cdots I^g_{A1}
\\\nonumber
&&=\Str_A \, P_{Ak}\, S_{A,k-1}\cdots  S_{A1}\,I^g_{A, k-1}\cdots
I^g_{A1}\cdot W_k\, S_{kN}\cdots S_{k,k+1}\cdot I^g_{A N}\cdots
I^g_{Ak} \,.~~~~~~~~~ \eea Now we use that \bea\nonumber
S_{kN}\cdots S_{k,k+1}\cdot I^g_{A N}\cdots I^g_{Ak} =  I^g_{A
N}\cdots I^g_{Ak} \cdot S_{kN}\cdots S_{k,k+1} \eea to get
\bea\nonumber T(p_k) =  \Str_A \, S_{k,k-1}\cdots  S_{k1}\,I^g_{k,
k-1}\cdots I^g_{k1} I^g_{k N}\cdots I^g_{k,k+1}\, P_{Ak}I^g_{Ak}\,
W_k\, S_{kN}\cdots S_{k,k+1} \,.~~~~ \eea The supertrace can be
easily taken \bea\nonumber && \Str_A \, P_{Ak}I^g_{Ak} =\Tr_2
\left( (-1)^{\epsilon_c}I\otimes  E^c_c\right)\left(  E_b^a\otimes
E^b_a  \right)  \left((-1)^{\epsilon_f\epsilon_g} E^f_f\otimes
E_g^g\right) \\\nonumber &&= (-1)^{\epsilon_a + \epsilon_a^2}
E^a_a = I\,, \eea and, therefore, we show that $T(p_k)=T_k$. Since
$T(u)T(v)=T(v)T(u)$ for any $u$ and $v$, we have shown that  the
periodicity equations (\ref{pc3}) are consistent.\footnote{In
framework of the algebraic Bethe Ansatz twisted boundary
conditions for Hubbard-like models have been studied in
\cite{RM2}. }

\subsubsection{Two-particle Bethe equations}\label{bae2}

To see how the formulas of the previous subsection  work let us consider a two-particle wave function given by
\bea\la{wf2p}
\Psi_{ij}(x_1,x_2) =\left\{ \begin{array}{c}
\cA^{12|12}_{ij}\, e^{i\, p_{1}x_1+i\, p_{2}x_2}  +\cA^{21|12}_{ij}\, e^{i\, p_{2}x_{1}+i\, p_{1}x_{2}}\quad \makebox{if}\quad x_1<x_2\\
\cA^{12|21}_{ij}\, e^{i\, p_{1}x_{2}+i\, p_{2}x_{1}}  +\cA^{21|21}_{ij}\, e^{i\, p_{2}x_{2}+i\, p_{1}x_{1}}\quad \makebox{if}\quad x_2<x_1
\end{array}
\right.\,. \eea According to (\ref{am}), we can identify
\bea\nonumber \cA^{12|12}_{ij} \sim
A_{i}^\dagger(p_{1})A_{j}^\dagger(p_2)\,,\quad \cA^{21|21}_{ij}
\sim (-)^{\epsilon_{i}\epsilon_{j}}
A_{j}^\dagger(p_{2})A_{i}^\dagger(p_1)\,. \eea It is clear that
the amplitudes $\cA^{12|12}_{ij}$ and $ \cA^{21|21}_{ij}$
correspond to the in- and out-states, respectively. By using the
ZF algebra we find \bea\nonumber
A_{j}^\dagger(p_{2})A_{i}^\dagger(p_1) =
S_{ji}^{lk}(p_2,p_1)A_{k}^\dagger(p_{1})A_{l}^\dagger(p_2)\quad
\Rightarrow\quad \cA^{21|21}_{ij} = (-)^{\epsilon_{i}\epsilon_{j}}
S_{ji}^{lk}(p_2,p_1)\cA^{12|12}_{kl} \,.~~~~~~~ \eea In a similar
way we get \bea\nonumber \cA^{21|12}_{ij} \sim
A_{i}^\dagger(p_{2})A_{j}^\dagger(p_1)\,,\quad \cA^{12|21}_{ij}
\sim
(-)^{\epsilon_{i}\epsilon_{j}}A_{j}^\dagger(p_{1})A_{i}^\dagger(p_2)\,,
\eea and \bea\nonumber A_{j}^\dagger(p_{1})A_{i}^\dagger(p_2) =
S_{ji}^{lk}(p_1,p_2)A_{k}^\dagger(p_{2})A_{l}^\dagger(p_1)\quad
\Rightarrow\quad \cA^{12|21}_{ij} = (-)^{\epsilon_{i}\epsilon_{j}}
S_{ji}^{lk}(p_1,p_2)\cA^{21|12}_{kl} \,.~~~~~~~ \eea The
amplitudes $\cA^{12|12}_{ij}$ and $\cA^{21|12}_{ij}$ are not
independent. By the symmetry condition (\ref{wsym})  they are
related to each other as follows \bea\nonumber \cA^{12|12}_{ij} =
(-)^{\epsilon_{i}\epsilon_{j}} \cA^{12|21}_{ji} =
S_{ij}^{lk}(p_1,p_2)\cA^{21|12}_{kl} \quad \Rightarrow\quad
\cA^{21|12}_{ij} =  S_{ij}^{lk}(p_2,p_1)\cA^{12|12}_{kl}\,.~~~
\eea The wave function (\ref{wf2p})  can be written in the matrix
form by multiplying it by the row $E^i\otimes E^j$ and summing
over $i,j$. Then we get \bea\la{wf2p2} \Psi(x_1,x_2) =\left\{
\begin{array}{c}
\cA\left( e^{i\, p_{1}x_1+i\, p_{2}x_2}  +S_{21}P_{12}\, e^{i\, p_{2}x_{1}+i\, p_{1}x_{2}}\right)\quad \makebox{if}\quad x_1<x_2\\
\cA\left( P_{12}^g e^{i\, p_{1}x_{2}+i\, p_{2}x_{1}}  +S_{21}^g\, e^{i\, p_{2}x_{2}+i\, p_{1}x_{1}}\right)\quad \makebox{if}\quad x_2<x_1
\end{array}
\right.\,,
\eea
where we recall that  $P_{12}^g=(-1)^{\epsilon_i\epsilon_j}E^i_j\otimes E^j_i$ is the graded permutation and $S_{21}^g$ is the graded S-matrix $S_{21}^g= P_{12}S(p_2,p_1)P_{12}I_{12}^g = P_{12}S(p_2,p_1)P_{12}^g$.

\smallskip

The (quasi)-periodicity condition can be easily imposed
\bea\nonumber \Psi(x_1,x_2) = \Psi(x_1+L,x_2) W_1\,,\quad
\Psi(x_1,x_2) =\Psi(x_1,x_2-L)W_2\,,\quad  x_1<x_2\,,~~~ \eea
where the matrix $W$ is equal to $I$ for periodic boundary
conditions and to $\Sigma$ for anti-periodic boundary conditions
for fermions. By using the wave $e^{ip_kx_k}$ this leads to the
following equations \bea\nonumber \cA\left( 1 - e^{ip_1
L}S_{21}^gW_1\right) = 0\,,\quad \cA\left( 1 - e^{-ip_2
L}S_{21}^gW_2\right) = 0\,, \eea or by using the wave
$e^{ip_2x_2+ip_1x_2}$ to \bea\nonumber \cA\left( S_{21}P_{12} -
e^{ip_2 L}P_{12}^gW_1\right) = 0\,,\quad \cA\left( S_{21}P_{12} -
e^{-ip_1 L}P_{12}^gW_2\right) = 0 \,. \eea These two sets of the
periodicity conditions are obviously equivalent because
$P_{12}^gW_1=W_2P_{12}^g$. Let us also mention that the equations
are compatible if the matrices $W_1S_{12}^g$ and $S_{21}^gW_2$
commute, and this follows from unitarity $S_{12}^gS_{21}^g=I$ and
\bea\nonumber W_1W_2S_{12}^g=S_{12}^gW_1W_2\,. \eea

\smallskip
Let us now see how the nesting procedure works for the case of one
$A_1^\dagger$ boson and one $A_3^\dagger$  fermion. Consider the
system of equations \bea\begin{aligned}
\cA^{21|12}_{13}&=&S_{13}^{13}(p_2,p_1)\cA^{12|12}_{31}+
S_{13}^{31}(p_2,p_1)\cA^{12|12}_{13}\, , \\
\cA^{21|12}_{31}&=&S_{31}^{13}(p_2,p_1)\cA^{12|12}_{31}+
S_{31}^{31}(p_2,p_1)\cA^{12|12}_{13}\, .
\end{aligned}\label{1imp}\eea
Assuming that $S_{ij}^{kl}$ are matrix elements of the string
S-matrix ${\cal S}$, we get
 \bea\begin{aligned}
\cA^{21|12}_{13}&=S_0(p_2,p_1)\Big[\frac{x_1^--x_2^-}{x_1^+-x_2^-}e^{\frac{i}{2}p_1}\cA^{12|12}_{31}-
\frac{x_1^+-x_1^-}{x_1^+-x_2^-}\frac{\eta(p_2)}{\eta(p_1)}\frac{e^{\frac{i}{2}p_1}}{e^{\frac{i}{2}p_2}}\cA^{12|12}_{13}\Big]\, , \\
\cA^{21|12}_{31}&=S_0(p_2,p_1)\Big[\frac{x_2^+-x_2^-}{x_2^--x_1^+}\frac{\eta(p_1)}{\eta(p_2)}\cA^{12|12}_{31}+
\frac{x_2^+-x_1^+}{x_2^--x_1^+}e^{-\frac{i}{2}p_2}\cA^{12|12}_{13}\Big]\,
,\end{aligned}\label{1impurity}\eea where $S_0(p_1,p_2)$ is the
scalar prefactor.

\smallskip
For the amplitudes of interest the general Bethe equations \bea
e^{-ip_1L}\cA^{12|12}_{ij}=(-1)^{\epsilon\epsilon_i+\epsilon_i\epsilon_j}S_{ji}^{lk}(p_2,p_1)\cA^{12|12}_{kl}
\la{genBethe} \eea read as follows \bea
\begin{aligned}
e^{-ip_1L}\cA^{12|12}_{13}&=S_{31}^{13}(p_2,p_1)\cA^{12|12}_{31}+
S_{31}^{31}(p_2,p_1)\cA^{12|12}_{13}=\cA^{21|12}_{31}\, ,\\
e^{-ip_1L}\cA^{12|12}_{31}&=(-1)^{\epsilon}\Big[S_{13}^{13}(p_2,p_1)\cA^{12|12}_{31}+
S_{13}^{31}(p_2,p_1)\cA^{12|12}_{13}\Big]=(-1)^{\epsilon}
\cA^{21|12}_{13}\, ,
\end{aligned} \label{ba1im}\eea
where we have used eqs.(\ref{1imp}). Note that in
eq.(\ref{genBethe}) the multiplier $(-1)^{\epsilon\epsilon_i}$
takes into account the boundary conditions for fermions:
$\epsilon=0$ for periodic fermions and $\epsilon=1$ for
anti-periodic ones, respectively.

\smallskip

The system (\ref{1impurity}) can be solved in two different ways
depending on the choice of the first level vacuum \cite{B}. Below
we present both solutions.

\begin{itemize}
\item Regarding $A_1\ldots A_1$ as the first level vacuum, we
first choose the following ansatz \bea\begin{aligned}
\cA^{12|12}_{13}&=f(p_2)S(p_1)\, , ~~~~~~~~~~~~~~~~~~~~~\cA^{12|12}_{31}=f(p_1)\, ,\\
\cA^{21|12}_{13}&=S_{11}^{11}(p_2,p_1)f(p_1)S(p_2)\, ,
~~~~~~~~\cA^{21|12}_{31}=S_{11}^{11}(p_2,p_1)f(p_2)\, ,
\end{aligned}\label{ansatz1}
\eea where $S_{11}^{11}(p_1,p_2)$ is the corresponding element of
the string S-matrix. One can easily show that this ansatz indeed
solves the system (\ref{1impurity}) provided we take \bea\nonumber
f(p)=\frac{e^{i\frac{p}{2}}}{\eta(p)}\frac{x^+-x^-}{y-x^-}\, ,
~~~~~~~~ S(p)=e^{i\frac{p}{2}}\frac{y-x^-}{y-x^+}\, . \eea
According to eqs.(\ref{ansatz1}), the last formulae give \bea
\begin{aligned}
e^{-ip_1L}f(p_2)S(p_1)&={\cal S}_{11}^{11}(p_2,p_1)f(p_2)\, ,\\
e^{-ip_1L}f(p_1)&=(-1)^{\epsilon}{\cal
S}_{11}^{11}(p_2,p_1)f(p_1)S(p_2)\,
\end{aligned} \nonumber\eea
and we derive the corresponding Bethe equations \bea\nonumber
\begin{aligned}
& e^{ip_1L}
=S_{11}^{11}(p_1,p_2)S(p_1)\, ,\\
& (-1)^{\epsilon}=S(p_1)S(p_2)\, .
\end{aligned}
\eea

\item If we choose $A_3\ldots A_3$ as the first level vacuum, we
modify the ansatz for the corresponding amplitudes as
follows\bea\begin{aligned}
\cA^{12|12}_{13}&=f(p_1)\, , ~~~~~~~~~~~~~~~~~~~~~\cA^{12|12}_{31}=f(p_2)S(p_1)\, .\\
\cA^{21|12}_{13}&=S_{33}^{33}(p_2,p_1)f(p_2)\, ,
~~~~~~\cA^{21|12}_{31}=S_{33}^{33}(p_2,p_1)f(p_1)S(p_2)\, .
\end{aligned}\label{ansatz2}\eea
Note that $S_{33}^{33}(p_1,p_2)=-S_0(p_1,p_2)$. This time
satisfaction of eqs.(\ref{1impurity}) requires one to choose
$$
f(p)=\eta(p)e^{-i\frac{p}{2}}\frac{y}{y-x^-}\, , ~~~~~~~~
S(p)=-e^{-i\frac{p}{2}}\frac{y-x^+}{y-x^-}\, .
$$
The Bethe equations (\ref{ba1im}) read
 \bea
\begin{aligned}
 e^{-ip_1L}f(p_1)
&=S_{33}^{33}(p_2,p_1)f(p_1)S(p_2)\, ,\\
 e^{-ip_1L} f(p_2)S(p_1)&=(-1)^{\epsilon}S_{33}^{33}(p_2,p_1)f(p_2)  \, ,
\end{aligned}\nonumber
\eea and, therefore, we find \bea
\begin{aligned}
& e^{ip_1L}
=(-1)^{\epsilon}S_{33}^{33}(p_1,p_2)S(p_1)\equiv (-1)^{\epsilon}S_0(p_1,p_2)\frac{x^+_1-y}{x^-_1-y}e^{-i\frac{p_1}{2}}\, ,\\
& (-1)^{\epsilon}=S(p_1)S(p_2)\, .
\end{aligned}\nonumber
\eea

\end{itemize}
This completes consideration of our simple example illustrating
the dependence of the Bethe equations on the periodicity
conditions for fermions.

\subsection{Large/small $g$ expansions of solutions to the bound state equation}\la{expan}
The four general solutions of the bound state equation (\ref{4thpol}) are
\bea\la{solp}
e^q=\frac{ \left(\sqrt{g^2 \sin ^2\frac{p}{2}+1}+1\right)
   \left( \cos \frac{p}{2}\,\sqrt{g^2 \sin ^2\frac{p}{2}+1}\pm\sqrt{\cos
   ^2\frac{p}{2}-g^2 \sin ^4\frac{p}{2}}\right)}{g^2\sin^2\frac{p}{2}}\,,~~~~~
   \eea
\bea\la{solm}
e^q=\frac{ \left(\sqrt{g^2 \sin ^2\frac{p}{2}+1}-1\right)
   \left( \cos \frac{p}{2}\,\sqrt{g^2 \sin ^2\frac{p}{2}+1}\pm\sqrt{\cos
   ^2\frac{p}{2}-g^2 \sin ^4\frac{p}{2}}\right)}{g^2\sin^2\frac{p}{2}}\,,~~~~~
\eea
where only the first two solutions (\ref{solp}) correspond to states with positive energy.

\smallskip

The large $g$ dependence of $q$ of the bound state solutions with
momentum exceeding $p_{\rm cr}$ is obtained by expanding
(\ref{solp}) in powers of $1/g$ with the bound state momentum $p$
kept fixed\footnote{We assume here that $p\in (0,\pi)$.}
\bea\la{bexa1} q_\pm = {1\ov g \sin{p\ov 2}}-{1\ov 6 g^3
\sin^3{p\ov 2}}\pm i\left({p\ov 2} - {\cos{p\ov 2}\ov 2 g^2
\sin^3{p\ov 2}}\right) +{\cal O}({1\ov g^4})\,.~~~~~ \eea

\smallskip

To find the large $g$ dependence of $q$ of the  bound state
solutions with momentum smaller than $p_{\rm cr}$ one should take
into account that $p_{\rm cr}\to2/\sqrt g$ as  $g\to\infty$, and
therefore one should consider a bound state with momentum $p$ of
the order $1/\sqrt g$ and keep the product $p \sqrt g$ fixed in
the large $g$ expansion \bea\la{bexa2} q_\pm = 2  {1 \pm \sqrt{1 -
{p^4g^2\ov 16 }}\ov gp} - {4\ov 3g^3 p^3} \left(1 - {p^4g^2\ov 16}
\pm{1\ov \sqrt{1 - {p^4g^2\ov 16} }} \right)  \,.~~~~~ \eea

\smallskip

The small $g$ dependence of $q$ of the  bound state  solutions
with momentum smaller than $p_{\rm cr}$  is obtained by expanding
(\ref{solp}) at small $g$ with the bound state momentum $p$ kept
fixed \bea\la{bexa3p}
&&q_+ = -2\log g + \log {4\cos{p\ov 2}\ov  \sin^2{p\ov 2}} + {g^2\ov 8} (1 + 3\cos p)\tan^2 {p\ov 2} + {\cal O}( g^4)\,,\quad \\
\la{bexa3m}
&&q_- = -\log\cos{p\ov 2} + {g^2\ov 4} \sin^2{p\ov 2}\tan^2{p\ov 2} + {\cal O}( g^4) \,.~~~~~
\eea

\smallskip

To find the small $g$ dependence of $q$ of the  bound state
solutions with the momentum exceeding $p_{\rm cr}$, one should
take into account that $p_{\rm cr}\to\pi - 2g$ as  $g\to 0$. Then,
one can parametrize $p$ as follows \bea\nonumber p = \pi - 2 g
\cos \a\,, \eea and keep $\a$ fixed in the expansion. Then we get
\bea\la{bexa4} q_\pm = -\log {g\ov 2} + {g^2\ov 4}(2 + \cos2\a )
\pm i\left(\a + {g^2\ov 6}\sin2\a - {5g^2\ov 6}\cot\a \right) +
{\cal O}( g^4)\,.~~~~~~~~~~~ \eea





\begin{thebibliography}{20}

\bibitem{M}
  J.~M.~Maldacena,
  ``The large N limit of superconformal field theories and supergravity,''
  Adv.\ Theor.\ Math.\ Phys.\  {\bf 2} (1998) 231
  [Int.\ J.\ Theor.\ Phys.\  {\bf 38} (1999) 1113], hep-th/9711200.

\bibitem{MZ}
  J.~A.~Minahan and K.~Zarembo,
  ``The Bethe-ansatz for N = 4 super Yang-Mills,''
  JHEP {\bf 0303} (2003) 013, hep-th/0212208.

\bibitem{BPR}
I.~Bena, J.~Polchinski and R.~Roiban, ``Hidden symmetries of the
$\AdS$ superstring,'' Phys.\ Rev.\ D {\bf 69} (2004) 046002,
hep-th/0305116.

\bibitem{Kazakov:2004qf}
  V.~A.~Kazakov, A.~Marshakov, J.~A.~Minahan and K.~Zarembo,
  ``Classical / quantum integrability in AdS/CFT,''
  JHEP {\bf 0405} (2004) 024, hep-th/0402207.


  \bibitem{BDS}
  N.~Beisert, V.~Dippel and M.~Staudacher,
  ``A novel long range spin chain and planar N = 4 super Yang-Mills,''
  JHEP {\bf 0407} (2004) 075, hep-th/0405001.

\bibitem{AFS}
  G.~Arutyunov, S.~Frolov and M.~Staudacher,
   ``Bethe ansatz for quantum strings,''
  JHEP {\bf 0410}, 016 (2004), hep-th/0406256;

\bibitem{S}
  M.~Staudacher,
  ``The factorized S-matrix of CFT/AdS,''
  JHEP {\bf 0505} (2005) 054, hep-th/0412188;

\bibitem{BS}
  N.~Beisert and M.~Staudacher,
  ``Long-range PSU(2,2$|$4) Bethe ansaetze for gauge theory and
strings,'' hep-th/0504190.

\bibitem{B}
  N.~Beisert,
  ``The $\su(2|2)$ dynamic S-matrix,''
  hep-th/0511082.


 \bibitem{AFPZ}
  G.~Arutyunov, S.~Frolov, J.~Plefka and M.~Zamaklar,
  ``The off-shell symmetry algebra of the light-cone $\AdS$
  superstring,''
  hep-th/0609157.


  \bibitem{Bn}
  N.~Beisert,
  ``The Analytic Bethe Ansatz for a Chain with Centrally Extended
  $\su(2|2)$
  Symmetry,''
  J.\ Stat.\ Mech.\  {\bf 0701} (2007) P017, nlin.si/0610017.

\bibitem{Janik}
  R.~A.~Janik,
  ``The $\AdS$ superstring worldsheet S-matrix and crossing symmetry,''
  Phys.\ Rev.\ D {\bf 73} (2006) 086006,
  hep-th/0603038.


\bibitem{BHL}
  N.~Beisert, R.~Hernandez and E.~Lopez,
  ``A crossing-symmetric phase for $\ads$ strings,''
  hep-th/0609044.


\bibitem{BES}
  N.~Beisert, B.~Eden and M.~Staudacher,
  ``Transcendentality and crossing,''
  hep-th/0610251.


\bibitem{Beisert:2005cw}
  N.~Beisert and A.~A.~Tseytlin,
  ``On quantum corrections to spinning strings and Bethe equations,''
  Phys.\ Lett.\ B {\bf 629} (2005) 102, hep-th/0509084.


\bibitem{HL}
  R.~Hernandez and E.~Lopez,
  ``Quantum corrections to the string Bethe ansatz,''
  JHEP {\bf 0607} (2006) 004, hep-th/0603204.

\bibitem{Freyhult:2006vr}
  L.~Freyhult and C.~Kristjansen,
  ``A universality test of the quantum string Bethe ansatz,''
  Phys.\ Lett.\  B {\bf 638} (2006) 258, hep-th/0604069.


\bibitem{AF06}
  G.~Arutyunov and S.~Frolov,
  ``On $\ads$ string S-matrix,''
  Phys.\ Lett.\ B {\bf 639} (2006) 378, hep-th/0604043.



\bibitem{Bern}
  Z.~Bern, M.~Czakon, L.~J.~Dixon, D.~A.~Kosower and V.~A.~Smirnov,
  ``The four-loop planar amplitude and cusp anomalous dimension in maximally
  supersymmetric Yang-Mills theory,''
  hep-th/0610248.

\bibitem{Benna:2006nd}
  M.~K.~Benna, S.~Benvenuti, I.~R.~Klebanov and A.~Scardicchio,
  ``A test of the AdS/CFT correspondence using high-spin operators,''
  hep-th/0611135.

\bibitem{Kotikov:2006ts}
  A.~V.~Kotikov and L.~N.~Lipatov,
  ``On the highest transcendentality in N = 4 SUSY,''
  hep-th/0611204.

\bibitem{Alday:2007qf}
  L.~F.~Alday, G.~Arutyunov, M.~K.~Benna, B.~Eden and I.~R.~Klebanov,
  ``On the strong coupling scaling dimension of high spin operators,''
  JHEP {\bf 0704} (2007) 082, hep-th/0702028.

\bibitem{Kostov:2007kx}
  I.~Kostov, D.~Serban and D.~Volin,
  ``Strong coupling limit of Bethe ansatz equations,''
hep-th/0703031.

\bibitem{Casteill:2007ct}
  P.~Y.~Casteill and C.~Kristjansen,
  ``The Strong Coupling Limit of the Scaling Function from the Quantum   String
  Bethe Ansatz,'' hep-th/0705.0890.

\bibitem{Maldacena:2006rv}
  J.~Maldacena and I.~Swanson,
  ``Connecting giant magnons to the pp-wave: An interpolating limit of $\AdS$,''
  hep-th/0612079.

 \bibitem{KMRZ}
  T.~Klose, T.~McLoughlin, R.~Roiban and K.~Zarembo,
  ``Worldsheet scattering in $\AdS$,''
  hep-th/0611169.

\bibitem{Klose:2007wq}
  T.~Klose and K.~Zarembo,
  ``Reduced sigma-model on $\AdS$: one-loop scattering amplitudes,''
  JHEP {\bf 0702} (2007) 071, hep-th/0701240.


\bibitem{Gromov:2007cd}
  N.~Gromov and P.~Vieira,
  ``Constructing the AdS/CFT dressing factor,''
  hep-th/0703266.

\bibitem{Roiban:2007jf}
  R.~Roiban, A.~Tirziu and A.~A.~Tseytlin,
  ``Two-loop world-sheet corrections in $\AdS$ superstring,''
  JHEP {\bf 0707} (2007) 056, arXiv:0704.3638 [hep-th].

\bibitem{Klose:2007rz}
  T.~Klose, T.~McLoughlin, J.~A.~Minahan and K.~Zarembo,
  ``World-sheet scattering in $\AdS$ at two loops,''
hep-th/0704.3891.

\bibitem{Puletti:2007hq}
  V.~Giangreco Marotta Puletti, T.~Klose and O.~Ohlsson Sax,
  ``Factorized world-sheet scattering in near-flat $\AdS$,''
  arXiv:0707.2082 [hep-th].

\bibitem{JanikTBA}
  R.~A.~Janik and T.~Lukowski,
 ``Wrapping interactions at strong coupling -- the giant magnon,''
  arXiv:0708.2208 [hep-th].

\bibitem{Roiban:2007dq}
  R.~Roiban and A.~A.~Tseytlin,
 ``Strong-coupling expansion of cusp anomaly from quantum superstring,''
  arXiv:0709.0681 [hep-th].

\bibitem{Korch}
  B.~Basso, G.~P.~Korchemsky and J.~Kotanski,
  ``Cusp anomalous dimension in maximally supersymmetric Yang-Mills theory at
  strong coupling,''
  arXiv:0708.3933 [hep-th].


\bibitem{SZZ}
  S.~Schafer-Nameki, M.~Zamaklar and K.~Zarembo,
  ``Quantum corrections to spinning strings in $\AdS$ and Bethe  ansatz:
  A comparative study,''
  JHEP {\bf 0509} (2005) 051;
 S.~Schafer-Nameki and M.~Zamaklar,
  ``Stringy sums and corrections to the quantum string Bethe ansatz,''
  JHEP {\bf 0510} (2005) 044;
  S.~Schafer-Nameki,
   Exact expressions for quantum corrections to spinning strings,
   Phys.\ Lett.\  B {\bf 639}, 571 (2006), hep-th/0602214;
   S.~Schafer-Nameki, M.~Zamaklar and K.~Zarembo,
  ``How accurate is the quantum string Bethe ansatz?,''
  hep-th/0610250.



\bibitem{Kotikov:2007cy}
  A.~V.~Kotikov, L.~N.~Lipatov, A.~Rej, M.~Staudacher and V.~N.~Velizhanin,
  ``Dressing and Wrapping,''
  hep-th/0704.3586.


 \bibitem{HM}
  D.~M.~Hofman and J.~M.~Maldacena,
  ``Giant magnons,''
  hep-th/0604135.


\bibitem{magnon}
G.~Arutyunov, S.~Frolov and M.~Zamaklar, ``Finite-size effects
from giant magnons,'' hep-th/0606126.


\bibitem{AJK}
  J.~Ambjorn, R.~A.~Janik and C.~Kristjansen,
  ``Wrapping interactions and a new source of corrections to the spin-chain  /
  string duality,''
  Nucl.\ Phys.\  B {\bf 736} (2006) 288, hep-th/0510171.

\bibitem{Teschner:2007ng}
  J.~Teschner,
  ``On the spectrum of the Sinh-Gordon model in finite volume,''
hep-th/0702214.


\bibitem{za}
  A.~B.~Zamolodchikov,
  ``Thermodynamic Bethe ansatz in relativistic models. Scaling three state Potts and Lee-Yang models,''
  Nucl.\ Phys.\ B {\bf 342} (1990) 695.


\bibitem{Destri:1992qk}
  C.~Destri and H.~J.~de Vega,
  Phys.\ Rev.\ Lett.\  {\bf 69} (1992) 2313.


\bibitem{Bazhanov:1996aq}
  V.~V.~Bazhanov, S.~L.~Lukyanov and A.~B.~Zamolodchikov,
  Nucl.\ Phys.\  B {\bf 489} (1997) 487, hep-th/9607099.


\bibitem{Yang:1968rm}
  C.~N.~Yang and C.~P.~Yang,
  ``Thermodynamics of a one-dimensional system of bosons with repulsive
  delta-function interaction,''
  J.\ Math.\ Phys.\  {\bf 10} (1969) 1115.

\bibitem{Dorey:1996re}
  P.~Dorey and R.~Tateo,
  ``Excited states by analytic continuation of TBA equations,''
  Nucl.\ Phys.\  B {\bf 482} (1996) 639, hep-th/9607167.

\bibitem{Martins:1991hw}
  M.~J.~Martins,
  ``Complex excitations in the thermodynamic Bethe ansatz approach,''
  Phys.\ Rev.\ Lett.\  {\bf 67} (1991) 419.



\bibitem{AFZ}
  G.~Arutyunov, S.~Frolov and M.~Zamaklar,
  ``The Zamolodchikov-Faddeev algebra for $\AdS$ superstring,''
  JHEP {\bf 0704} (2007) 002, hep-th/0612229.



\bibitem{Zam}
  A.~B.~Zamolodchikov and A.~B.~Zamolodchikov,
  ``Factorized S-matrices in two dimensions as the exact solutions of  certain
  relativistic quantum field models,''
  Annals Phys.\  {\bf 120} (1979) 253.

\bibitem{Fad}
L.~D.~Faddeev, Sov.Sci.Rev.Math.Phys. 1C(1980)~107.



  \bibitem{D}
  N.~Dorey,
  ``Magnon bound states and the AdS/CFT correspondence,''
  J.\ Phys.\ A  {\bf 39} (2006) 13119, hep-th/0604175.

  \bibitem{CDO}
  H.~Y.~Chen, N.~Dorey and K.~Okamura,
  ``The asymptotic spectrum of the N = 4 super Yang-Mills spin chain,''
  JHEP {\bf 0703} (2007) 005, hep-th/0610295.


  \bibitem{DHM}
  N.~Dorey, D.~M.~Hofman and J.~Maldacena,
  ``On the singularities of the magnon S-matrix,''
  hep-th/0703104.

\bibitem{Dorey2}
  H.~Y.~Chen, N.~Dorey and K.~Okamura,
  ``On the scattering of magnon boundstates,''
  JHEP {\bf 0611} (2006) 035, hep-th/0608047.

\bibitem{Roiban:2006gs}
  R.~Roiban,
  ``Magnon bound-state scattering in gauge and string theory,''
  JHEP {\bf 0704} (2007) 048,
  hep-th/0608049.

\bibitem{SF}
L.D.~Faddeev and A.A.~Slavnov, ``Gauge fields: an introduction to
quantum theory,'' 1991, Addison-Wesley PC, Redwood, CA, US, 217 pp

\bibitem{We}
  G.~Arutyunov and S.~Frolov,
  ``Uniform light-cone gauge for strings in $\AdS$: Solving
  $\su(1|1)$
  sector,''
  JHEP {\bf 0601} (2006) 055, hep-th/0510208.


\bibitem{FPZ}
  S.~Frolov, J.~Plefka and M.~Zamaklar,
  ``The $\AdS$ superstring in light-cone gauge and its Bethe
  equations,''
  J.\ Phys.\ A  {\bf 39} (2006) 13037, hep-th/0603008.

\bibitem{Witten}
  E.~Witten,
  ``Constraints On Supersymmetry Breaking,''
  \mbox{Nucl.\ Phys.\  B {\bf 202} (1982) 253.}




\bibitem{Arutyunov:2004yx}
  G.~Arutyunov and S.~Frolov,
  ``Integrable Hamiltonian for classical strings on $\AdS$,''
  JHEP {\bf 0502} (2005) 059, hep-th/0411089.


  \bibitem{MM}
  M.~J.~Martins and C.~S.~Melo,
  ``The Bethe ansatz approach for factorizable centrally extended S-matrices,''
  Nucl.\ Phys.\  B {\bf 785} (2007) 246,
  hep-th/0703086.


 \bibitem{Shastry} B.~S.~Shastry, ``Exact integrability of the
 one-dimensional Hubburd-model", Phys.Rev.Lett {\rm 56} (1986)
 2453.

  \bibitem{RM1}
  P.~B.~Ramos and M.~J.~Martins,
  ``Algebraic Bethe Ansatz Approach For The One-Dimensional Hubbard Model,''
  J.\ Phys.\ A  {\bf 30} (1997) L195, hep-th/9605141.

\bibitem{RM2}
P.~B.~Ramos and M.~J.~Martins,``The quantum inverse scattering
method for Hubbard-like models", Nucl.\ Phys.\  B {\bf 522} [FS]
(1998) 413-470.



\bibitem{EFGKK} F.~H.~L.~Essler, H.~Frahm, F.~G\"ohmann,
A.~Kl\"umper and V.~Korepin, ``The one-dimensional Hubbard model",
{\it Cambridge University Press}, 2005.

\bibitem{Gomez:2006va}
  C.~Gomez and R.~Hernandez,
  ``The magnon kinematics of the AdS/CFT correspondence,''
  hep-th/0608029;
  J.~Plefka, F.~Spill and A.~Torrielli,
  ``On the Hopf algebra structure of the AdS/CFT S-matrix,''
  hep-th/0608038.

\bibitem{sualg}
  A.~Torrielli,
  ``Classical r-matrix of the $\su(2|2)$
 SYM spin-chain,''
  Phys.\ Rev.\  D {\bf 75} (2007) 105020,
  hep-th/0701281;
N.~Beisert,
  ``The S-Matrix of AdS/CFT and Yangian Symmetry,''
  PoS {\bf SOLVAY} (2006) 002
  [arXiv:0704.0400 [nlin.SI]];
 T.~Matsumoto, S.~Moriyama and A.~Torrielli,
  ``A Secret Symmetry of the AdS/CFT S-matrix,''
  JHEP {\bf 0709} (2007) 099,
  [arXiv:0708.1285 [hep-th]];
  N.~Beisert and F.~Spill,
  ``The Classical r-matrix of AdS/CFT and its Lie Bialgebra Structure,''
  arXiv:0708.1762 [hep-th].


  \bibitem{Le}
  M.~de Leeuw,
  ``Coordinate Bethe Ansatz for the String S-Matrix,''
  hep-th/0705.2369.


 \bibitem{AAF1}
  L.~F.~Alday, G.~Arutyunov and S.~Frolov,
 ``New integrable system of 2dim fermions from strings on $\AdS$,''
  JHEP {\bf 0601} (2006) 078, hep-th/0508140;
 ``Green-Schwarz strings in TsT-transformed backgrounds,''
  JHEP {\bf 0606} (2006) 018, hep-th/0512253.

\bibitem{Faddeev:1996iy}
  L.~D.~Faddeev,
  ``How Algebraic Bethe Ansatz works for integrable model,''
  hep-th/9605187.

\bibitem{St}
  A.~Rej, M.~Staudacher and S.~Zieme,
 ``Nesting and dressing,''  J.\ Stat.\ Mech.\  {\bf 0708} (2007) P08006, hep-th/0702151.

  \bibitem{st2}
  M.~Beccaria and V.~Forini,
  ``Anomalous dimensions of finite size field strength operators in N=4 SYM,''
  arXiv:0710.0217 [hep-th].

\bibitem{Kazakov:2007fy}
  V.~Kazakov, A.~Sorin and A.~Zabrodin,
  ``Supersymmetric Bethe ansatz and Baxter equations from discrete Hirota
  dynamics,''
hep-th/0703147.

\bibitem{kor}
  F.~G\"ohmann and V.~E.~Korepin,
  ``Solution of the quantum inverse problem,''
  J.\ Phys.\ A  {\bf 33} (2000) 1199, hep-th/9910253.


\end{thebibliography}
\end{document}